\documentclass[aps,pra,superscriptaddress,numerical,showkeys,floatfix,longbibliography,preprint,footinbib]{revtex4-2}
\usepackage{epsfig,amsmath,amssymb,txfonts,hyperref,multirow, bbold} 
\usepackage[version=4]{mhchem}
\usepackage{xcolor}
\usepackage[normalem]{ulem}
\usepackage[figuresleft]{rotating}
\usepackage{bm}
\usepackage{pdfpages} 

\makeatletter 
\AtBeginDocument{\let\LS@rot\@undefined}
\makeatother

\newcommand{\be}{\begin{equation}}
\newcommand{\ee}{\end{equation}}

\newcommand{\Fig}[1]{Fig.~\ref{fig:#1}}

\newcommand{\Tab}[1]{Table~\ref{tab:#1}}

\newcommand{\Eqn}[1]{Eqn.~\ref{eqn:#1}}

\newlength{\wholefigwidth}
\newlength{\halffigwidth}
\begin{document}

\title{Explainable Machine Learning for Hydrogen Diffusion in Metals and Random Binary Alloys}
\author{Grace M. Lu}
\affiliation{Department of Materials Science and Engineering, University of Illinois at Urbana-Champaign, Urbana, Illinois 61801, USA}
\author{Matthew Witman}
\affiliation{Sandia National Laboratories, Livermore, California 94551, USA}
\author{Sapan Agarwal}
\affiliation{Sandia National Laboratories, Livermore, California 94551, USA}
\author{Vitalie Stavila}
\affiliation{Sandia National Laboratories, Livermore, California 94551, USA}
\author{Dallas R. Trinkle}
\email{dtrinkle@illinois.edu}
\affiliation{Department of Materials Science and Engineering, University of Illinois at Urbana-Champaign, Urbana, Illinois 61801, USA}

\date{\today}
\begin{abstract}
Hydrogen diffusion in metals and alloys plays an important role in the discovery of new materials for fuel cell and energy storage technology. While analytic models use hand-selected features that have clear physical ties to hydrogen diffusion, they often lack accuracy when making quantitative predictions. Machine learning models are capable of making accurate predictions, but their inner workings are obscured, rendering it unclear which physical features are truly important. To develop interpretable machine learning models to predict the activation energies of hydrogen diffusion in metals and random binary alloys, we create a database for physical and chemical properties of the species and use it to fit six machine learning models. Our models achieve root-mean-squared-errors between 98--119 meV on the testing data and accurately predict that elemental Ru has a large activation energy, while elemental Cr and Fe have small activation energies.
By analyzing the feature importances of these fitted models, we identify relevant physical properties for predicting hydrogen diffusivity. While metrics for measuring the individual feature importances for machine learning models exist, correlations between the features lead to disagreement between models and limit the conclusions that can be drawn.
Instead grouped feature importances, formed by combining the features via their correlations, agree across the six models and reveal that the two groups containing the packing factor and electronic specific heat are particularly significant for predicting hydrogen diffusion in metals and random binary alloys. 
This framework allows us to interpret machine learning models and enables rapid screening of new materials with the desired rates of hydrogen diffusion.
\end{abstract}


\maketitle

\section{Introduction}
\label{sec:intro}
At present, there is a critical need for sustainable carbon-free energy storage technologies that can address the intermittent and “non-dispatchable” 
character of renewable energy resources. Hydrogen-based technologies satisfy such requirements and are attainable solutions with a potential for zero-carbon emissions. Hydrogen has the highest gravimetric energy density (121 MJ kg\textsuperscript{--1}) of any fuel, is naturally abundant and can be converted into electrical energy via high-efficiency fuel cells \cite{Zuttel2010}. A major hurdle for the use of hydrogen as a clean and efficient energy carrier is developing ways to store and transport it safely and economically. Hydrogen can be stored in gas or liquid form inside of high-pressure cylinders \cite{Irani2002}, primarily for use as a fuel for vehicles, and new materials for these cylinders require slow diffusion rates of hydrogen to both limit leakage and the effects of hydrogen embrittlement  \cite{Song2013}. Currently, these tanks are often made of aluminum and its alloys or steels \cite{Irani2002}. Hydrogen can also be chemically stored in metal hydrides in solid form \cite{schapbach2001, Schneemann2018}, but efficient hydrogen absorption and desorption requires fast hydrogen transport through the bulk metal. Two common options are Mg hydrides \cite{Sakintuna2007, fernandez2010, webb2015, webb2019}, which have high hydrogen storage capabilities but extremely slow kinetics, and Pd hydrides, which have been highly studied because they absorb hydrogen at room temperature \cite{aicheng2011}. For all of these storage/transportation options, a quantitative understanding of the interactions of hydrogen with various containment vessels and storage media is essential in the ongoing efforts aimed at rational design of new materials.

When hydrogen diffuses through metals and metal alloys, hydrogen atoms migrate through the lattice of a host material. At the initial stage, when hydrogen is in contact with a metal surface, dihydrogen molecules absorb to the surface through weak van der Waals interactions, with a low absorption energy of 3--5 kJ mol\textsuperscript{--1} 
 \cite{Kirchheim2014}. The second step is chemical absorption, which occurs with a relatively high enthalpy of tens or hundreds of kJ mol\textsuperscript{--1}, depending on the metal/alloy \cite{Wipf2001}. Following chemical absorption, hydrogen atoms diffuse sub-surface due to large concentration gradients. Hydrogen in crystalline metals and alloys diffuses through the lattice through interstitial diffusion, in which hydrogen atoms move through the interstitial sites of the host metal. Hydrogen atoms are precisely located in interstitials and in the diffusion process they pass from one interstitial site to another. In metal lattices, hydrogen atoms tend to be preferentially located at BCC tetrahedral or FCC octahedral interstitial sites \cite{Kirchheim2014}. 

Traditional experimental and computational modeling approaches to determining hydrogen diffusion in metals and metal alloys often consume tremendous time and resources \cite{Kamakoti2003, Baykara2004, Jiang2004,Bhatia2005,Wimmer2008,Duan2010, Connetable2011,Liu2012,Lu2013, Fernandez2015,Tafen2015,Han2016, Yang2016, Liu2017, Liu2018,Oliveira2019,Yang2019, Connetable2019, Onwudinanti2020}. Random alloys, in particular, are costly to study using density-functional theory (DFT) calculations because they require multiple calculations to thoroughly sample all of the possible configurations. To streamline the discovery of new materials for hydrogen containment or transportation, accurate analytic models can be used to connect hydrogen diffusion to other material properties that are easier to measure experimentally or using DFT. These models can be used for rapid screening of new metals and alloys. Unfortunately, while analytic models for hydrogen diffusion exist, they lack quantitative accuracy. An elastic model for BCC metals by Ferro \cite{ferro1957} calculated the expected energy needed to distort the lattice sufficiently for an atom to travel between neighboring interstitial sites and predicted that the activation energy $Q$ could be described by:

\be\label{eqn:elastic} Q=1.3 NAGa(d-\lambda)^2\left(1-\frac{\beta n T_D}{10 T_m}\right), \ee
where $N$ is the number of atoms, $A$ is a constant that converts the units from mechanical energy to heat, $G$ is the shear modulus, $a$ is the lattice constant of the host, $d$ is the diameter of the interstitial atom, and $\lambda$ is the diameter of the interstitial cavity. The last term in the parenthesis in \Eqn{elastic} describes a correction factor for the elastic modulus where $n$ is the number of atoms in the host that are displaced, $T_D$ is the Debye temperature, $T_m$ is the melting temperature, and $\beta=\frac{T_m}{G}\frac{dG}{dT}$. Unfortunately, this simple elastic description was found to have at best only qualitative agreement with experimental activation energies \cite{mcneil1965}. A later model by Flynn and Stoneham \cite{Flynn1970} used quantum theory to calculate the transition rate of localized eigenstates between interstitial sites where $Q$ is calculated by:

\be\label{eqn:flynn} Q = \frac{M\omega_D^2 d^2}{360}\left(\frac{1+\nu}{1-\nu}\right)^2\left(\frac{\delta V}{\Omega}\right)^2 \Phi(q_m d,\eta),\ee

\noindent where $M$ is the mass of a host atom, $\omega_D$ is the Debye frequency, $d$ is the jump distance, $\nu$ is Poisson's ratio, $\delta V$ is the lattice dilatation caused by each defect, and $\Omega$ is the impurity atomic volume. The function $\Phi$ is specified by the parameters $q_m d$ ($q_m$ is the radius of the Debye sphere in reciprocal space) and $\eta$, which is the fraction of the volume change from isotropic dilatation. While this model worked well for BCC metals, the activation energies for FCC metals were underestimated 
because they only considered the lattice distortions at the interstitial sites and ignored any distortions that may occur during the transition.

Due to the lack of accurate theoretical models for material transport properties, machine learning approaches have been introduced recently and have successfully been used to predict light element diffusion in FCC, BCC, and HCP metals \cite{Zeng2018}, solute diffusion in FCC metals \cite{Wu2017}, and solute diffusion in FCC, BCC, and HCP metals \cite{Wei2021, He2020}. However, existing machine-learning methods fail to accurately predict hydrogen diffusion due to the scarcity of data and the lack of a suitable set of features to capture the complex behavior of hydrogen in metals. Machine learning methods have instead been used in conjunction with kinetic Monte Carlo \cite{Zhou2022} and path-integral molecular dynamics \cite{Kimizuka2022} to calculate hydrogen diffusion activation energies for a single material. 

The first key requirement for successful machine learning-based materials modeling is target property data in sufficient quantity and accuracy. For other applications involving hydrogen-metal interactions, e.g., storage or compression with alloys, it can become an intractable task to obtain the figures of merit (hydrogen plateau pressures) with first-principles accuracy at the scale needed for training ML models; consequently, several studies have relied on a database of accumulated experimental metal hydride data to train thermodynamic models and facilitate discovery of promising alloys \cite{Hattrick-Simpers2018,Witman2020,Witman2021}. Therefore, the first key contribution of this work is the construction of a database containing activation energies for hydrogen diffusion in metals and binary alloys that we extracted from previous work. This will provide a useful community tool moving forward to build upon the current work. 

A second key requirement is generating a sufficiently representative featurization that captures the key physical, chemical, or electronic material characteristics needed for successful learning of the target property. One powerful yet simple approach is the development of so-called compositional ML models, where all features are derived purely from a composition's element's properties and molar fractions, which are combined through simple mathematical operations.\cite{Ward2018} 
Such a featurization combined with standard ML approach can accurately model properties like formation energies of inorganic compounds,\cite{Ward2016} but are not always sufficient. Other more complex, crystal-structure based featurization strategies like graph neural networks\cite{Xie2018} are powerful but more difficult to implement when crystal structures inherently contain substitutional disorder, e.g., alloys vs.~intermetallic compounds. Therefore, the second key contribution of this work is to augment this typical compositional ML feature vector with domain-specific features (including manually selected crystal structure-related information) that are needed to improve predictions of hydrogen diffusion activation energies.

After meeting these two requirements, we investigate the performance of a variety of model types ($k$-nearest neighbors, Gaussian process, random forest, gradient boosting, positive ElasticNet, and Bayesian Ridge) and measure their performance with different train-test split strategies to provide a comprehensive outlook on predictive capabilities. Using our final models, we calculate importances of each feature. In this paper, to both improve agreement between the models and allow for interpretability, we group our features into correlated feature groups. Thus, as the final key contribution in this work, we obtain critical physical insights on hydrogen diffusion in metals and binary alloys using these correlated feature groups. Within this framework, future additions/improvements to our database, featurization strategies, model development, and interpretability analyses will provide new tools moving forward to understand and design materials with targeted hydrogen diffusion properties. 

\section{Methods}
\label{sec:methods}

\subsection{Database and Features}

\begin{figure*}[htb]
 \centering\includegraphics[width=\halffigwidth]{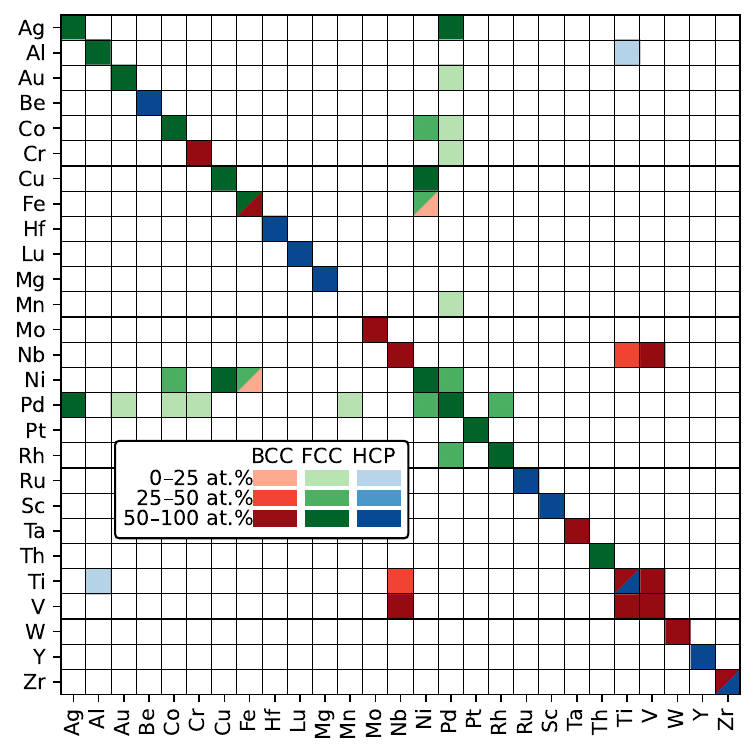}
 \caption{Elements and binary solid solutions included in the hydrogen diffusion database. The saturation shows the range of compositions present in the database, and entries with two colors have data corresponding to different crystal structures. Data for Mn are only present in Pd-Mn alloys; no elemental data are available. All of the data come from experimental measurements, except elemental Rh and Ru, which are from DFT. The majority of alloy data are Pd-based alloys.}
 \label{fig:database}
\end{figure*}

The database used in this work to fit machine learning models for hydrogen diffusion consists of 28 metals and 14 different elemental combinations of binary alloys, as shown in \Fig{database}. For our database entries, we choose to use only data measured experimentally at high temperatures above 250 K to reduce quantum mechanical effects like tunneling and to exclude any non-Arrhenius behavior. Because experimental data was chosen, small concentrations of crystallographic defects are implicitly included. Since experimental data are unavailable for Rh and Ru, we include the activation energies for these two metals from density functional theory (DFT) calculations to improve coverage of high activation barriers. The DFT calculations included have used techniques that have agreed with previous experimental measurements done for different metals, even though these calculations were done for perfect crystals with no defects. Since the activation energies are significantly larger than the other metals, it may be possible that there is some deviation from the experimental results due to the lack of crystallographic defects. No other theoretical calculations were included.
For metals with more than one source, we use the average value of the reported activation energies. The alloy compositions in our database, which make up 72 total entries, span a range of 37 atomic percent for each elemental combination on average, with a maximum of 89 atomic percent for TiV. While all other alloys have at least two compositions, there is only one for TiAl (\ce{Ti88Al12}). Pd-based binary alloys dominate the alloy data with 42 Pd alloys in the dataset, of which 40 contain at least 50 atomic \% Pd. The next most prevalent elements are Fe, Ni, and Ti, which have 15 entries each. Therefore, the database is biased towards transition state metals and our models, while still applicable to other metals, should have smaller errors for the transition state metals. 

\Tab{metals} lists the activation energies for the pure metals organized by crystal structure.
 The activation energies for the metals depend heavily on the crystal structures, and the BCC metals have 0.24 eV smaller activation energies on average than the close-packed FCC and HCP metals. For HCP metals where the reported diffusion was anisotropic, we use the effective activation energy by taking an average over the activation energies in directions parallel and perpendicular to the basal plane. Note that Fe has activation energies for hydrogen diffusion in both BCC and FCC crystal structures, while Ti and Zr have activation energies in both BCC and HCP crystal structures. The mean activation energy is 0.31 eV with a standard deviation of 0.15 eV. While the pure elements shown in \Tab{metals} have a larger mean of 0.38 eV and a larger standard deviation of 0.20 eV, the opposite is true for the binary alloys, which have a mean of 0.29 eV and a standard deviation of 0.11 eV, so the alloys not only cover a smaller composition range, but also a smaller range of activation energies. 

\begin{table}[htb]
\centering
\caption{Activation energy for hydrogen diffusion in pure metals. The values are sorted in order of increasing activation energy. In the case of multiple measurements of activation energy, we provide the average value. As mentioned in \Fig{database}, both Rh and Ru values come from density-functional theory.}
\label{tab:metals}
\begin{tabular}{ll@{\quad}ll@{\quad}ll}
\hline\hline
BCC & $Q$ (eV) & FCC & $Q$ (eV) & HCP & $Q$ (eV) \\ \hline 
Cr \nocite{Stover1986, Jost1940, Toda1958, Volkl1971, Holleck1970, Nishimura1999, Ebisuzaki1968, Katsuta1979, Kearns1972, Gelezunas1963, Gulbransen1954, Someno1960, Han1987, Raczynski1978,Quick1978, Ishikawa1985, Kunz1983, Naito1990, Bell1983, Herro1982, Qi1983, Schaumann1970, Bauer1978, Hampele1989, Heidemann1976, WasilewskiRJ;Kehl1954, Kunz1983, Katsuta1982, Tanabe1992, Katsuta1964, Katsuta1983, Zvezdin1968, Caskey1977, Magnusson2017, Hagi1986,Katz1971, Yamakawa1979, Cummings1987, Atrens1980, Furuya1984, Yamanishi1983} & 0.020 \cite{Stover1986} & Pd & 0.236 \cite{Jost1940, Toda1958,  Volkl1971, Holleck1970} & Mg & 0.250 \cite{Nishimura1999} \\
V & 0.046 \cite{Herro1982, Schaumann1970, Qi1983} & Pt & 0.264 \cite{Ebisuzaki1968, Katsuta1979} & Zr & 0.427\cite{Kearns1972, Gelezunas1963, Gulbransen1954, Someno1960} \\
Fe & 0.069 \cite{Raczynski1978,Quick1978} & Au & 0.273 \cite{Zvezdin1968, Ishikawa1985} & Hf & 0.464 \cite{Kunz1983, Naito1990} \\
Nb & 0.098 \cite{Herro1982, Bell1983, Qi1983, Schaumann1970, Bauer1978} & Ag & 0.312 \cite{Katsuta1979a} & Y & 0.496 \cite{Anderson1989, Maeda1993, Buxbaum1985} \\
Ta & 0.152 \cite{Qi1983, Hampele1989, Schaumann1970, Bauer1978, Heidemann1976} & Cu & 0.375 \cite{Zvezdin1968, Caskey1977, Katz1971, Magnusson2017, Hagi1986} & Ti & 0.508 \cite{WasilewskiRJ;Kehl1954, Kunz1983} \\
Mo & 0.185 \cite{Katsuta1982, Tanabe1992, Katsuta1964, Katsuta1983} & Ni & 0.413 \cite{Yamakawa1979, Cummings1987, Hagi1986, Katz1971, Atrens1980, Furuya1984, Yamanishi1983} & Sc & 0.540 \cite{Han1987} \\
Ti & 0.273 \cite{Naito1998, WasilewskiRJ;Kehl1954} & Th & 0.419 \cite{Westlake1959} & Lu & 0.575 \cite{Volkl1987} \\
Zr & 0.361 \cite{Gelezunas1963} & Al & 0.452 \cite{Outlaw1982, Papp1981} & Be & 0.611 \cite{Kizu1995} \\
W & 0.387 \cite{Frauenfelder1969} & Fe & 0.460 \cite{Mehrer1990} & Ru & 0.763 \cite{Onwudinanti2020} \\
 &  & Co & 0.507 \cite{Caskey1974} & &  \\
 &  & Rh & 0.892 \cite{Baykara2004} & & \\ 
 \hline\hline
\end{tabular}
\end{table} 

\begin{figure*}[htb]
\includegraphics[width=\wholefigwidth]{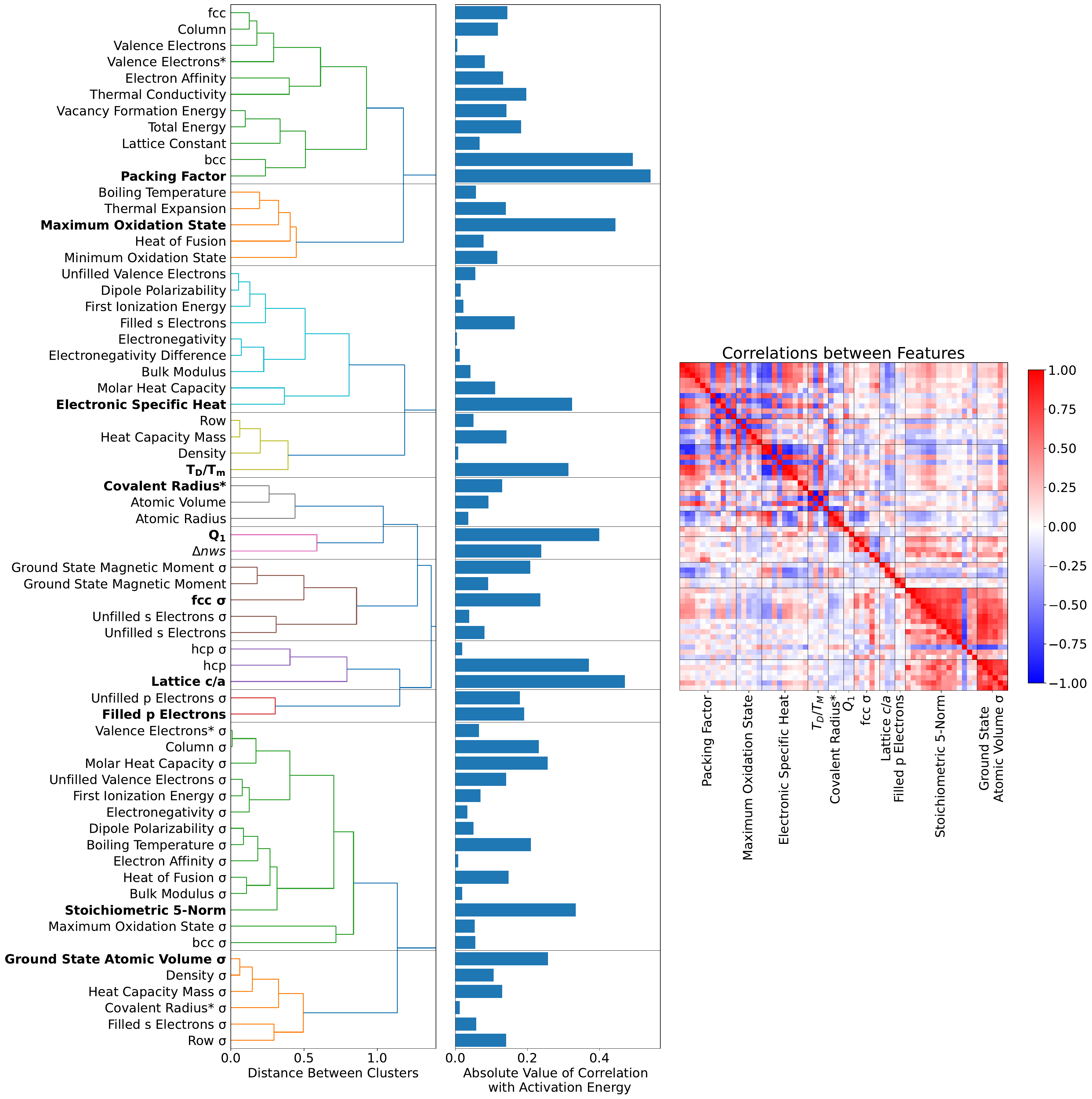}
\caption{Hierarchical grouping of features using the Ward variance minimization. For the binary alloys, instead of alloy properties, we use the weighted average of the elemental features by atomic composition and the standard deviations $\sigma$ of the MAGPIE features. For metals, the standard deviations are set to 0. When features from MAGPIE and different sources differ in value, we include both and denote the MAGPIE version with an asterisk. Spearman rank correlations define the distance between individual features, which are shown in the upper triangle of the right panel; the lower triangle provides the Pearson correlations. Using a greedy algorithm that groups the two closest features at every step, Ward variance minimization creates groups of correlated features, shown in the left panel. We choose a distance threshold of 1 to define groups, which are separated by black lines in both the left and middle panels. After grouping is complete, we name each group with the feature with the highest absolute value of the correlation with the activation energy as shown in the middle panel; these representative features are bolded in the labels on the left and used as labels in the correlation plot.}
\label{fig:corr}
\end{figure*}

\Fig{corr} shows the properties that are used as input into our machine learning models and their correlations with the hydrogen activation energy. For the binary alloys, no alloy properties are included, but we use the averages of the elemental properties weighted by the atomic percentages. These properties come from two sources: theoretical models of interstitial diffusion and broader elemental and stoichiometric properties from the MAGPIE database \cite{Ward2016}. For the MAGPIE features we also include the standard deviations $\sigma$ of the features, and for metals, the standard deviations are set to 0. By including these two types of features, we allow our model to use known proponents to hydrogen diffusion, while also picking out new features. We include features from models of hydrogen diffusion: the electronic specific heat \cite{kittel1996}, thermal conductivity \cite{kittel1996}, and the 15 features \cite{li2001correlation, wiki:List_of_elements_by_atomic_properties, bakker1998enthalpies, knowledgedoor, knowledgedoor1, shang2016, Wang2004, Villars1987, Ranganathan2006} that were used to model C, O, and B interstitial diffusion in metals \cite{Zeng2018}. The electronic specific heat coefficient of the metal is included because it has been found by Arnoult and McLellan \cite{Arnoult1973} to have an empirical linear relationship with the relative partial enthalpy of hydrogen solubility. The thermal conductivity allows us to take into account phonon modes in the metal. A detailed description of the MAGPIE features and their sources is found in the original paper \cite{Ward2016}. The method of counting valence electrons differs between MAGPIE and Zeng \textit{et al.} \cite{Zeng2018}, so we have included both, and the MAGPIE version is denoted with an asterisk. The MAGPIE version defines the valence electrons by excluding any noble gas electron configurations from the total electrons, while Zeng \textit{et al.} only count the outermost shell. While we have renamed most features so that their names are intuitive, four features are more difficult to understand. 
The stoichiometric 5-norm is defined as $\sum_{i=0}^{n}(x_i^5)^{1/5}$ where $x_i$ is the atomic fraction of each element, $\Delta nws$ is the absolute difference of the electron density at the boundary of the Wigner-Seitz cell between the host and hydrogen, $T_d/T_m$ is the ratio of the Debye temperature to the melting temperature, and $Q_1$ is obtained from an empirical elastic model \cite{vykhodets2005}, which is a simplified version of \Eqn{elastic}:
 \be \label{eqn:q1} Q_1 = Ga(d-\lambda)^2,\ee where all the variables are the same as was defined in \Eqn{elastic}. To reduce the impact of multicollinearity on our models, we remove 11 features with correlations greater than 0.95 with another feature. When two features were correlated strongly, we kept the feature with the strongest correlation with the activation energy. We also remove constant features that are the same for all entries in the database. In the middle panel of \Fig{corr}, we show the Pearson correlation between each feature and the activation energy. The packing factor has the largest correlation with the activation energy with a correlation value of 0.54. For input into our machine learning models, we scale all features so that they have a mean of 0 and a standard deviation of 1.

We combine our features into ten groups using their correlations in \Fig{corr}, and the three largest groups (packing factor, electronic specific heat, and stoichiometric 5-norm) roughly split the features into structural and energetic, electronic, and alloy standard deviation property groups. The hierarchical tree shown in the left panel forms correlated feature groups using Ward's minimum variance method to minimize the variance between elements in the same cluster \cite{ward1963}. Groups are formed by choosing a threshold, which we choose to be 1 (the maximum value of the distance metric between two elements).
This distance threshold creates a suitable number of groups to improve interpretability of our features. Having too many groups is difficult to interpret, while too few groups makes each group too broad.
To define the distance, we used the Spearman rank correlation to take into account any nonlinear, monotonic relationships between the features as the majority of our models are nonlinear. Shown in the right panel, the Spearman rank correlations in the upper right triangle are also stronger than their Pearson counterparts in the bottom left, which allows us to form groups with stronger intra-group correlations. We named each of the 11 groups after the representative feature with the strongest absolute value of the correlation with the activation energy. The three largest groups are ``stoichiometric 5-fold" (14 features), ``packing factor" (11), and ``electronic specific heat" (9). The stoichiometric 5-fold group contains the standard deviations of the features that measure how the elemental properties vary, the packing factor group contains energetic and structural information, and the electronic specific heat group contains electronic properties. The standard deviations of the properties are spread over 5 out of the 11 groups (fcc $\sigma$, lattice $c/a$, filled $p$ electrons, stoichiometric 5-fold, and ground state atomic volume $\sigma$) in the bottom half of \Fig{corr}. These groups improve interpretability of our features by reducing the feature set.

\subsection{Machine Learning Models}

In this paper, we use six different machine learning models that work well on small data sets: $k$-nearest neighbors, Gaussian process, random forest, gradient boosting, positive ElasticNet, and Bayesian Ridge. All models are trained using scikit-learn \cite{scikit-learn}, a Python package for machine learning.

\subsubsection{\texorpdfstring{$k$}{K}-Nearest Neighbors}

The $k$-nearest neighbors models find the closest $k$ training samples from an input data point, and then assign the new point a linear combination of its neighbors' values \cite{Cover1967, Altman1992}. The distance between points used is the Euclidean distance. Because it is an algorithm that uses all of its training data and does not build a model, $k$-nearest neighbors models do not generalize well to unseen data. They also suffer from the curse of dimensionality, where the amount of data needed to get accurate predictions increases exponentially with the dimensionality of the data \cite{murphy2012}. We choose to optimize two hyperparameters: weighting scheme and number of neighbors. The two weighting schemes, uniform or distance, determine how to weigh the $k$ neighbors' activation energies when assigning one to a new point. The uniform weighting scheme gives equal weights to all neighbors, while the distance weighting scheme assigns a weight that is proportional to the distance to the new point.

\subsubsection{Gaussian Process}

The Gaussian process model creates a prior distribution over possible functions, then uses Bayesian interference to update the posterior based on new training data \cite{Rasmussen2006, Deringer2021}. These models define the covariance of their prior distribution using kernels, and train a Gaussian posterior distribution over functions whose mean is used for predictions. We assume that the prior distributions for the noise is the Gaussian $\mathcal{N}(0, \alpha^{-1})$ and the prior distribution for the weights $\mathbf{w}$ is $\mathcal{N}(0, \lambda^{-1})$ where $\alpha^{-1}$ is the variance, $\mathbb{1}$ is the identity matrix and $\lambda^{-1}$ is a covariance matrix. Given a function $\phi(\mathbf{x})$ where $\mathbf{x}$ is an input data point, the model is described as: 
$$f(\mathbf{x}) = \phi(\mathbf{x})^\intercal \mathbf{w}.$$ We then combine the $\phi(\mathbf{x})$ into a matrix $\Phi(X)$ that includes all the training data and then define the posterior function for the weights as: \be \label{eqn:gp} p(\mathbf{w}|\Phi(X), \mathbf{y}) \sim \mathcal{N}(
\alpha A^{-1}\Phi(X) \mathbf{y}, A^{-1}),\ee where  $A = \alpha \Phi(X) \Phi(X)^{\intercal} + \lambda\mathbb{1}$ and $\mathbf{y}$ is the target output. Instead of explicitly defining the basis functions, we use the kernel trick to make the calculations faster. For our models, 
we choose to use two different types of kernels: white kernels and radial basis function (RBF) kernels. The noise level on the white kernel denotes the size of the random noise fluctuations in the data, while the RBF kernel's length scale denotes the length scales of the fluctuations. While the actual value of these hyperparameters is optimized by the model, we choose a starting value with nested cross validation to reduce the chances of falling into a local minimum where all deviation in the activation energies is considered random noise.

\subsubsection{Random Forest}
Random forest models are ensemble methods that take an average over independent decision trees \cite{Breiman2001}. Each tree is trained on a random sample of the complete dataset and acts as a weak learner. Even though each individual decision tree often overfits, using random samples of the data and averaging over the trees allow errors to cancel out. The hyperparameters of a random forest model are the total number of trees  and the maximum depth of each tree.

\subsubsection{Gradient Boosting}
Another ensemble method, gradient boosting models build each tree sequentially, so that each tree reduces the bias of the final combined estimator \cite{Friedman2001, Friedman2002}. This allows it to perform better than the random forest model where the trees are independent. Gradient boosting models, which work well on small data sets, have been used to design explainable models for other material properties like thermodynamic properties of hydride formation in alloys and intermetallics \cite{Witman2020} and for interstitial diffusion, not including hydrogen, in metals \cite{Zeng2018}. We optimized the learning rate of each tree and the complexity of the trees through the number of estimators and maximum depth. 

\subsubsection{Positive ElasticNet}

The positive ElasticNet model is an ordinary least squares estimator with an additional Ridge and lasso penalty \cite{Zou2005}. It assumes that the models are linear, while also penalizing the size and number of non-zero coefficients, which makes it more stable than a simple least squares estimator. To make sure that all predicted activation energies are positive and therefore physical, we further constrain all coefficients to be positive, scale all the features between 0--1, and include a reversed version of all the features so that for each feature $x$, we include a new feature $x'$ where $x' = 1 - x$. This allows us to include negative correlations between our features and the activation energy, while still constraining the activation energies to be positive. We choose to enforce positive results by using positive coefficients instead of using a non-linear transformation to maintain its interpretability. A log transformation would turn the ElasticNet model into a product of terms, instead of a sum, reducing the physical relevance.
We optimize the size of the Ridge penalty $\alpha$ and the $l_1$ ratio between the Ridge and lasso penalty.

\subsubsection{Bayesian Ridge}

Bayesian Ridge models are probabilistic, linear models that tune the weights and regularization parameters from an uninformative prior \cite{Tipping2001} where the prior distributions for the noise is a Gaussian $\mathcal{N}(0, \alpha^{-1})$ and the prior distribution for the weights is $\mathcal{N}(0, \lambda^{-1}\mathbb{1})$ where $\lambda^{-1}$ is the variance and $\mathbb{1}$ is the identity matrix. Given an input matrix $X$ of features and a desired output $\mathbf{y}$, the posterior distribution for the weights $\mathbf{w}$ is defined as \be \label{eqn:br} p(\mathbf{w}|X, \mathbf{y}) \sim \mathcal{N}(
\alpha A^{-1}X \mathbf{y}, A^{-1}),\ee where  $A = \alpha X X^{\intercal} + \lambda\mathbb{1}$. Note that these are identical to the Gaussian process except that the matrix of basis functions $\Phi(X)$ is replaced with simply $X$. The four hyperparameters $\alpha_1$, $\alpha_2$, $\lambda_1$, and $\lambda_2$ define the prior Gamma distributions from which the regularization parameters are chosen. In this work, we use $10^{-6}$ for the four hyperparameters to guarantee a flat, uninformative prior distribution so that the expectation-maximization algorithm can adjust the precision of the weights and noise without bias. Because no negative predictions were made, we did not add any additional constraints to guarantee positive activation energy predictions for the Bayesian Ridge models.

\section{Results}
\subsection{Model Fitting}
\label{sec:model}

\begin{table*}[htb]
\centering
\caption{Hyperparameters used for the models. We choose the optimal values for each parameter using cross validated grid-search. The splitting scheme for the cross validation uses the leave-one-group-out method, acting on the element groups, which are shown as the rows and columns in \Fig{database}.}
\label{tab:hyperparameters}
\begin{tabular}{llllll}
\hline\hline
Model & Hyperparameter & Start & End & Step Size & Optimal value\\
\hline
\multirow{2}{*}{K-Nearest Neighbors}
& Number of Neighbors & 1 & 8 & 1 & 2\\
& Weighting Scheme & \multicolumn{3}{l}{Distance, Uniform} & Distance \\
\hline
\multirow{2}{*}{Gaussian Process}
& Length Scale (RBF) & 5 & 50 & 5 & 35 \\
& Noise Level (white) & $10^{-5}$ & $1$ & $10\times$ & $10^{-2}$ \\
\hline
\multirow{2}{*}{Random Forest}
& Number of Estimators & 40 & 90 & 10 & 70\\
& Maximum Depth & 3 & 14 & 1 & 10\\
\hline
\multirow{3}{*}{Gradient Boosting}
& Learning Rate & 0.1 & 1 & 0.1 & 0.3\\
& Number of Estimators & 100 & 170 & 10 & 150\\
& Maximum Depth & 2 & 8 & 1 & 2\\
\hline
\multirow{2}{*}{Positive ElasticNet}
& Ridge Penalty $\alpha$ & $0.001$ & 1 & $\approx 2 \times$ & $0.01$\\
& $l_1$ ratio & 0.025 & 1.0 & 0.0125 & 0.05\\
\hline
\multirow{4}{*}{Bayesian Ridge}
& $\alpha_1$ & & & & $10^{-6}$\\
& $\alpha_2$ & & & & $10^{-6}$\\
& $\lambda_1$ & & & & $10^{-6}$ \\
& $\lambda_2$ & & & & $10^{-6}$ \\
\hline\hline
\end{tabular}%
\end{table*}
In order to ensure that our results are independent of the chemistry of our training set, we tune the hyperparameters (see \Tab{hyperparameters}) via nested cross validation by leaving out all data that contain a certain element (alloy split) \cite{Lu2019} at each iteration and then choosing the optimal hyperparameter as the median over all iterations. Within each iteration, we perform a 80--20\% split and perform a grid search to find the hyperparameters that lead to the smallest mean squared error between the predicted and experimental activation energies. The start and end values are the upper and lower bounds, and each hyperparameter varies by step size increments. Because the Bayesian Ridge's hyperparameters define the prior Gamma distributions from which the optimal values are chosen, we choose the values manually instead of through nested cross validation. We also performed a sensitivity analysis of our hyperparameters shown as Table S1 of the Supplemental Material \footnote{See Supplemental Material at [URL will be inserted by publisher] for hyperparameter sensitivity analysis, linear models for deuterium
and tritium activation energies, and the grouping analysis applied to the SHAP values}, and other than the learning rate for the gradient boosting tree model, the RMSEs on the testing data changed by less than 10 meV when all other hyperparameters were increased or decreased by one step size. Decreasing the learning rate by a single step (0.1) led to the RMSE decreasing by 19 meV as smaller learning rates can reduce the amount of overfitting, thus improving predictions on the test set. To make sure that no positive bias is introduced to our error predictions, the original hyperparameter is used. As an additional measure of the possible bias in our hyperparameter choice, we recalculated the test RMSEs for the alloy split so that all test data is not used to also select the hyperparameter and found a less than 7 meV deviation between the two RMSEs as shown in Table S2 of the Supplemental Material \cite{Note1}. This implies that there is no positive bias introduced from reusing data, and that our models are not overfitting to the hyperparameters.

\begin{figure*}[htb]
\centering\includegraphics[width=\halffigwidth]{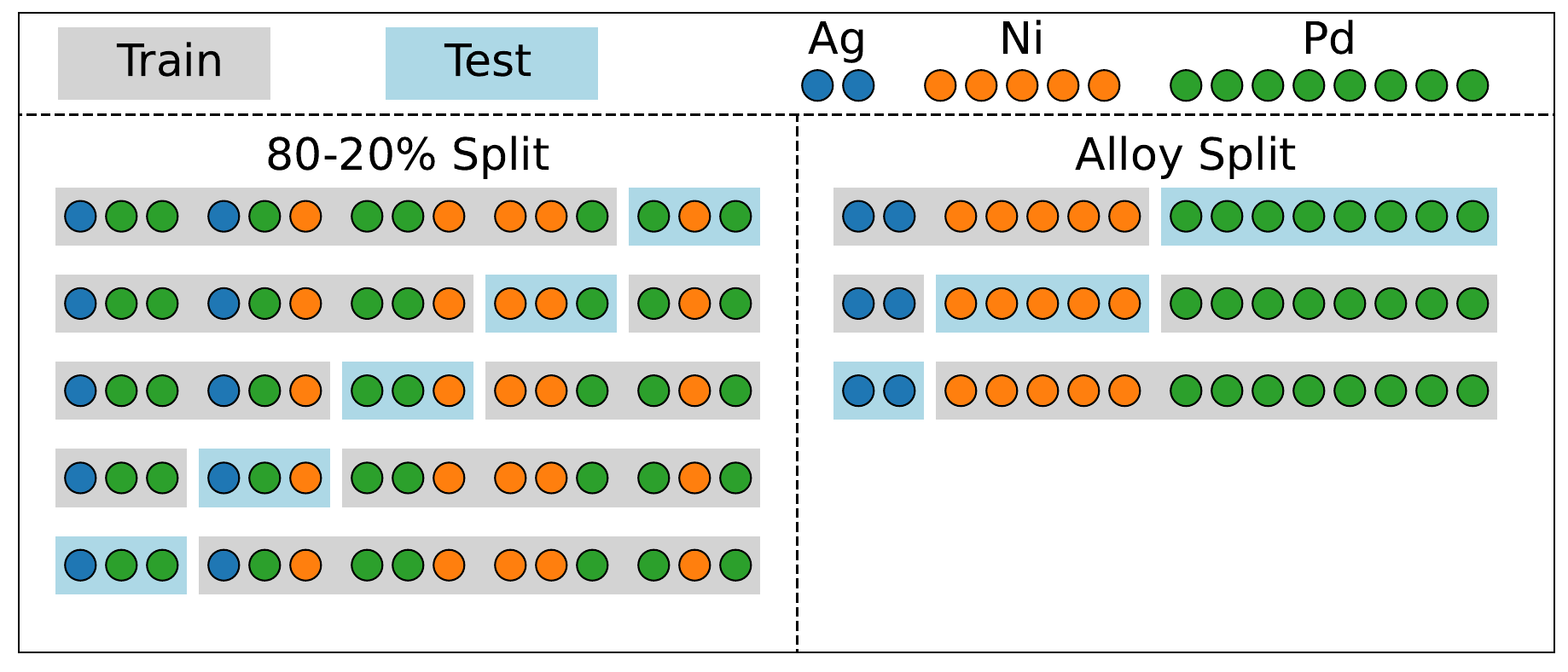}
  \caption{Schematic showing the difference between the random 80--20\% train-test split and the alloy split. The alloy train-test split (shown on the right) works by grouping every entry in the database that contains a certain element, then using one of these groups as the testing data. Binary alloys are included in two groups. These alloy groups drastically differ in size, as the Pd group contains 43 entries while others (Be, Hf, Lu, Mg, Mo, Pt, Ru, Sc, Ta, Th, Y, and W) only have one.}
  \label{fig:val}
\end{figure*}

\begin{figure*}[htb]
 \centering\includegraphics[width=\halffigwidth]{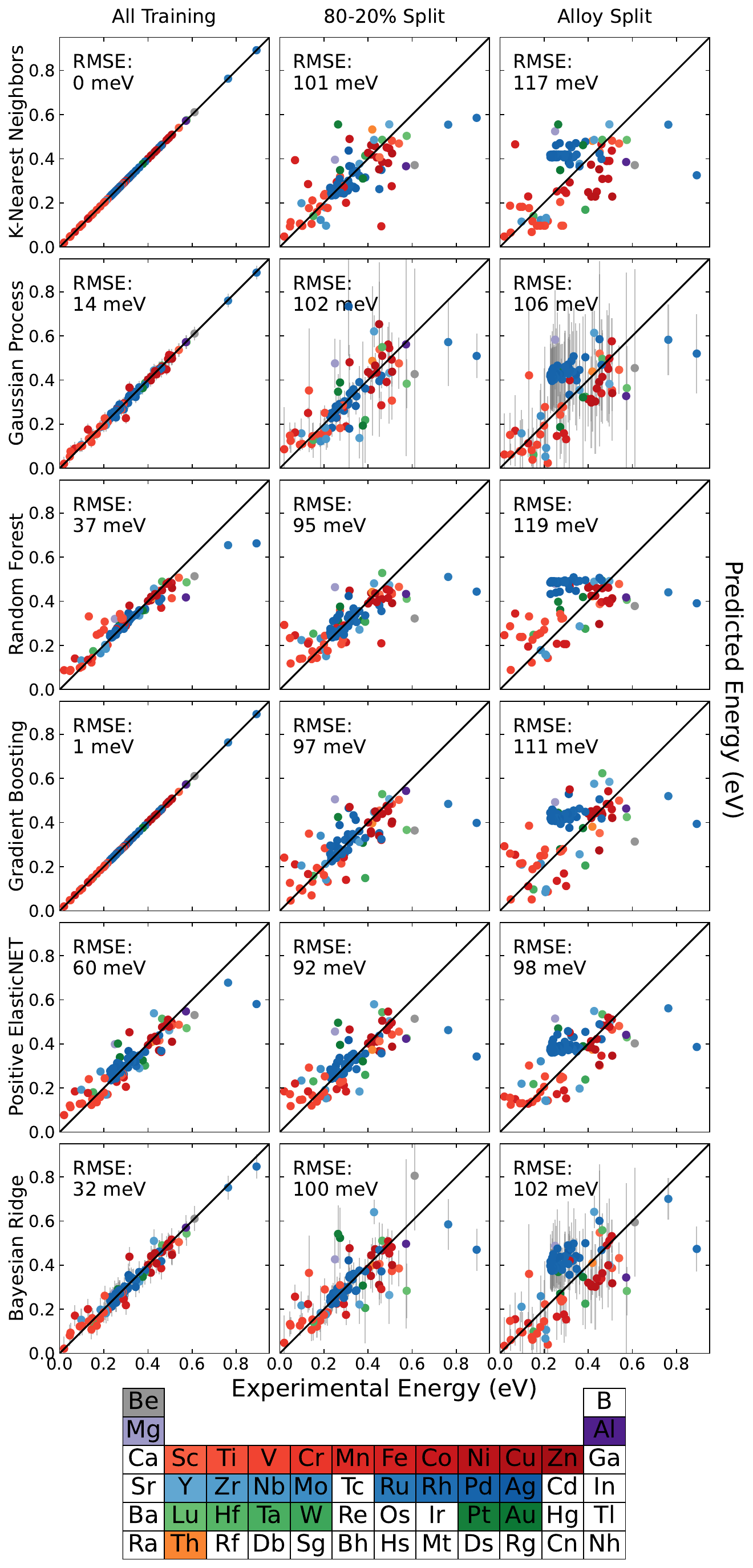}
 \caption{Predicted vs.~experimental activation energies for hydrogen diffusion from the six models. Element placement in the periodic table determines the color-coding scheme, with the colors denoting rows, and the saturation denoting columns. The left column shows models trained on the entire dataset. In the middle column, we show validation sets from models trained with random 80--20\% training-validation samples split. On the right, we plot validation sets from models trained while removing all metals and alloys that contain a certain element at a time. The blue points in the experimental activation energy range of 0.22--0.46 eV are Pd alloys and elemental Ag, and the largest outliers are Ru (0.76 eV) and Rh (0.89 eV). The alloy train-test split performs worse than the random train-test split for all models, but the smallest RMSE difference between the train-test splits occurs for the Gaussian process and Bayesian Ridge models.}
 \label{fig:models}
\end{figure*}

Using our optimal hyperparameters, we train six models with three different train-test splits and obtain similar root-mean-squared-errors (RMSEs) for the 80--20\% split and alloy split; the test RMSE of the alloy split is, on average, 18 meV higher than the 80--20\% split. A schematic detailing the two train-test splits is shown in \Fig{val}. Each model is trained 100 times (with different random splits for the 80--20\% train-test split), and the average testing RMSEs are reported. The error of the training predictions (leftmost column of \Fig{models}) shows the ideal performance of our models when it is trained on the entire dataset. In the middle and right columns of \Fig{models}, we plot only predictions on the test set, and no training data is shown.
For the alloy split, shown in the right column of \Fig{models}, we choose to plot only the binary alloy predictions from the model where the dominant element is removed from training (or the average of the two predictions for equiatomic systems) instead of plotting both predictions from each of the alloy splits. 
The 80--20\% train-test split has smaller RMSEs than the alloy split for all six models due to significantly more accurate alloy predictions because of similar training and testing sets. Alloy activation energy predictions have 85 meV smaller RMSEs on average when we use the 80--20\% split instead of the alloy split, while elemental predictions have 6 meV larger RMSEs. 
The Gaussian proces, positive ElasticNet, and Bayesian Ridge model have two smallest RMSEs for the alloy split and are the best at generalizing to unseen data. This is because these three models are the only continuous models that have in-built regularization and reduce overfitting, especially with the small database used in this study.
The Gaussian process model also deals effectively with noisy data by controlling the length scale of the RBF kernel, which ensures that small fluctuations below this length scale in the training data are ignored \cite{Tipping2001}. These average RMSEs are greater than the errors from typical DFT calculations for hydrogen diffusion which usually lie within a few 10s of meV from experimental values \cite{Kamakoti2003, Baykara2004, Jiang2004,Bhatia2005,Wimmer2008,Duan2010, Connetable2011,Liu2012,Lu2013, Fernandez2015,Tafen2015,Han2016, Yang2016, Liu2017,   Liu2018,Oliveira2019,Yang2019, Connetable2019}. Because the average RMSEs for the models are greater than the isotope effect, our models are unable to accurately predict activation energies for deuterium and tritium. Because more complex models will overfit to the small amount of deuterium and tritium data, we instead fit a separate linear model with a single feature in Section S2 of the Supplemental Material \cite{Note1}.

Even though the activation energies are predicted with relatively small RMSEs, our machine learning models are limited because they overemphasize crystal-dependent features over element-dependent features for the alloy train-test split (right column of \Fig{models}). This is most obvious for the FCC Pd alloys and elemental Rh, which are predicted to have activation energies close to the mean for their crystal structure (0.34 eV) but in reality, their activation energies are the lower and upper bounds. Pd and its alloys, which consist of a third of the total database, have activation energies that are systematically overestimated especially on the Pd-rich side with smaller activation energies. Therefore, when the Pd alloys are excluded from training in the alloy-split validation method, the models learn to predict values that are closer to the average for their crystal structure even though it leads to incorrect predictions on the Pd test set. This leads to the prominent `clump' of blue Pd alloys in the range of 0.36--0.46 eV in \Fig{models}. Rh, an outlier with the largest activation energy, is drastically underestimated and has the largest errors for all six models. 
This behavior is because strong correlations between the activation energy and the packing factor can dominate effects that are only significant for a few specific elemental systems, such as Pd and Rh having smaller atomic radii compared to other elements in their row, which has been shown to have an effect on the speed of solute diffusion in Ni \cite{Janotti2004}. This limitation of our machine learning models is likely due to a lack of other metals or alloys in our training database with such extreme activation energies.

\subsection{Uncertainty Quantification}
\begin{figure}[htb]
\includegraphics[width=\halffigwidth]{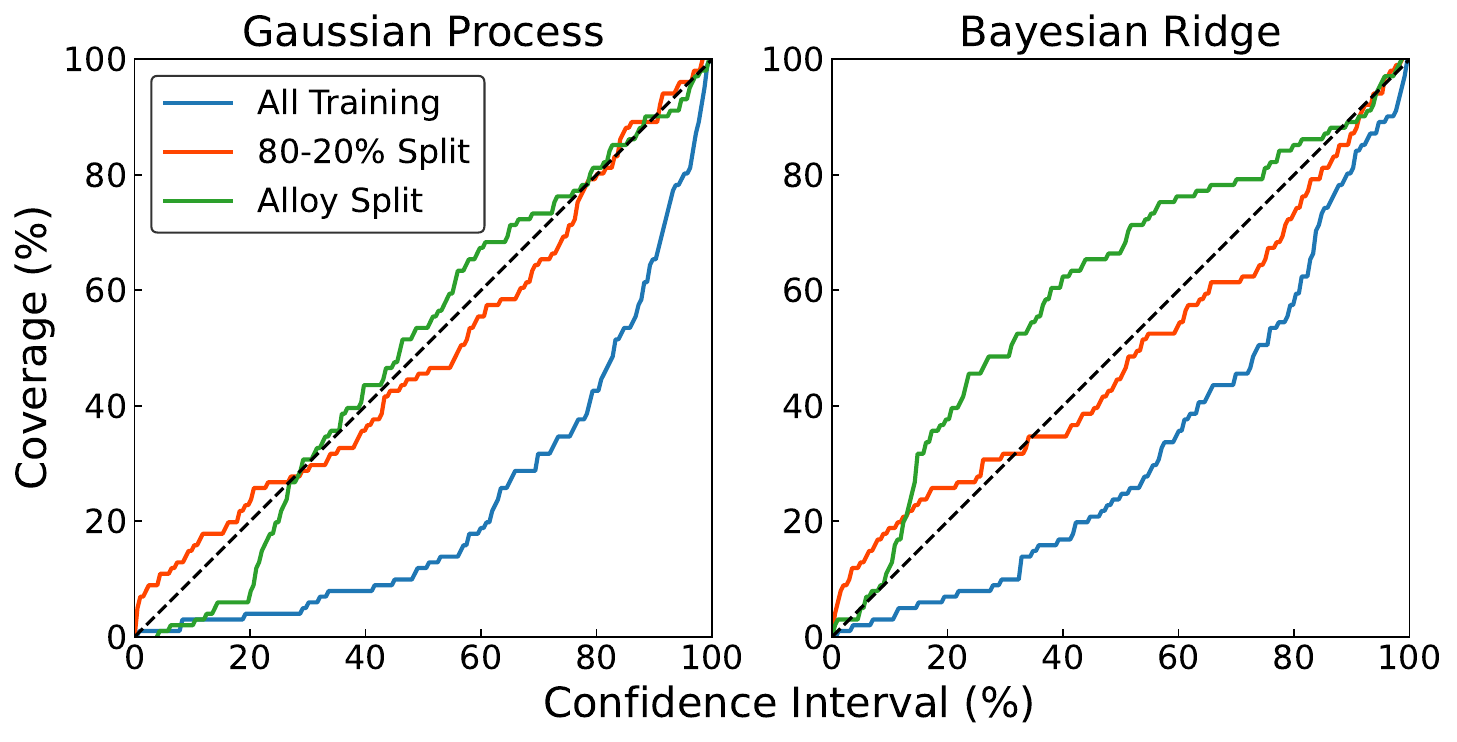}
\caption{Verification of the uncertainty quantification on the testing data for the Gaussian process and Bayesian Ridge models from coverage percent. The coverage percent is the fraction of observed model errors within a given confidence interval. The dashed line corresponds to coverage percent that matches the confidence interval: curves that lie below the line overestimate the error while curves that lie above the line underestimate the errors. For the Gaussian process all training model, 3\% of the observed errors lie in a 20\% confidence interval while for the Bayesian Ridge, 7\% of the observed errors lie in the same interval. While both models overestimate the errors for the all training train-test split due to high variance but low bias in the models, the Bayesian Ridge also underestimates errors in the alloy train-test split due to worse performance on unseen data. 
}
\label{fig:cumulative}
\end{figure}

For models which explicitly quantify uncertainty (Gaussian Process and Bayesian Ridge), it is important to understand the reliability of these uncertainty estimates, especially if one is predicting new materials containing previously seen elements (80--20\% test split) or entirely unseen elements (alloy test split). The uncertainty quantification in \Fig{cumulative} reveals that the predicted standard deviations for the Gaussian process model accurately predicts the observed errors, but the Bayesian Ridge model's predicted standard deviations are only reliable on testing data that is similar to its training data. Instead of a single solution, the models fit a posterior distribution described in \Eqn{gp} and \Eqn{br} whose standard deviations depend heavily on hyperparameters that are learned through the data---either the noise level on the white kernel for the Gaussian process or the precision of the weights and noise for the Bayesian Ridge. 
The average sizes of the predicted standard deviation for the Gaussian process model are 27, 80, and 172 meV for the all training, 80--20\% split, and alloy split respectively, while for the Bayesian Ridge model, the average standard deviations are 48, 72, and 120 meV for the three train-test splits. As expected, the uncertainty increases for train-test splits with different elemental compositions than the training data.
In \Fig{cumulative}, we perform this analysis by comparing the predicted standard deviations from the model uncertainty quantification with the observed errors by plotting the coverage percent, which indicates the fraction of data points within a given confidence interval. The confidence intervals are calculated assuming that the predicted standard deviations are accurate. If the p-values are uniformly distributed (the dashed line in \Fig{cumulative}), then the observed errors sample the same Gaussian distribution as the predicted standard deviations. Curves that lie above the line suggest that observed errors sample a broader Gaussian than the predicted standard deviations, so the predicted error bars are too small. For curves that lie below the line, the observed errors are too large, and the predicted standard deviations are too conservative.
The all training models both overestimate the error, and in both models, less than 10\% of the predicted values (3\% for Gaussian process and 7\% for Bayesian Ridge) fall more than 1.28 standard deviations away (the 20\% confidence interval) from the true experimental value.
This likely arises because the models have large amounts of variance, and there is a wide distribution of possible functions with low bias, so the mean ends up being close to the true value \cite{murphy2012}. 
While both models' predicted standard deviations deviate from the observed errors for the alloy split case, the predicted standard deviations on the 80--20\% split for both the Gaussian process and Bayesian Ridge agree strongly. 
The Gaussian process slightly overestimates the error, and almost all of its observed errors are within one predicted standard deviation, but it recovers to the ideal distribution after the 30\% confidence interval.
On the other hand, the Bayesian Ridge model heavily underestimates the observed errors for the alloy split. This can be understood by interpreting the Bayesian Ridge model as a simpler Gaussian process model whose kernel choice forces it to only use a limited prior. This makes the posterior distribution too concentrated, which leads to overconfidence as it cannot recognize that the new data are different than its training data \cite{murphy2012, bishop2006}. Therefore, the Gaussian process model's predicted standard deviations can be trusted with new elemental compositions, but the Bayesian Ridge's predictions can only be trusted for similar data.

\subsection{Model Explainability}
\label{sec:model_explain}

\begin{figure*}[htb]
 \centering\includegraphics[width=\wholefigwidth]{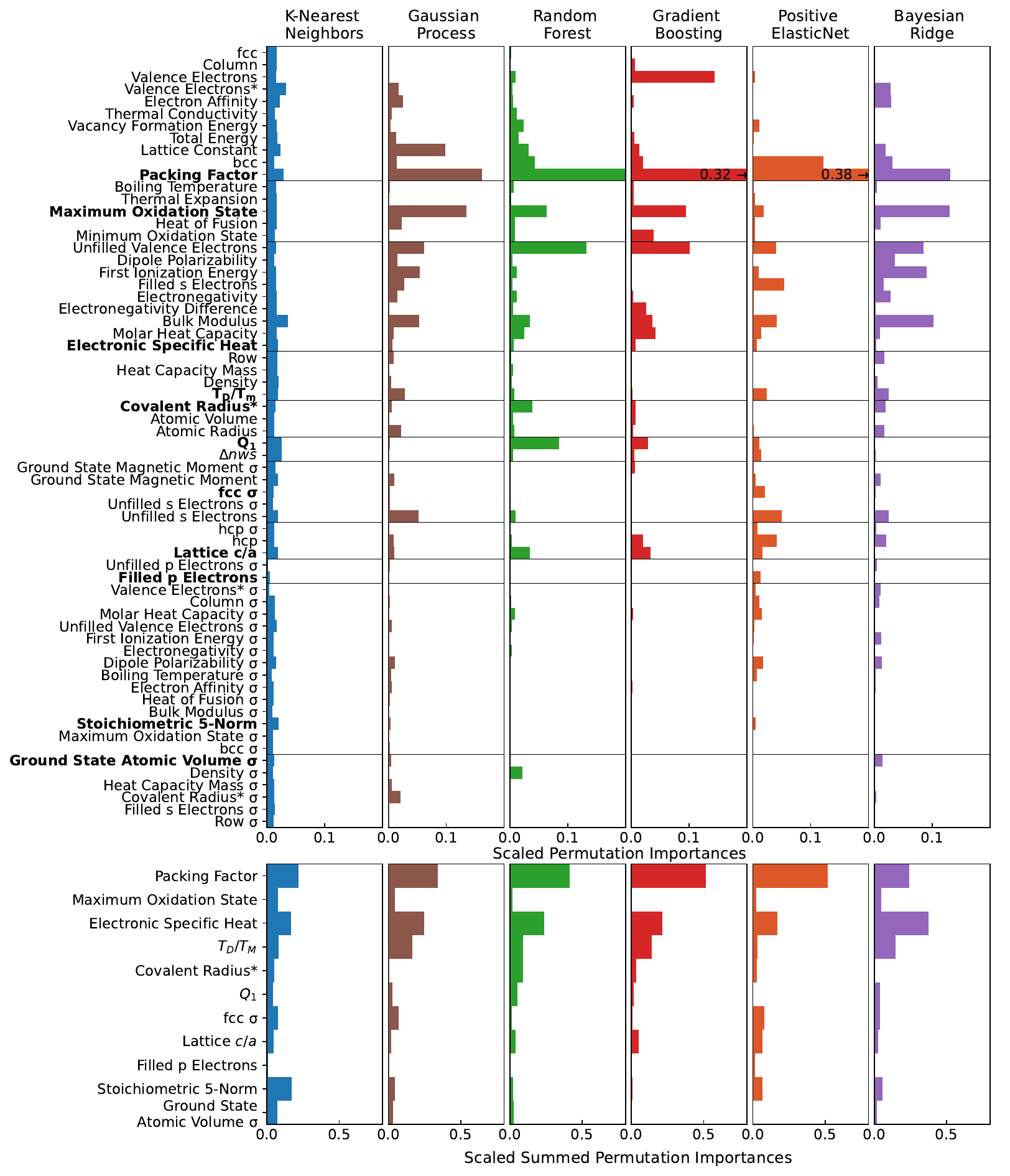}
 \caption{The permutation importances for the six models averaged over 200 repetitions, shown for all features. The raw permutation importances are shown in the top row. While the packing factor is the most important feature for 4 of the 6 models, there is wide deviation between the rankings. Using the feature groups as shown in \Fig{corr}, we sum permutation importances between feature groups as shown in the bottom subfigure. Permutation importances are scaled so that they sum up to 1. The two most important feature groups are the packing factor and electronic specific heat.}
 \label{fig:all_notoverlapping}
\end{figure*}

Strong correlations between features lead to wide variations of feature importances calculated for different models, making physical interpretation of the models difficult as shown in \Fig{all_notoverlapping}.
 We use the permutation importance, which measures how much the $R^2$ value changes after shuffling the features, to measure the importance of each feature. Scaling the permutation importances within each model to sum to 1 enables comparisons between the models. Even though the importance is measured using the same metric for all six models, we find large variation in the feature importances; just listing the top three features from each model leads to a total of ten different features -- none of which are consistent for all the models. The models agree most when identifying the importance of the packing factor, as it is the most important feature for all except for the $k$-nearest neighbors model. The $k$-nearest neighbors model assigns nearly equal importances to all features due to it being a non-parametric model that only takes into account distances between points rather than a model-building method. 
This wide variation in the importance of individual features arises from correlated material properties and makes interpretation difficult, but after grouping the features using the greedy algorithm shown in \Fig{corr}, we find that specific feature groups are selected. 
 Similar behavior can be seen in previous work on modeling solute diffusion in metals \cite{Wu2017, Lu2019}, which saw varying feature importances by using different models and cross validation techniques. Thus, it becomes clear that using the individual permutation importances (or other metrics of feature importances like the Shapley additive explanations \cite{NIPS2017_7062} shown in S3 of the supplementary information \cite{Note1}) for individual features do not offer as much information on important features as grouping analyses do. 

The feature groups in \Fig{all_notoverlapping} lead to stronger agreement between the models, and when we then sum the importances to rank the groups, the two most important ones are the packing factor and electronic specific heat. Here, we ignore the $k$-nearest neighbor model because its grouped importance is ultimately just a measure of the group size, but the other five models agree that the packing factor and electronic specific heat are the two most important groups. 
Agreement depends on sparsity as more sparse models, like the gradient boosting tree and positive ElasticNet models, conclude that the packing factor group is more important, while less sparse models, like the Gaussian process and Bayesian Ridge models, assign the packing factor and electronic specific heat more similar importances. The Bayesian Ridge model is the only model to swap the ranks of the packing factor and the electronic specific heat groups because it assigns a large importance to many features in the same group. The packing factor group contains energetic and structure information like the vacancy formation energy and lattice constant, which are directly related to hydrogen diffusion. While the packing factor and crystal structure determine what kind of interstitial sites are available to form the diffusion pathways, the vacancy formation energy has an strong empirical correlation with the activation energy \cite{shang2016}. The electronic specific heat feature group, which contains electronic properties and the bulk modulus, is dominated by features that measure hydrogen's interaction with the bulk. The top two groups are followed distantly by the maximum oxidation state, then the $Q_1$, lattice $c/a$, stoichiometric 5-norm, and fcc $\sigma$ groups, which all have effectively the same importances.

\begin{figure*}[htb]
 \centering\includegraphics[width=\halffigwidth]{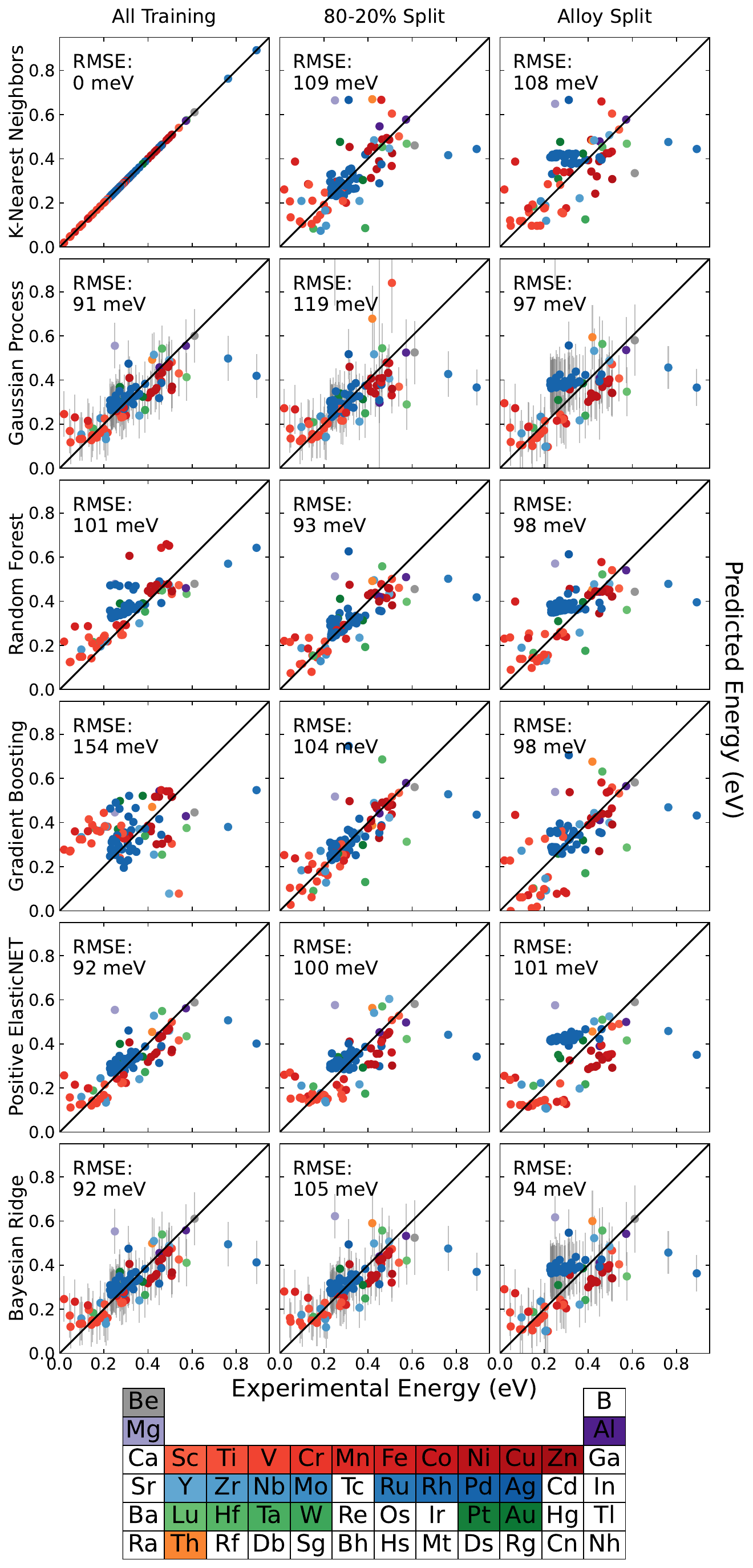}
 \caption{Predicted and experimental activation energies for diffusion using only the representative feature from each group. The RMSEs for the alloy split test-train scheme are smaller than those from the complete model, implying that there is some overfitting with the larger feature set and that fitting with the representative feature set gives the best transferability.}
 \label{fig:models_9}
\end{figure*}

\begin{figure*}[htb]
 \centering\includegraphics[width=\halffigwidth]{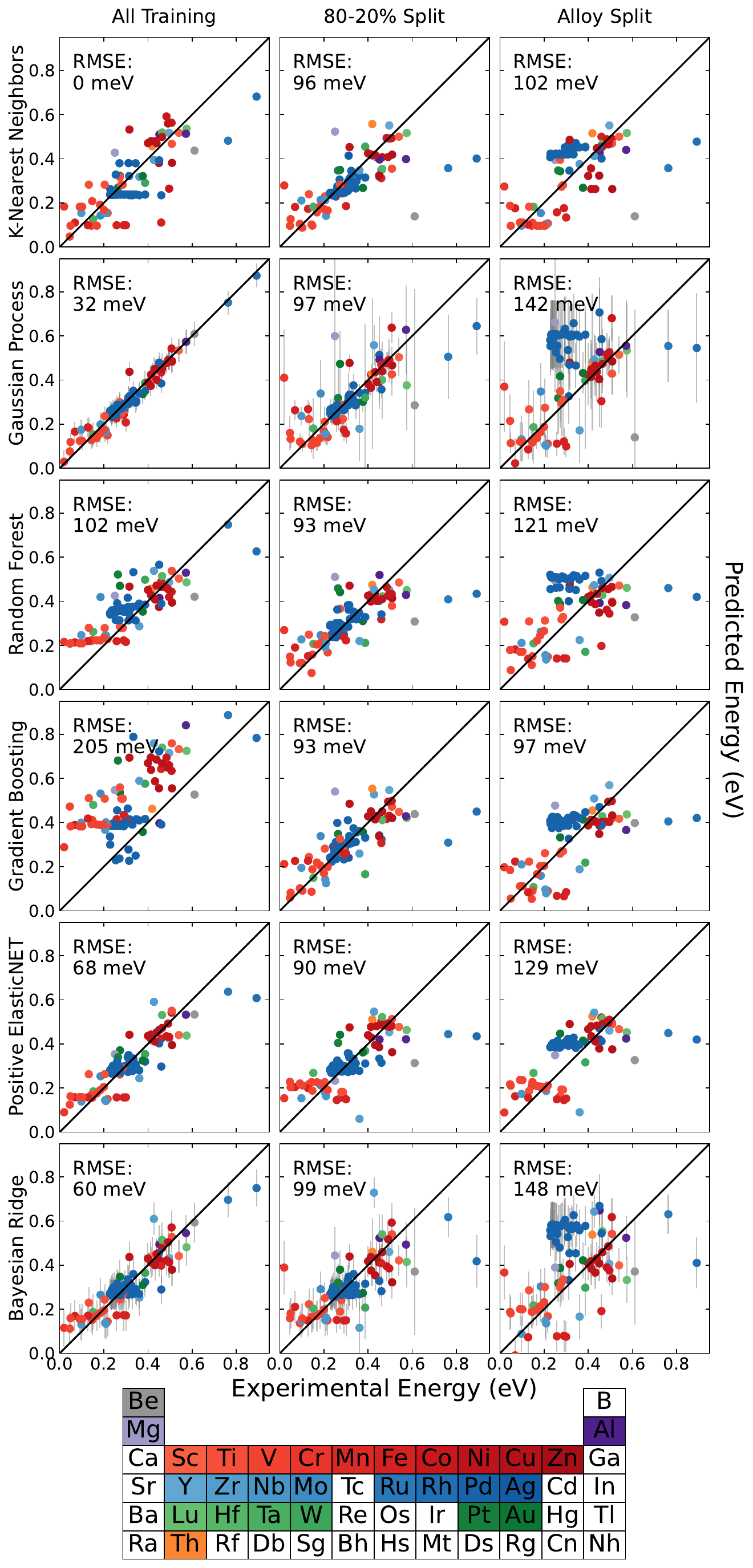}
 \caption{Predicted and experimental activation energies for diffusion using only the top two feature groups, packing factor and electronic specific heat. The Bayesian Ridge and Gaussian process models are the most negatively impacted by the smaller feature set, which is a result of these two models having larger importances for the maximum oxidation state group shown in  \Fig{all_notoverlapping}.}
 \label{fig:smaller_models}
\end{figure*}

A reduced feature set with only each group's representative feature (\Fig{models_9}) has RMSEs similar to the fits using the whole feature set while using only features in the top two groups (\Fig{smaller_models}) has significantly larger RMSEs.
Out of the three feature sets, trained models using the representative feature set have the best transferability because the RMSEs for the alloy split are the smallest. This is because while the representative feature set is the smallest, the removed features are correlated with the representative feature that is kept, so little information is lost. 
The top two groups feature set, on the other hand, has significantly worse predictions on the alloy split, especially for the Gaussian process, positive ElasticNet, and Bayesian Ridge models. This is a result of extremely poor predictions of the activation energies for the Pd alloys. Due to the larger importances for the maximum oxidation state group for the Gaussian process and Bayesian Ridge models, shown in \Fig{all_notoverlapping}, they rely more strongly on the removed information than the others, so they are more negatively impacted by reducing the feature set to only the top two groups.
For the positive ElasticNet model, we reoptimized the hyperparameters using the same nested cross validation technique for the two smaller feature subsets. If we use the same hyperparameters as we do for the full feature set, we observe significant banding of alloy predictions for both the 80--20\% train-test split and the alloy split due to overzealous regularization as the size of the Ridge and Lasso penalties are too large for the smaller feature set, which leads to too many ignored features and small coefficients.
This suggests that any model for the activation energy of hydrogen in metals should contain at least one feature from each group.

\section{Discussion}
\label{sec:discussion}
In this paper, we create a database and fit six different machine learning models to predict hydrogen diffusion activation energy in metals and random binary alloys and achieve smaller errors than a similar, earlier machine learning model for interstitial diffusion~\cite{Zeng2018}.
On the testing data, these models obtain a 92--102 meV RMSE for a random 80--20\% train-test split, and a larger 105--124 meV RMSE for an alloy leave-one-group-out train-test split, which measures how the model performs on unseen elements. These errors are smaller in magnitude than a gradient boosting model for interstitial diffusion (C, O, B, and N) by Zeng \textit{et al.} \cite{Zeng2018} who obtained a final RMSE of 311 meV on testing data even though we have similar elemental compositions and the sizes of our databases were similar. 
Because Zeng \textit{et al.} only used random train-test splits, which we find to have smaller RMSEs than the grouped alloy split by the host metal, and used a smaller test set, we expect this RMSE to be a underestimation of the error when applying the model to materials with elements not existing in the training data.
This difference in the RMSEs is likely because the range of their activation energies 0.19--2.45 eV was approximately 3 times larger than our range of 0.02--0.89 eV because they were dealing with multiple solute types.

Comparisons between our importance analysis and other analytic and machine learning models show that due to hydrogen's small size, features that are included in Ferro's elastic model \cite{ferro1957} are missing; otherwise, the involved material properties agree with the other models.
By training our machine learning models on the downselected feature sets, our machine learning models behave akin to analytic models and can be analyzed similarly.
The interstitial model in Zeng \textit{et al.} \cite{Zeng2018} obtained four important features for predicting non-hydrogen interstitial diffusion: electronegativity difference, total energy, $Q_1$ (see \Eqn{q1}) and thermal expansion. Other than $Q_1$, these features fall in the three most important feature groups for our hydrogen diffusion model (electronic specific heat, packing factor, and maximum oxidation state), pointing to fundamental similarities between hydrogen diffusion and that of other small interstitial atoms. However, $Q_1$ contains 3\% of the overall importance, and the other material property that appears in both our machine learning model and Ferro's elastic model \cite{ferro1957} shown in \Eqn{elastic}, $T_D/T_m$, is equally unimportant and has less than 3\% of the overall importance. This implies that Ferro's elastic model is less applicable to predicting hydrogen diffusion, possibly due to its small size leading to smaller lattice distortions. 
The quantum mechanical model by Flynn and Stoneham \cite{Flynn1970} in \Eqn{flynn} has fewer directly overlapping features with our machine learning models, but similar physical and electronic features appear.
The lattice constant determines the jump distance $d$ in Flynn's model, and the radius of the Debye sphere $q_m$, or screening length, is directly tied to the strength of electronic interactions between the hydrogen and the host metal --- similar to the features found in the electronic specific heat feature group, which contains electronic properties and the bulk modulus.
Within this group, the electronegativity appears in the Miedema model \cite{DeBoer1988}, which has been used to calculate the enthalpy of formation \cite{Bouten1980, VanMal1974, Miedema1976} and the hydrogen content \cite{Herbst2002} for metal hydrides, and the bulk modulus measures the compressibility of a hydrogen atom into a crystal.
Therefore, while electronic interactions between the hydrogen and the bulk remain important for predicting hydrogen diffusion, due to hydrogen's small size, elastic models are less applicable.

In this work we construct a database of material properties for predicting hydrogen diffusivity, and grouped features to interpret ML models on this task. Not only will our machine learning models enable predictions of hydrogen diffusion in metals or random alloys that have not previously been measured in experiments due to the wide variety of elements included in our database, but the new physical insights into the activation energies of hydrogen diffusion in metals and binary alloys will help streamline the process for materials discovery for fast-diffusing or slow-diffusing materials for hydrogen containment or transportation. For ternary or quaternary alloys, even though they were not directly included in the training data, we anticipate that our machine learning models will work within the estimated error bars, but more importantly, the same features that were important for the metal and binary alloys should still be physically meaningful to screen for ideal hydrogen diffusivities. Because the material properties used in this study are easy to measure either experimentally or using DFT, the results will enable data-driven screening of materials without needing explicit measurements of the hydrogen diffusion. Additionally, the technique of grouping the features into correlated feature groups before analyzing feature importances can be applied to other material property predictions, namely for diffusion and other temporal properties, and enable rapid screening for other desired properties.

\nocite{Katz1971, Kunz1983, Qi1983, Buxbaum1985}

\nocite{Stover1986, Caskey1977,Raczynski1978, Quick1978,Naito1990,Volkl1987, Herro1982, Qi1983, Katz1971,Yamakawa1979,Jost1940, Volkl1971,Bauer1978,Naito1998,Schaumann1970,Maeda1993,Gulbransen1954}

\begin{acknowledgments}
This work was supported by by the National Science Foundation Graduate Research Fellowship under Grant No. 1746047 (G.L.). The authors gratefully thank Dr. Rick Karnesky, Josh Vita, and Luke Wirth for the helpful comments and discussions. This work was supported by the Laboratory Directed Research and Development (LDRD) program at Sandia National Laboratories. Sandia National Laboratories is a multimission laboratory managed and operated by National Technology \& Engineering Solutions of Sandia, LLC, a wholly owned subsidiary of Honeywell International Inc., for the U.S. Department of Energy's National Nuclear Security Administration (DOE/NNSA) under contract No. DE-NA0003525. This written work is authored by an employee of NTESS. The employee, not NTESS, owns the right, title and interest in and to the written work and is responsible for its contents. Any subjective views or opinions that might be expressed in the written work do not necessarily represent the views of the U.S. Government. 
The DOE will provide public access to results of federally sponsored research in accordance with the DOE Public Access Plan.
\end{acknowledgments}

\bibliography{hydrogendiffusion}

\begin{thebibliography}{133}%
\makeatletter
\providecommand \@ifxundefined [1]{%
 \@ifx{#1\undefined}
}%
\providecommand \@ifnum [1]{%
 \ifnum #1\expandafter \@firstoftwo
 \else \expandafter \@secondoftwo
 \fi
}%
\providecommand \@ifx [1]{%
 \ifx #1\expandafter \@firstoftwo
 \else \expandafter \@secondoftwo
 \fi
}%
\providecommand \natexlab [1]{#1}%
\providecommand \enquote  [1]{``#1''}%
\providecommand \bibnamefont  [1]{#1}%
\providecommand \bibfnamefont [1]{#1}%
\providecommand \citenamefont [1]{#1}%
\providecommand \href@noop [0]{\@secondoftwo}%
\providecommand \href [0]{\begingroup \@sanitize@url \@href}%
\providecommand \@href[1]{\@@startlink{#1}\@@href}%
\providecommand \@@href[1]{\endgroup#1\@@endlink}%
\providecommand \@sanitize@url [0]{\catcode `\\12\catcode `\$12\catcode
  `\&12\catcode `\#12\catcode `\^12\catcode `\_12\catcode `\%12\relax}%
\providecommand \@@startlink[1]{}%
\providecommand \@@endlink[0]{}%
\providecommand \url  [0]{\begingroup\@sanitize@url \@url }%
\providecommand \@url [1]{\endgroup\@href {#1}{\urlprefix }}%
\providecommand \urlprefix  [0]{URL }%
\providecommand \Eprint [0]{\href }%
\providecommand \doibase [0]{https://doi.org/}%
\providecommand \selectlanguage [0]{\@gobble}%
\providecommand \bibinfo  [0]{\@secondoftwo}%
\providecommand \bibfield  [0]{\@secondoftwo}%
\providecommand \translation [1]{[#1]}%
\providecommand \BibitemOpen [0]{}%
\providecommand \bibitemStop [0]{}%
\providecommand \bibitemNoStop [0]{.\EOS\space}%
\providecommand \EOS [0]{\spacefactor3000\relax}%
\providecommand \BibitemShut  [1]{\csname bibitem#1\endcsname}%
\let\auto@bib@innerbib\@empty
\bibitem [{\citenamefont {Z{\"{u}}ttel}\ \emph {et~al.}(2010)\citenamefont
  {Z{\"{u}}ttel}, \citenamefont {Remhof}, \citenamefont {Borgschulte},\ and\
  \citenamefont {Friedrichs}}]{Zuttel2010}%
  \BibitemOpen
  \bibfield  {author} {\bibinfo {author} {\bibfnamefont {A.}~\bibnamefont
  {Z{\"{u}}ttel}}, \bibinfo {author} {\bibfnamefont {A.}~\bibnamefont
  {Remhof}}, \bibinfo {author} {\bibfnamefont {A.}~\bibnamefont
  {Borgschulte}},\ and\ \bibinfo {author} {\bibfnamefont {O.}~\bibnamefont
  {Friedrichs}},\ }\bibfield  {title} {\bibinfo {title} {{Hydrogen: the future
  energy carrier}},\ }\href {https://doi.org/10.1098/rsta.2010.0113} {\bibfield
   {journal} {\bibinfo  {journal} {Philosophical Transactions of the Royal
  Society A: Mathematical, Physical and Engineering Sciences}\ }\textbf
  {\bibinfo {volume} {368}},\ \bibinfo {pages} {3329} (\bibinfo {year}
  {2010})}\BibitemShut {NoStop}%
\bibitem [{\citenamefont {Irani}(2002)}]{Irani2002}%
  \BibitemOpen
  \bibfield  {author} {\bibinfo {author} {\bibfnamefont {R.}~\bibnamefont
  {Irani}},\ }\bibfield  {title} {\bibinfo {title} {Hydrogen storage:
  High-pressure gas containment.},\ }\href@noop {} {\bibfield  {journal}
  {\bibinfo  {journal} {MRS Bulletin}\ }\textbf {\bibinfo {volume} {27}},\
  \bibinfo {pages} {680} (\bibinfo {year} {2002})}\BibitemShut {NoStop}%
\bibitem [{\citenamefont {Song}\ and\ \citenamefont {Curtin}(2013)}]{Song2013}%
  \BibitemOpen
  \bibfield  {author} {\bibinfo {author} {\bibfnamefont {J.}~\bibnamefont
  {Song}}\ and\ \bibinfo {author} {\bibfnamefont {W.~A.}\ \bibnamefont
  {Curtin}},\ }\bibfield  {title} {\bibinfo {title} {{Atomic mechanism and
  prediction of hydrogen embrittlement in iron}},\ }\href
  {https://doi.org/10.1038/nmat3479} {\bibfield  {journal} {\bibinfo  {journal}
  {Nature Materials}\ }\textbf {\bibinfo {volume} {12}},\ \bibinfo {pages}
  {145} (\bibinfo {year} {2013})}\BibitemShut {NoStop}%
\bibitem [{\citenamefont {Schapbach}\ and\ \citenamefont
  {Zuttel}(2001)}]{schapbach2001}%
  \BibitemOpen
  \bibfield  {author} {\bibinfo {author} {\bibfnamefont {L.}~\bibnamefont
  {Schapbach}}\ and\ \bibinfo {author} {\bibfnamefont {A.}~\bibnamefont
  {Zuttel}},\ }\bibfield  {title} {\bibinfo {title} {Hydrogen-storage materials
  for mobile applications.},\ }\href@noop {} {\bibfield  {journal} {\bibinfo
  {journal} {Nature}\ }\textbf {\bibinfo {volume} {414}},\ \bibinfo {pages}
  {353} (\bibinfo {year} {2001})}\BibitemShut {NoStop}%
\bibitem [{\citenamefont {Schneemann}\ \emph {et~al.}(2018)\citenamefont
  {Schneemann}, \citenamefont {White}, \citenamefont {Kang}, \citenamefont
  {Jeong}, \citenamefont {Wan}, \citenamefont {Cho}, \citenamefont {Heo},
  \citenamefont {Prendergast}, \citenamefont {Urban}, \citenamefont {Wood},
  \citenamefont {Allendorf},\ and\ \citenamefont {Stavila}}]{Schneemann2018}%
  \BibitemOpen
  \bibfield  {author} {\bibinfo {author} {\bibfnamefont {A.}~\bibnamefont
  {Schneemann}}, \bibinfo {author} {\bibfnamefont {J.~L.}\ \bibnamefont
  {White}}, \bibinfo {author} {\bibfnamefont {S.}~\bibnamefont {Kang}},
  \bibinfo {author} {\bibfnamefont {S.}~\bibnamefont {Jeong}}, \bibinfo
  {author} {\bibfnamefont {L.~F.}\ \bibnamefont {Wan}}, \bibinfo {author}
  {\bibfnamefont {E.~S.}\ \bibnamefont {Cho}}, \bibinfo {author} {\bibfnamefont
  {T.~W.}\ \bibnamefont {Heo}}, \bibinfo {author} {\bibfnamefont
  {D.}~\bibnamefont {Prendergast}}, \bibinfo {author} {\bibfnamefont {J.~J.}\
  \bibnamefont {Urban}}, \bibinfo {author} {\bibfnamefont {B.~C.}\ \bibnamefont
  {Wood}}, \bibinfo {author} {\bibfnamefont {M.~D.}\ \bibnamefont
  {Allendorf}},\ and\ \bibinfo {author} {\bibfnamefont {V.}~\bibnamefont
  {Stavila}},\ }\bibfield  {title} {\bibinfo {title} {{Nanostructured Metal
  Hydrides for Hydrogen Storage}},\ }\href
  {https://doi.org/10.1021/acs.chemrev.8b00313} {\bibfield  {journal} {\bibinfo
   {journal} {Chemical Reviews}\ }\textbf {\bibinfo {volume} {118}},\ \bibinfo
  {pages} {10775} (\bibinfo {year} {2018})}\BibitemShut {NoStop}%
\bibitem [{\citenamefont {Sakintuna}\ \emph {et~al.}(2007)\citenamefont
  {Sakintuna}, \citenamefont {Lamari-Darkrim},\ and\ \citenamefont
  {Hirscher}}]{Sakintuna2007}%
  \BibitemOpen
  \bibfield  {author} {\bibinfo {author} {\bibfnamefont {B.}~\bibnamefont
  {Sakintuna}}, \bibinfo {author} {\bibfnamefont {F.}~\bibnamefont
  {Lamari-Darkrim}},\ and\ \bibinfo {author} {\bibfnamefont {M.}~\bibnamefont
  {Hirscher}},\ }\bibfield  {title} {\bibinfo {title} {Metal hydride materials
  for solid hydrogen storage: A review},\ }\href
  {https://doi.org/10.1016/j.ijhydene.2006.11.022} {\bibfield  {journal}
  {\bibinfo  {journal} {International Journal of Hydrogen Energy}\ }\textbf
  {\bibinfo {volume} {32}},\ \bibinfo {pages} {1121} (\bibinfo {year}
  {2007})}\BibitemShut {NoStop}%
\bibitem [{\citenamefont {Aguey-Zinsou}\ and\ \citenamefont
  {Ares-Fernández}(2010)}]{fernandez2010}%
  \BibitemOpen
  \bibfield  {author} {\bibinfo {author} {\bibfnamefont {K.-F.}\ \bibnamefont
  {Aguey-Zinsou}}\ and\ \bibinfo {author} {\bibfnamefont {J.-R.}\ \bibnamefont
  {Ares-Fernández}},\ }\bibfield  {title} {\bibinfo {title} {Hydrogen in
  magnesium: New perspectives toward functional stores.},\ }\href
  {https://doi.org/10.1039/B921645F} {\bibfield  {journal} {\bibinfo  {journal}
  {Energy and Environmental Science}\ }\textbf {\bibinfo {volume} {3}},\
  \bibinfo {pages} {526} (\bibinfo {year} {2010})}\BibitemShut {NoStop}%
\bibitem [{\citenamefont {Webb}(2015)}]{webb2015}%
  \BibitemOpen
  \bibfield  {author} {\bibinfo {author} {\bibfnamefont {C.}~\bibnamefont
  {Webb}},\ }\bibfield  {title} {\bibinfo {title} {A review of
  catalyst-enhanced magnesium hydride as a hydrogen storage material.},\ }\href
  {https://doi.org/10.1016/j.jpcs.2014.06.014} {\bibfield  {journal} {\bibinfo
  {journal} {Journal of Physics and Chemistry of Solids}\ }\textbf {\bibinfo
  {volume} {84}},\ \bibinfo {pages} {96} (\bibinfo {year} {2015})}\BibitemShut
  {NoStop}%
\bibitem [{\citenamefont {Yartys}\ \emph {et~al.}(2019)\citenamefont {Yartys},
  \citenamefont {Denys}, \citenamefont {Lototskyy}, \citenamefont {Akiba},
  \citenamefont {Albert}, \citenamefont {Felderhoff}, \citenamefont {Antonov},
  \citenamefont {Kuzovnikov}, \citenamefont {Ares}, \citenamefont {Baricco},
  \citenamefont {Bourgeois}, \citenamefont {Crivello}, \citenamefont {Cuevas},
  \citenamefont {Joubert}, \citenamefont {Latroche}, \citenamefont {Buckley},
  \citenamefont {Humphries}, \citenamefont {Paskevicius}, \citenamefont
  {Sofianos}, \citenamefont {Bellosta~von Colbe}, \citenamefont {Dornheim},
  \citenamefont {Grant}, \citenamefont {Stuart}, \citenamefont {Walker},
  \citenamefont {Jacob}, \citenamefont {Jensen}, \citenamefont {de~Jongh},
  \citenamefont {Popilevsky}, \citenamefont {Skripnyuk}, \citenamefont
  {Rabkin}, \citenamefont {Wang}, \citenamefont {Zhu},\ and\ \citenamefont
  {Webb}}]{webb2019}%
  \BibitemOpen
  \bibfield  {author} {\bibinfo {author} {\bibfnamefont {V.}~\bibnamefont
  {Yartys}}, \bibinfo {author} {\bibfnamefont {B.}~\bibnamefont {Denys},
  \bibfnamefont {R.V.and~Hauback}}, \bibinfo {author} {\bibfnamefont
  {M.}~\bibnamefont {Lototskyy}}, \bibinfo {author} {\bibfnamefont
  {E.}~\bibnamefont {Akiba}}, \bibinfo {author} {\bibfnamefont
  {R.}~\bibnamefont {Albert}}, \bibinfo {author} {\bibfnamefont
  {M.}~\bibnamefont {Felderhoff}}, \bibinfo {author} {\bibfnamefont
  {V.}~\bibnamefont {Antonov}}, \bibinfo {author} {\bibfnamefont
  {M.}~\bibnamefont {Kuzovnikov}}, \bibinfo {author} {\bibfnamefont
  {J.}~\bibnamefont {Ares}}, \bibinfo {author} {\bibfnamefont {M.}~\bibnamefont
  {Baricco}}, \bibinfo {author} {\bibfnamefont {N.}~\bibnamefont {Bourgeois}},
  \bibinfo {author} {\bibfnamefont {J.-C.}\ \bibnamefont {Crivello}}, \bibinfo
  {author} {\bibfnamefont {F.}~\bibnamefont {Cuevas}}, \bibinfo {author}
  {\bibfnamefont {J.-M.}\ \bibnamefont {Joubert}}, \bibinfo {author}
  {\bibfnamefont {M.}~\bibnamefont {Latroche}}, \bibinfo {author}
  {\bibfnamefont {C.}~\bibnamefont {Buckley}}, \bibinfo {author} {\bibfnamefont
  {T.}~\bibnamefont {Humphries}}, \bibinfo {author} {\bibfnamefont
  {M.}~\bibnamefont {Paskevicius}}, \bibinfo {author} {\bibfnamefont
  {M.}~\bibnamefont {Sofianos}}, \bibinfo {author} {\bibfnamefont
  {J.}~\bibnamefont {Bellosta~von Colbe}}, \bibinfo {author} {\bibfnamefont
  {M.}~\bibnamefont {Dornheim}}, \bibinfo {author} {\bibfnamefont
  {D.}~\bibnamefont {Grant}}, \bibinfo {author} {\bibfnamefont
  {A.}~\bibnamefont {Stuart}}, \bibinfo {author} {\bibfnamefont
  {G.}~\bibnamefont {Walker}}, \bibinfo {author} {\bibfnamefont
  {I.}~\bibnamefont {Jacob}}, \bibinfo {author} {\bibfnamefont
  {T.}~\bibnamefont {Jensen}}, \bibinfo {author} {\bibfnamefont
  {L.}~\bibnamefont {de~Jongh}, \bibfnamefont {P.E.and~Pasquini}}, \bibinfo
  {author} {\bibfnamefont {L.}~\bibnamefont {Popilevsky}}, \bibinfo {author}
  {\bibfnamefont {V.}~\bibnamefont {Skripnyuk}}, \bibinfo {author}
  {\bibfnamefont {E.}~\bibnamefont {Rabkin}}, \bibinfo {author} {\bibfnamefont
  {H.}~\bibnamefont {Wang}}, \bibinfo {author} {\bibfnamefont {M.}~\bibnamefont
  {Zhu}},\ and\ \bibinfo {author} {\bibfnamefont {C.}~\bibnamefont {Webb}},\
  }\bibfield  {title} {\bibinfo {title} {Magnesium based materials for hydrogen
  based energy storage: Past, present and future.},\ }\href
  {https://doi.org/10.1016/j.ijhydene.2018.12.212} {\bibfield  {journal}
  {\bibinfo  {journal} {International Journal of Hydrogen Energy}\ }\textbf
  {\bibinfo {volume} {44}},\ \bibinfo {pages} {7809} (\bibinfo {year}
  {2019})}\BibitemShut {NoStop}%
\bibitem [{\citenamefont {Adams}\ and\ \citenamefont
  {Chen}(2011)}]{aicheng2011}%
  \BibitemOpen
  \bibfield  {author} {\bibinfo {author} {\bibfnamefont {B.~D.}\ \bibnamefont
  {Adams}}\ and\ \bibinfo {author} {\bibfnamefont {A.}~\bibnamefont {Chen}},\
  }\bibfield  {title} {\bibinfo {title} {The role of palladium in a hydrogen
  economy.},\ }\href {https://doi.org/10.1016/S1369-7021(11)70143-2} {\bibfield
   {journal} {\bibinfo  {journal} {Materials Today}\ }\textbf {\bibinfo
  {volume} {14}},\ \bibinfo {pages} {282 } (\bibinfo {year}
  {2011})}\BibitemShut {NoStop}%
\bibitem [{\citenamefont {Kirchheim}\ and\ \citenamefont
  {Pundt}(2014)}]{Kirchheim2014}%
  \BibitemOpen
  \bibfield  {author} {\bibinfo {author} {\bibfnamefont {R.}~\bibnamefont
  {Kirchheim}}\ and\ \bibinfo {author} {\bibfnamefont {A.}~\bibnamefont
  {Pundt}},\ }\bibfield  {title} {\bibinfo {title} {{Hydrogen in Metals}},\
  }in\ \href {https://doi.org/10.1016/B978-0-444-53770-6.00025-3} {\emph
  {\bibinfo {booktitle} {Physical Metallurgy}}}\ (\bibinfo  {publisher}
  {Elsevier},\ \bibinfo {year} {2014})\ pp.\ \bibinfo {pages}
  {2597--2705}\BibitemShut {NoStop}%
\bibitem [{\citenamefont {Wipf}(2001)}]{Wipf2001}%
  \BibitemOpen
  \bibfield  {author} {\bibinfo {author} {\bibfnamefont {H.}~\bibnamefont
  {Wipf}},\ }\bibfield  {title} {\bibinfo {title} {{Solubility and Diffusion of
  Hydrogen in Pure Metals and Alloys}},\ }\href
  {https://doi.org/10.1238/Physica.Topical.094a00043} {\bibfield  {journal}
  {\bibinfo  {journal} {Physica Scripta}\ }\textbf {\bibinfo {volume} {T94}},\
  \bibinfo {pages} {43} (\bibinfo {year} {2001})}\BibitemShut {NoStop}%
\bibitem [{\citenamefont {Kamakoti}(2003)}]{Kamakoti2003}%
  \BibitemOpen
  \bibfield  {author} {\bibinfo {author} {\bibfnamefont {P.}~\bibnamefont
  {Kamakoti}},\ }\bibfield  {title} {\bibinfo {title} {{A comparison of
  hydrogen diffusivities in Pd and CuPd alloys using density functional
  theory}},\ }\href {https://doi.org/10.1016/j.memsci.2003.07.008} {\bibfield
  {journal} {\bibinfo  {journal} {Journal of Membrane Science}\ }\textbf
  {\bibinfo {volume} {225}},\ \bibinfo {pages} {145} (\bibinfo {year}
  {2003})}\BibitemShut {NoStop}%
\bibitem [{\citenamefont {Baykara}(2004)}]{Baykara2004}%
  \BibitemOpen
  \bibfield  {author} {\bibinfo {author} {\bibfnamefont {S.}~\bibnamefont
  {Baykara}},\ }\bibfield  {title} {\bibinfo {title} {{Theoretical evaluation
  of diffusivity of hydrogen in palladium and rhodium}},\ }\href
  {https://doi.org/10.1016/j.ijhydene.2004.02.015} {\bibfield  {journal}
  {\bibinfo  {journal} {International Journal of Hydrogen Energy}\ }\textbf
  {\bibinfo {volume} {29}},\ \bibinfo {pages} {1631} (\bibinfo {year}
  {2004})}\BibitemShut {NoStop}%
\bibitem [{\citenamefont {Jiang}\ and\ \citenamefont
  {Carter}(2004)}]{Jiang2004}%
  \BibitemOpen
  \bibfield  {author} {\bibinfo {author} {\bibfnamefont {D.~E.}\ \bibnamefont
  {Jiang}}\ and\ \bibinfo {author} {\bibfnamefont {E.~A.}\ \bibnamefont
  {Carter}},\ }\bibfield  {title} {\bibinfo {title} {{Diffusion of interstitial
  hydrogen into and through bcc Fe from first principles}},\ }\href
  {https://doi.org/10.1103/PhysRevB.70.064102} {\bibfield  {journal} {\bibinfo
  {journal} {Physical Review B}\ }\textbf {\bibinfo {volume} {70}},\ \bibinfo
  {pages} {064102} (\bibinfo {year} {2004})}\BibitemShut {NoStop}%
\bibitem [{\citenamefont {Bhatia}\ and\ \citenamefont
  {Sholl}(2005)}]{Bhatia2005}%
  \BibitemOpen
  \bibfield  {author} {\bibinfo {author} {\bibfnamefont {B.}~\bibnamefont
  {Bhatia}}\ and\ \bibinfo {author} {\bibfnamefont {D.~S.}\ \bibnamefont
  {Sholl}},\ }\bibfield  {title} {\bibinfo {title} {{Quantitative assessment of
  hydrogen diffusion by activated hopping and quantum tunneling in ordered
  intermetallics}},\ }\href {https://doi.org/10.1103/PhysRevB.72.224302}
  {\bibfield  {journal} {\bibinfo  {journal} {Physical Review B}\ }\textbf
  {\bibinfo {volume} {72}},\ \bibinfo {pages} {224302} (\bibinfo {year}
  {2005})}\BibitemShut {NoStop}%
\bibitem [{\citenamefont {Wimmer}\ \emph {et~al.}(2008)\citenamefont {Wimmer},
  \citenamefont {Wolf}, \citenamefont {Sticht}, \citenamefont {Saxe},
  \citenamefont {Geller}, \citenamefont {Najafabadi},\ and\ \citenamefont
  {Young}}]{Wimmer2008}%
  \BibitemOpen
  \bibfield  {author} {\bibinfo {author} {\bibfnamefont {E.}~\bibnamefont
  {Wimmer}}, \bibinfo {author} {\bibfnamefont {W.}~\bibnamefont {Wolf}},
  \bibinfo {author} {\bibfnamefont {J.}~\bibnamefont {Sticht}}, \bibinfo
  {author} {\bibfnamefont {P.}~\bibnamefont {Saxe}}, \bibinfo {author}
  {\bibfnamefont {C.~B.}\ \bibnamefont {Geller}}, \bibinfo {author}
  {\bibfnamefont {R.}~\bibnamefont {Najafabadi}},\ and\ \bibinfo {author}
  {\bibfnamefont {G.~A.}\ \bibnamefont {Young}},\ }\bibfield  {title} {\bibinfo
  {title} {{Temperature-dependent diffusion coefficients from ab initio
  computations: Hydrogen, deuterium, and tritium in nickel}},\ }\href
  {https://doi.org/10.1103/PhysRevB.77.134305} {\bibfield  {journal} {\bibinfo
  {journal} {Physical Review B}\ }\textbf {\bibinfo {volume} {77}},\ \bibinfo
  {pages} {134305} (\bibinfo {year} {2008})}\BibitemShut {NoStop}%
\bibitem [{\citenamefont {Duan}\ \emph {et~al.}(2010)\citenamefont {Duan},
  \citenamefont {Liu}, \citenamefont {Zhou}, \citenamefont {Zhang},
  \citenamefont {Jin}, \citenamefont {Lu},\ and\ \citenamefont
  {Luo}}]{Duan2010}%
  \BibitemOpen
  \bibfield  {author} {\bibinfo {author} {\bibfnamefont {C.}~\bibnamefont
  {Duan}}, \bibinfo {author} {\bibfnamefont {Y.-L.}\ \bibnamefont {Liu}},
  \bibinfo {author} {\bibfnamefont {H.-B.}\ \bibnamefont {Zhou}}, \bibinfo
  {author} {\bibfnamefont {Y.}~\bibnamefont {Zhang}}, \bibinfo {author}
  {\bibfnamefont {S.}~\bibnamefont {Jin}}, \bibinfo {author} {\bibfnamefont
  {G.-H.}\ \bibnamefont {Lu}},\ and\ \bibinfo {author} {\bibfnamefont {G.-N.}\
  \bibnamefont {Luo}},\ }\bibfield  {title} {\bibinfo {title}
  {{First-principles study on dissolution and diffusion properties of hydrogen
  in molybdenum}},\ }\href {https://doi.org/10.1016/j.jnucmat.2010.06.029}
  {\bibfield  {journal} {\bibinfo  {journal} {Journal of Nuclear Materials}\
  }\textbf {\bibinfo {volume} {404}},\ \bibinfo {pages} {109} (\bibinfo {year}
  {2010})}\BibitemShut {NoStop}%
\bibitem [{\citenamefont {Conn{\'{e}}table}\ \emph {et~al.}(2011)\citenamefont
  {Conn{\'{e}}table}, \citenamefont {Huez}, \citenamefont {Andrieu},\ and\
  \citenamefont {Mijoule}}]{Connetable2011}%
  \BibitemOpen
  \bibfield  {author} {\bibinfo {author} {\bibfnamefont {D.}~\bibnamefont
  {Conn{\'{e}}table}}, \bibinfo {author} {\bibfnamefont {J.}~\bibnamefont
  {Huez}}, \bibinfo {author} {\bibfnamefont {{\'{E}}.}~\bibnamefont
  {Andrieu}},\ and\ \bibinfo {author} {\bibfnamefont {C.}~\bibnamefont
  {Mijoule}},\ }\bibfield  {title} {\bibinfo {title} {{First-principles study
  of diffusion and interactions of vacancies and hydrogen in hcp-titanium}},\
  }\href {https://doi.org/10.1088/0953-8984/23/40/405401} {\bibfield  {journal}
  {\bibinfo  {journal} {Journal of Physics: Condensed Matter}\ }\textbf
  {\bibinfo {volume} {23}},\ \bibinfo {pages} {405401} (\bibinfo {year}
  {2011})}\BibitemShut {NoStop}%
\bibitem [{\citenamefont {Liu}\ \emph {et~al.}(2012)\citenamefont {Liu},
  \citenamefont {Jin}, \citenamefont {Sun},\ and\ \citenamefont
  {Duan}}]{Liu2012}%
  \BibitemOpen
  \bibfield  {author} {\bibinfo {author} {\bibfnamefont {Y.-L.}\ \bibnamefont
  {Liu}}, \bibinfo {author} {\bibfnamefont {S.}~\bibnamefont {Jin}}, \bibinfo
  {author} {\bibfnamefont {L.}~\bibnamefont {Sun}},\ and\ \bibinfo {author}
  {\bibfnamefont {C.}~\bibnamefont {Duan}},\ }\bibfield  {title} {\bibinfo
  {title} {{First-principles investigation on diffusion and permeation
  behaviors of hydrogen isotopes in molybdenum}},\ }\href
  {https://doi.org/10.1016/j.commatsci.2011.11.002} {\bibfield  {journal}
  {\bibinfo  {journal} {Computational Materials Science}\ }\textbf {\bibinfo
  {volume} {54}},\ \bibinfo {pages} {32} (\bibinfo {year} {2012})}\BibitemShut
  {NoStop}%
\bibitem [{\citenamefont {Lu}\ and\ \citenamefont {Zhang}(2013)}]{Lu2013}%
  \BibitemOpen
  \bibfield  {author} {\bibinfo {author} {\bibfnamefont {Y.}~\bibnamefont
  {Lu}}\ and\ \bibinfo {author} {\bibfnamefont {P.}~\bibnamefont {Zhang}},\
  }\bibfield  {title} {\bibinfo {title} {{First-principles study of
  temperature-dependent diffusion coefficients: Hydrogen, deuterium, and
  tritium in $\alpha$-Ti}},\ }\href {https://doi.org/10.1063/1.4805362}
  {\bibfield  {journal} {\bibinfo  {journal} {Journal of Applied Physics}\
  }\textbf {\bibinfo {volume} {113}},\ \bibinfo {pages} {193502} (\bibinfo
  {year} {2013})}\BibitemShut {NoStop}%
\bibitem [{\citenamefont {Fernandez}\ \emph {et~al.}(2015)\citenamefont
  {Fernandez}, \citenamefont {Ferro},\ and\ \citenamefont
  {Kato}}]{Fernandez2015}%
  \BibitemOpen
  \bibfield  {author} {\bibinfo {author} {\bibfnamefont {N.}~\bibnamefont
  {Fernandez}}, \bibinfo {author} {\bibfnamefont {Y.}~\bibnamefont {Ferro}},\
  and\ \bibinfo {author} {\bibfnamefont {D.}~\bibnamefont {Kato}},\ }\bibfield
  {title} {\bibinfo {title} {{Hydrogen diffusion and vacancies formation in
  tungsten: Density Functional Theory calculations and statistical models}},\
  }\href {https://doi.org/10.1016/j.actamat.2015.04.052} {\bibfield  {journal}
  {\bibinfo  {journal} {Acta Materialia}\ }\textbf {\bibinfo {volume} {94}},\
  \bibinfo {pages} {307} (\bibinfo {year} {2015})}\BibitemShut {NoStop}%
\bibitem [{\citenamefont {Tafen}(2015)}]{Tafen2015}%
  \BibitemOpen
  \bibfield  {author} {\bibinfo {author} {\bibfnamefont {D.~N.}\ \bibnamefont
  {Tafen}},\ }\bibfield  {title} {\bibinfo {title} {{First-principles-based
  kinetic Monte Carlo studies of diffusion of hydrogen in Ni–Al and Ni–Fe
  binary alloys}},\ }\href {https://doi.org/10.1007/s10853-015-8885-4}
  {\bibfield  {journal} {\bibinfo  {journal} {Journal of Materials Science}\
  }\textbf {\bibinfo {volume} {50}},\ \bibinfo {pages} {3361} (\bibinfo {year}
  {2015})}\BibitemShut {NoStop}%
\bibitem [{\citenamefont {Han}\ \emph {et~al.}(2016)\citenamefont {Han},
  \citenamefont {Zhou}, \citenamefont {Ma},\ and\ \citenamefont
  {Liu}}]{Han2016}%
  \BibitemOpen
  \bibfield  {author} {\bibinfo {author} {\bibfnamefont {Q.-F.}\ \bibnamefont
  {Han}}, \bibinfo {author} {\bibfnamefont {Z.-Y.}\ \bibnamefont {Zhou}},
  \bibinfo {author} {\bibfnamefont {Y.}~\bibnamefont {Ma}},\ and\ \bibinfo
  {author} {\bibfnamefont {Y.-L.}\ \bibnamefont {Liu}},\ }\bibfield  {title}
  {\bibinfo {title} {{A comparative investigation of the behaviors of H in Au
  and Ag from first principles}},\ }\href
  {https://doi.org/10.1088/0965-0393/24/4/045009} {\bibfield  {journal}
  {\bibinfo  {journal} {Modelling and Simulation in Materials Science and
  Engineering}\ }\textbf {\bibinfo {volume} {24}},\ \bibinfo {pages} {045009}
  (\bibinfo {year} {2016})}\BibitemShut {NoStop}%
\bibitem [{\citenamefont {Yang}\ \emph {et~al.}(2016)\citenamefont {Yang},
  \citenamefont {Lu},\ and\ \citenamefont {Zhang}}]{Yang2016}%
  \BibitemOpen
  \bibfield  {author} {\bibinfo {author} {\bibfnamefont {X.-Y.}\ \bibnamefont
  {Yang}}, \bibinfo {author} {\bibfnamefont {Y.}~\bibnamefont {Lu}},\ and\
  \bibinfo {author} {\bibfnamefont {P.}~\bibnamefont {Zhang}},\ }\bibfield
  {title} {\bibinfo {title} {{First-principles study of the stability and
  diffusion properties of hydrogen in zirconium carbide}},\ }\href
  {https://doi.org/10.1016/j.jnucmat.2016.07.008} {\bibfield  {journal}
  {\bibinfo  {journal} {Journal of Nuclear Materials}\ }\textbf {\bibinfo
  {volume} {479}},\ \bibinfo {pages} {130} (\bibinfo {year}
  {2016})}\BibitemShut {NoStop}%
\bibitem [{\citenamefont {Liu}\ \emph {et~al.}(2017)\citenamefont {Liu},
  \citenamefont {Wang}, \citenamefont {He},\ and\ \citenamefont
  {Gong}}]{Liu2017}%
  \BibitemOpen
  \bibfield  {author} {\bibinfo {author} {\bibfnamefont {L.}~\bibnamefont
  {Liu}}, \bibinfo {author} {\bibfnamefont {J.}~\bibnamefont {Wang}}, \bibinfo
  {author} {\bibfnamefont {Y.}~\bibnamefont {He}},\ and\ \bibinfo {author}
  {\bibfnamefont {H.}~\bibnamefont {Gong}},\ }\bibfield  {title} {\bibinfo
  {title} {{Solubility, diffusivity, and permeability of hydrogen at PdCu
  phases}},\ }\href {https://doi.org/10.1016/j.memsci.2017.07.057} {\bibfield
  {journal} {\bibinfo  {journal} {Journal of Membrane Science}\ }\textbf
  {\bibinfo {volume} {542}},\ \bibinfo {pages} {24} (\bibinfo {year}
  {2017})}\BibitemShut {NoStop}%
\bibitem [{\citenamefont {Liu}\ \emph {et~al.}(2018)\citenamefont {Liu},
  \citenamefont {Wang}, \citenamefont {Qian}, \citenamefont {He}, \citenamefont
  {Gong}, \citenamefont {Liang},\ and\ \citenamefont {Zhou}}]{Liu2018}%
  \BibitemOpen
  \bibfield  {author} {\bibinfo {author} {\bibfnamefont {L.}~\bibnamefont
  {Liu}}, \bibinfo {author} {\bibfnamefont {J.}~\bibnamefont {Wang}}, \bibinfo
  {author} {\bibfnamefont {J.}~\bibnamefont {Qian}}, \bibinfo {author}
  {\bibfnamefont {Y.}~\bibnamefont {He}}, \bibinfo {author} {\bibfnamefont
  {H.}~\bibnamefont {Gong}}, \bibinfo {author} {\bibfnamefont {C.}~\bibnamefont
  {Liang}},\ and\ \bibinfo {author} {\bibfnamefont {S.}~\bibnamefont {Zhou}},\
  }\bibfield  {title} {\bibinfo {title} {{Fundamental effects of Ag alloying on
  hydrogen behaviors in PdCu}},\ }\href
  {https://doi.org/10.1016/j.memsci.2017.12.078} {\bibfield  {journal}
  {\bibinfo  {journal} {Journal of Membrane Science}\ }\textbf {\bibinfo
  {volume} {550}},\ \bibinfo {pages} {230} (\bibinfo {year}
  {2018})}\BibitemShut {NoStop}%
\bibitem [{\citenamefont {Oliveira}\ \emph {et~al.}(2019)\citenamefont
  {Oliveira}, \citenamefont {Tenorio},\ and\ \citenamefont
  {Rocha}}]{Oliveira2019}%
  \BibitemOpen
  \bibfield  {author} {\bibinfo {author} {\bibfnamefont {R.~R.}\ \bibnamefont
  {Oliveira}}, \bibinfo {author} {\bibfnamefont {B.~N.}\ \bibnamefont
  {Tenorio}},\ and\ \bibinfo {author} {\bibfnamefont {A.~B.}\ \bibnamefont
  {Rocha}},\ }\bibfield  {title} {\bibinfo {title} {{Ab initio study of
  diffusion of hydrogen, silver and lithium in PbS and Ag2S}},\ }\href
  {https://doi.org/10.1016/j.commatsci.2019.04.046} {\bibfield  {journal}
  {\bibinfo  {journal} {Computational Materials Science}\ }\textbf {\bibinfo
  {volume} {166}},\ \bibinfo {pages} {75} (\bibinfo {year} {2019})}\BibitemShut
  {NoStop}%
\bibitem [{\citenamefont {Yang}\ and\ \citenamefont {Wirth}(2019)}]{Yang2019}%
  \BibitemOpen
  \bibfield  {author} {\bibinfo {author} {\bibfnamefont {L.}~\bibnamefont
  {Yang}}\ and\ \bibinfo {author} {\bibfnamefont {B.~D.}\ \bibnamefont
  {Wirth}},\ }\bibfield  {title} {\bibinfo {title} {{First-principles study of
  hydrogen diffusion and self-clustering below tungsten surfaces}},\ }\href
  {https://doi.org/10.1063/1.5092595} {\bibfield  {journal} {\bibinfo
  {journal} {Journal of Applied Physics}\ }\textbf {\bibinfo {volume} {125}},\
  \bibinfo {pages} {165105} (\bibinfo {year} {2019})}\BibitemShut {NoStop}%
\bibitem [{\citenamefont {Conn{\'{e}}table}(2019)}]{Connetable2019}%
  \BibitemOpen
  \bibfield  {author} {\bibinfo {author} {\bibfnamefont {D.}~\bibnamefont
  {Conn{\'{e}}table}},\ }\bibfield  {title} {\bibinfo {title} {{Theoretical
  study of the insertion and diffusivity of hydrogen in the Ti3Al-D019 system:
  Comparison with Ti-hcp and TiAl-L10 systems}},\ }\href
  {https://doi.org/10.1016/j.ijhydene.2019.10.095} {\bibfield  {journal}
  {\bibinfo  {journal} {International Journal of Hydrogen Energy}\ }\textbf
  {\bibinfo {volume} {44}},\ \bibinfo {pages} {32307} (\bibinfo {year}
  {2019})}\BibitemShut {NoStop}%
\bibitem [{\citenamefont {Onwudinanti}\ \emph {et~al.}(2020)\citenamefont
  {Onwudinanti}, \citenamefont {Brocks}, \citenamefont {Koelman}, \citenamefont
  {Morgan},\ and\ \citenamefont {Tao}}]{Onwudinanti2020}%
  \BibitemOpen
  \bibfield  {author} {\bibinfo {author} {\bibfnamefont {C.}~\bibnamefont
  {Onwudinanti}}, \bibinfo {author} {\bibfnamefont {G.}~\bibnamefont {Brocks}},
  \bibinfo {author} {\bibfnamefont {V.}~\bibnamefont {Koelman}}, \bibinfo
  {author} {\bibfnamefont {T.}~\bibnamefont {Morgan}},\ and\ \bibinfo {author}
  {\bibfnamefont {S.}~\bibnamefont {Tao}},\ }\bibfield  {title} {\bibinfo
  {title} {{Hydrogen diffusion out of ruthenium—an ab initio study of the
  role of adsorbates}},\ }\href {https://doi.org/10.1039/D0CP00448K} {\bibfield
   {journal} {\bibinfo  {journal} {Physical Chemistry Chemical Physics}\
  }\textbf {\bibinfo {volume} {22}},\ \bibinfo {pages} {7935} (\bibinfo {year}
  {2020})}\BibitemShut {NoStop}%
\bibitem [{\citenamefont {Ferro}(1957)}]{ferro1957}%
  \BibitemOpen
  \bibfield  {author} {\bibinfo {author} {\bibfnamefont {A.}~\bibnamefont
  {Ferro}},\ }\bibfield  {title} {\bibinfo {title} {Theory of diffusion
  constants in interstitial solid solutions of b.c.c. metals},\ }\href
  {https://doi.org/10.1063/1.1722883} {\bibfield  {journal} {\bibinfo
  {journal} {Journal of Applied Physics}\ }\textbf {\bibinfo {volume} {28}},\
  \bibinfo {pages} {895} (\bibinfo {year} {1957})}\BibitemShut {NoStop}%
\bibitem [{\citenamefont {McNeil}(1965)}]{mcneil1965}%
  \BibitemOpen
  \bibfield  {author} {\bibinfo {author} {\bibfnamefont {M.~B.}\ \bibnamefont
  {McNeil}},\ }\bibfield  {title} {\bibinfo {title} {{Calculation of Activation
  Energies for Interstitial Diffusion by a Simple Elastic Model}},\ }\href
  {https://doi.org/10.1063/1.1714481} {\bibfield  {journal} {\bibinfo
  {journal} {Journal of Applied Physics}\ }\textbf {\bibinfo {volume} {36}},\
  \bibinfo {pages} {2328} (\bibinfo {year} {1965})}\BibitemShut {NoStop}%
\bibitem [{\citenamefont {Flynn}\ and\ \citenamefont
  {Stoneham}(1970)}]{Flynn1970}%
  \BibitemOpen
  \bibfield  {author} {\bibinfo {author} {\bibfnamefont {C.~P.}\ \bibnamefont
  {Flynn}}\ and\ \bibinfo {author} {\bibfnamefont {A.~M.}\ \bibnamefont
  {Stoneham}},\ }\bibfield  {title} {\bibinfo {title} {{Quantum Theory of
  Diffusion with Application to Light Interstitials in Metals}},\ }\href
  {https://doi.org/10.1103/PhysRevB.1.3966} {\bibfield  {journal} {\bibinfo
  {journal} {Physical Review B}\ }\textbf {\bibinfo {volume} {1}},\ \bibinfo
  {pages} {3966} (\bibinfo {year} {1970})}\BibitemShut {NoStop}%
\bibitem [{\citenamefont {Zeng}\ \emph {et~al.}(2018)\citenamefont {Zeng},
  \citenamefont {Li},\ and\ \citenamefont {Bai}}]{Zeng2018}%
  \BibitemOpen
  \bibfield  {author} {\bibinfo {author} {\bibfnamefont {Y.}~\bibnamefont
  {Zeng}}, \bibinfo {author} {\bibfnamefont {Q.}~\bibnamefont {Li}},\ and\
  \bibinfo {author} {\bibfnamefont {K.}~\bibnamefont {Bai}},\ }\bibfield
  {title} {\bibinfo {title} {{Prediction of interstitial diffusion activation
  energies of nitrogen, oxygen, boron and carbon in bcc, fcc, and hcp metals
  using machine learning}},\ }\href
  {https://doi.org/10.1016/j.commatsci.2017.12.030} {\bibfield  {journal}
  {\bibinfo  {journal} {Computational Materials Science}\ }\textbf {\bibinfo
  {volume} {144}},\ \bibinfo {pages} {232} (\bibinfo {year}
  {2018})}\BibitemShut {NoStop}%
\bibitem [{\citenamefont {Wu}\ \emph {et~al.}(2017)\citenamefont {Wu},
  \citenamefont {Lorenson}, \citenamefont {Anderson}, \citenamefont {Witteman},
  \citenamefont {Wu}, \citenamefont {Meredig},\ and\ \citenamefont
  {Morgan}}]{Wu2017}%
  \BibitemOpen
  \bibfield  {author} {\bibinfo {author} {\bibfnamefont {H.}~\bibnamefont
  {Wu}}, \bibinfo {author} {\bibfnamefont {A.}~\bibnamefont {Lorenson}},
  \bibinfo {author} {\bibfnamefont {B.}~\bibnamefont {Anderson}}, \bibinfo
  {author} {\bibfnamefont {L.}~\bibnamefont {Witteman}}, \bibinfo {author}
  {\bibfnamefont {H.}~\bibnamefont {Wu}}, \bibinfo {author} {\bibfnamefont
  {B.}~\bibnamefont {Meredig}},\ and\ \bibinfo {author} {\bibfnamefont
  {D.}~\bibnamefont {Morgan}},\ }\bibfield  {title} {\bibinfo {title} {{Robust
  FCC solute diffusion predictions from ab-initio machine learning methods}},\
  }\href {https://doi.org/10.1016/j.commatsci.2017.03.052} {\bibfield
  {journal} {\bibinfo  {journal} {Computational Materials Science}\ }\textbf
  {\bibinfo {volume} {134}},\ \bibinfo {pages} {160} (\bibinfo {year}
  {2017})}\BibitemShut {NoStop}%
\bibitem [{\citenamefont {Wei}\ \emph {et~al.}(2021)\citenamefont {Wei},
  \citenamefont {Yu}, \citenamefont {Lu}, \citenamefont {Han}, \citenamefont
  {Wang},\ and\ \citenamefont {Liu}}]{Wei2021}%
  \BibitemOpen
  \bibfield  {author} {\bibinfo {author} {\bibfnamefont {Z.}~\bibnamefont
  {Wei}}, \bibinfo {author} {\bibfnamefont {J.}~\bibnamefont {Yu}}, \bibinfo
  {author} {\bibfnamefont {Y.}~\bibnamefont {Lu}}, \bibinfo {author}
  {\bibfnamefont {J.}~\bibnamefont {Han}}, \bibinfo {author} {\bibfnamefont
  {C.}~\bibnamefont {Wang}},\ and\ \bibinfo {author} {\bibfnamefont
  {X.}~\bibnamefont {Liu}},\ }\bibfield  {title} {\bibinfo {title} {{Prediction
  of diffusion coefficients in fcc, bcc and hcp phases remained stable or
  metastable by the machine-learning methods}},\ }\href
  {https://doi.org/10.1016/j.matdes.2020.109287} {\bibfield  {journal}
  {\bibinfo  {journal} {Materials \& Design}\ }\textbf {\bibinfo {volume}
  {198}},\ \bibinfo {pages} {109287} (\bibinfo {year} {2021})}\BibitemShut
  {NoStop}%
\bibitem [{\citenamefont {He}\ \emph {et~al.}(2020)\citenamefont {He},
  \citenamefont {Kong},\ and\ \citenamefont {Liu}}]{He2020}%
  \BibitemOpen
  \bibfield  {author} {\bibinfo {author} {\bibfnamefont {K.-n.}\ \bibnamefont
  {He}}, \bibinfo {author} {\bibfnamefont {X.-s.}\ \bibnamefont {Kong}},\ and\
  \bibinfo {author} {\bibfnamefont {C.}~\bibnamefont {Liu}},\ }\bibfield
  {title} {\bibinfo {title} {{Robust activation energy predictions of solute
  diffusion from machine learning method}},\ }\href
  {https://doi.org/10.1016/j.commatsci.2020.109948} {\bibfield  {journal}
  {\bibinfo  {journal} {Computational Materials Science}\ }\textbf {\bibinfo
  {volume} {184}},\ \bibinfo {pages} {109948} (\bibinfo {year}
  {2020})}\BibitemShut {NoStop}%
\bibitem [{\citenamefont {Zhou}\ \emph {et~al.}(2022)\citenamefont {Zhou},
  \citenamefont {Zhu}, \citenamefont {Wu}, \citenamefont {Yang}, \citenamefont
  {Lookman},\ and\ \citenamefont {Wu}}]{Zhou2022}%
  \BibitemOpen
  \bibfield  {author} {\bibinfo {author} {\bibfnamefont {X.-Y.}\ \bibnamefont
  {Zhou}}, \bibinfo {author} {\bibfnamefont {J.-H.}\ \bibnamefont {Zhu}},
  \bibinfo {author} {\bibfnamefont {Y.}~\bibnamefont {Wu}}, \bibinfo {author}
  {\bibfnamefont {X.-S.}\ \bibnamefont {Yang}}, \bibinfo {author}
  {\bibfnamefont {T.}~\bibnamefont {Lookman}},\ and\ \bibinfo {author}
  {\bibfnamefont {H.-H.}\ \bibnamefont {Wu}},\ }\bibfield  {title} {\bibinfo
  {title} {{Machine learning assisted design of FeCoNiCrMn high-entropy alloys
  with ultra-low hydrogen diffusion coefficients}},\ }\href
  {https://doi.org/10.1016/j.actamat.2021.117535} {\bibfield  {journal}
  {\bibinfo  {journal} {Acta Materialia}\ }\textbf {\bibinfo {volume} {224}},\
  \bibinfo {pages} {117535} (\bibinfo {year} {2022})}\BibitemShut {NoStop}%
\bibitem [{\citenamefont {Kimizuka}\ \emph {et~al.}(2022)\citenamefont
  {Kimizuka}, \citenamefont {Thomsen},\ and\ \citenamefont
  {Shiga}}]{Kimizuka2022}%
  \BibitemOpen
  \bibfield  {author} {\bibinfo {author} {\bibfnamefont {H.}~\bibnamefont
  {Kimizuka}}, \bibinfo {author} {\bibfnamefont {B.}~\bibnamefont {Thomsen}},\
  and\ \bibinfo {author} {\bibfnamefont {M.}~\bibnamefont {Shiga}},\ }\bibfield
   {title} {\bibinfo {title} {{Artificial neural network-based path integral
  simulations of hydrogen isotope diffusion in palladium}},\ }\href
  {https://doi.org/10.1088/2515-7655/ac7e6b} {\bibfield  {journal} {\bibinfo
  {journal} {Journal of Physics: Energy}\ }\textbf {\bibinfo {volume} {4}},\
  \bibinfo {pages} {034004} (\bibinfo {year} {2022})}\BibitemShut {NoStop}%
\bibitem [{\citenamefont {Hattrick-Simpers}\ \emph {et~al.}(2018)\citenamefont
  {Hattrick-Simpers}, \citenamefont {Choudhary},\ and\ \citenamefont
  {Corgnale}}]{Hattrick-Simpers2018}%
  \BibitemOpen
  \bibfield  {author} {\bibinfo {author} {\bibfnamefont {J.~R.}\ \bibnamefont
  {Hattrick-Simpers}}, \bibinfo {author} {\bibfnamefont {K.}~\bibnamefont
  {Choudhary}},\ and\ \bibinfo {author} {\bibfnamefont {C.}~\bibnamefont
  {Corgnale}},\ }\bibfield  {title} {\bibinfo {title} {{A simple constrained
  machine learning model for predicting high-pressure-hydrogen-compressor
  materials}},\ }\href {https://doi.org/10.1039/C8ME00005K} {\bibfield
  {journal} {\bibinfo  {journal} {Mol. Syst. Des. Eng.}\ }\textbf {\bibinfo
  {volume} {3}},\ \bibinfo {pages} {509} (\bibinfo {year} {2018})}\BibitemShut
  {NoStop}%
\bibitem [{\citenamefont {Witman}\ \emph {et~al.}(2020)\citenamefont {Witman},
  \citenamefont {Ling}, \citenamefont {Grant}, \citenamefont {Walker},
  \citenamefont {Agarwal}, \citenamefont {Stavila},\ and\ \citenamefont
  {Allendorf}}]{Witman2020}%
  \BibitemOpen
  \bibfield  {author} {\bibinfo {author} {\bibfnamefont {M.}~\bibnamefont
  {Witman}}, \bibinfo {author} {\bibfnamefont {S.}~\bibnamefont {Ling}},
  \bibinfo {author} {\bibfnamefont {D.~M.}\ \bibnamefont {Grant}}, \bibinfo
  {author} {\bibfnamefont {G.~S.}\ \bibnamefont {Walker}}, \bibinfo {author}
  {\bibfnamefont {S.}~\bibnamefont {Agarwal}}, \bibinfo {author} {\bibfnamefont
  {V.}~\bibnamefont {Stavila}},\ and\ \bibinfo {author} {\bibfnamefont {M.~D.}\
  \bibnamefont {Allendorf}},\ }\bibfield  {title} {\bibinfo {title}
  {{Extracting an Empirical Intermetallic Hydride Design Principle from Limited
  Data via Interpretable Machine Learning}},\ }\href
  {https://doi.org/10.1021/acs.jpclett.9b02971} {\bibfield  {journal} {\bibinfo
   {journal} {The Journal of Physical Chemistry Letters}\ }\textbf {\bibinfo
  {volume} {11}},\ \bibinfo {pages} {40} (\bibinfo {year} {2020})}\BibitemShut
  {NoStop}%
\bibitem [{\citenamefont {Witman}\ \emph {et~al.}(2021)\citenamefont {Witman},
  \citenamefont {Ek}, \citenamefont {Ling}, \citenamefont {Chames},
  \citenamefont {Agarwal}, \citenamefont {Wong}, \citenamefont {Allendorf},
  \citenamefont {Sahlberg},\ and\ \citenamefont {Stavila}}]{Witman2021}%
  \BibitemOpen
  \bibfield  {author} {\bibinfo {author} {\bibfnamefont {M.}~\bibnamefont
  {Witman}}, \bibinfo {author} {\bibfnamefont {G.}~\bibnamefont {Ek}}, \bibinfo
  {author} {\bibfnamefont {S.}~\bibnamefont {Ling}}, \bibinfo {author}
  {\bibfnamefont {J.}~\bibnamefont {Chames}}, \bibinfo {author} {\bibfnamefont
  {S.}~\bibnamefont {Agarwal}}, \bibinfo {author} {\bibfnamefont
  {J.}~\bibnamefont {Wong}}, \bibinfo {author} {\bibfnamefont {M.~D.}\
  \bibnamefont {Allendorf}}, \bibinfo {author} {\bibfnamefont {M.}~\bibnamefont
  {Sahlberg}},\ and\ \bibinfo {author} {\bibfnamefont {V.}~\bibnamefont
  {Stavila}},\ }\bibfield  {title} {\bibinfo {title} {{Data-Driven Discovery
  and Synthesis of High Entropy Alloy Hydrides with Targeted Thermodynamic
  Stability}},\ }\href {https://doi.org/10.1021/acs.chemmater.1c00647}
  {\bibfield  {journal} {\bibinfo  {journal} {Chem. Mater.}\ }\textbf {\bibinfo
  {volume} {33}},\ \bibinfo {pages} {4067} (\bibinfo {year}
  {2021})}\BibitemShut {NoStop}%
\bibitem [{\citenamefont {Ward}\ \emph {et~al.}(2018)\citenamefont {Ward},
  \citenamefont {Dunn}, \citenamefont {Faghaninia}, \citenamefont {Zimmermann},
  \citenamefont {Bajaj}, \citenamefont {Wang}, \citenamefont {Montoya},
  \citenamefont {Chen}, \citenamefont {Bystrom}, \citenamefont {Dylla},
  \citenamefont {Chard}, \citenamefont {Asta}, \citenamefont {Persson},
  \citenamefont {Snyder}, \citenamefont {Foster},\ and\ \citenamefont
  {Jain}}]{Ward2018}%
  \BibitemOpen
  \bibfield  {author} {\bibinfo {author} {\bibfnamefont {L.}~\bibnamefont
  {Ward}}, \bibinfo {author} {\bibfnamefont {A.}~\bibnamefont {Dunn}}, \bibinfo
  {author} {\bibfnamefont {A.}~\bibnamefont {Faghaninia}}, \bibinfo {author}
  {\bibfnamefont {N.~E.}\ \bibnamefont {Zimmermann}}, \bibinfo {author}
  {\bibfnamefont {S.}~\bibnamefont {Bajaj}}, \bibinfo {author} {\bibfnamefont
  {Q.}~\bibnamefont {Wang}}, \bibinfo {author} {\bibfnamefont {J.}~\bibnamefont
  {Montoya}}, \bibinfo {author} {\bibfnamefont {J.}~\bibnamefont {Chen}},
  \bibinfo {author} {\bibfnamefont {K.}~\bibnamefont {Bystrom}}, \bibinfo
  {author} {\bibfnamefont {M.}~\bibnamefont {Dylla}}, \bibinfo {author}
  {\bibfnamefont {K.}~\bibnamefont {Chard}}, \bibinfo {author} {\bibfnamefont
  {M.}~\bibnamefont {Asta}}, \bibinfo {author} {\bibfnamefont {K.~A.}\
  \bibnamefont {Persson}}, \bibinfo {author} {\bibfnamefont {G.~J.}\
  \bibnamefont {Snyder}}, \bibinfo {author} {\bibfnamefont {I.}~\bibnamefont
  {Foster}},\ and\ \bibinfo {author} {\bibfnamefont {A.}~\bibnamefont {Jain}},\
  }\bibfield  {title} {\bibinfo {title} {{Matminer: An open source toolkit for
  materials data mining}},\ }\href
  {https://doi.org/10.1016/j.commatsci.2018.05.018} {\bibfield  {journal}
  {\bibinfo  {journal} {Comput. Mater. Sci.}\ }\textbf {\bibinfo {volume}
  {152}},\ \bibinfo {pages} {60} (\bibinfo {year} {2018})}\BibitemShut
  {NoStop}%
\bibitem [{\citenamefont {Ward}\ \emph {et~al.}(2016)\citenamefont {Ward},
  \citenamefont {Agrawal}, \citenamefont {Choudhary},\ and\ \citenamefont
  {Wolverton}}]{Ward2016}%
  \BibitemOpen
  \bibfield  {author} {\bibinfo {author} {\bibfnamefont {L.}~\bibnamefont
  {Ward}}, \bibinfo {author} {\bibfnamefont {A.}~\bibnamefont {Agrawal}},
  \bibinfo {author} {\bibfnamefont {A.}~\bibnamefont {Choudhary}},\ and\
  \bibinfo {author} {\bibfnamefont {C.}~\bibnamefont {Wolverton}},\ }\bibfield
  {title} {\bibinfo {title} {{A general-purpose machine learning framework for
  predicting properties of inorganic materials}},\ }\href
  {https://doi.org/10.1038/npjcompumats.2016.28} {\bibfield  {journal}
  {\bibinfo  {journal} {npj Computational Materials}\ }\textbf {\bibinfo
  {volume} {2}},\ \bibinfo {pages} {16028} (\bibinfo {year}
  {2016})}\BibitemShut {NoStop}%
\bibitem [{\citenamefont {Xie}\ and\ \citenamefont {Grossman}(2018)}]{Xie2018}%
  \BibitemOpen
  \bibfield  {author} {\bibinfo {author} {\bibfnamefont {T.}~\bibnamefont
  {Xie}}\ and\ \bibinfo {author} {\bibfnamefont {J.~C.}\ \bibnamefont
  {Grossman}},\ }\bibfield  {title} {\bibinfo {title} {{Crystal Graph
  Convolutional Neural Networks for an Accurate and Interpretable Prediction of
  Material Properties}},\ }\href
  {https://doi.org/10.1103/PhysRevLett.120.145301} {\bibfield  {journal}
  {\bibinfo  {journal} {Phys. Rev. Lett.}\ }\textbf {\bibinfo {volume} {120}},\
  \bibinfo {pages} {145301} (\bibinfo {year} {2018})},\ \Eprint
  {https://arxiv.org/abs/1710.10324} {arXiv:1710.10324} \BibitemShut {NoStop}%
\bibitem [{\citenamefont {St{\"{o}}ver}\ \emph {et~al.}(1986)\citenamefont
  {St{\"{o}}ver}, \citenamefont {Buchkremer},\ and\ \citenamefont
  {Hecker}}]{Stover1986}%
  \BibitemOpen
  \bibfield  {author} {\bibinfo {author} {\bibfnamefont {D.}~\bibnamefont
  {St{\"{o}}ver}}, \bibinfo {author} {\bibfnamefont {H.}~\bibnamefont
  {Buchkremer}},\ and\ \bibinfo {author} {\bibfnamefont {R.}~\bibnamefont
  {Hecker}},\ }\bibfield  {title} {\bibinfo {title} {{Hydrogen and deuterium
  permeation through metallic and surface-oxidized chromium}},\ }\href
  {https://doi.org/10.1016/0257-8972(86)90085-X} {\bibfield  {journal}
  {\bibinfo  {journal} {Surface and Coatings Technology}\ }\textbf {\bibinfo
  {volume} {28}},\ \bibinfo {pages} {281} (\bibinfo {year} {1986})}\BibitemShut
  {NoStop}%
\bibitem [{\citenamefont {Jost}\ and\ \citenamefont
  {Widmann}(1940)}]{Jost1940}%
  \BibitemOpen
  \bibfield  {author} {\bibinfo {author} {\bibfnamefont {W.}~\bibnamefont
  {Jost}}\ and\ \bibinfo {author} {\bibfnamefont {A.}~\bibnamefont {Widmann}},\
  }\bibfield  {title} {\bibinfo {title} {{{\"{U}}ber die Diffusion von
  Wasserstoff und von Deuterium in Palladium. II}},\ }\href
  {https://doi.org/10.1515/zpch-1940-4523} {\bibfield  {journal} {\bibinfo
  {journal} {Zeitschrift f{\"{u}}r Physikalische Chemie}\ }\textbf {\bibinfo
  {volume} {45B}},\ \bibinfo {pages} {285} (\bibinfo {year}
  {1940})}\BibitemShut {NoStop}%
\bibitem [{\citenamefont {Toda}(1958)}]{Toda1958}%
  \BibitemOpen
  \bibfield  {author} {\bibinfo {author} {\bibfnamefont {G.}~\bibnamefont
  {Toda}},\ }\bibfield  {title} {\bibinfo {title} {{Rate of permeation and
  diffusion coefficient of hydrogen through palladium}},\ }\href
  {http://hdl.handle.net/2115/24667} {\bibfield  {journal} {\bibinfo  {journal}
  {Journal of the Research Institute for Catalysis Hokkaido University}\
  }\textbf {\bibinfo {volume} {6}},\ \bibinfo {pages} {13} (\bibinfo {year}
  {1958})}\BibitemShut {NoStop}%
\bibitem [{\citenamefont {V{\"{o}}lkl}\ \emph {et~al.}(1971)\citenamefont
  {V{\"{o}}lkl}, \citenamefont {Wollenweber}, \citenamefont {Klatt},\ and\
  \citenamefont {Alefeld}}]{Volkl1971}%
  \BibitemOpen
  \bibfield  {author} {\bibinfo {author} {\bibfnamefont {J.}~\bibnamefont
  {V{\"{o}}lkl}}, \bibinfo {author} {\bibfnamefont {G.}~\bibnamefont
  {Wollenweber}}, \bibinfo {author} {\bibfnamefont {K.-H.}\ \bibnamefont
  {Klatt}},\ and\ \bibinfo {author} {\bibfnamefont {G.}~\bibnamefont
  {Alefeld}},\ }\bibfield  {title} {\bibinfo {title} {{Notizen: Reversed
  Isotope Dependence for Hydrogen Diffusion in Palladium}},\ }\href
  {https://doi.org/10.1515/zna-1971-0522} {\bibfield  {journal} {\bibinfo
  {journal} {Zeitschrift f{\"{u}}r Naturforschung A}\ }\textbf {\bibinfo
  {volume} {26}},\ \bibinfo {pages} {922} (\bibinfo {year} {1971})}\BibitemShut
  {NoStop}%
\bibitem [{\citenamefont {Holleck}(1970)}]{Holleck1970}%
  \BibitemOpen
  \bibfield  {author} {\bibinfo {author} {\bibfnamefont {G.~L.}\ \bibnamefont
  {Holleck}},\ }\bibfield  {title} {\bibinfo {title} {{Diffusion and solubility
  of hydrogen in palladium and palladium--silver alloys}},\ }\href
  {https://doi.org/10.1021/j100698a005} {\bibfield  {journal} {\bibinfo
  {journal} {The Journal of Physical Chemistry}\ }\textbf {\bibinfo {volume}
  {74}},\ \bibinfo {pages} {503} (\bibinfo {year} {1970})}\BibitemShut
  {NoStop}%
\bibitem [{\citenamefont {Nishimura}\ \emph {et~al.}(1999)\citenamefont
  {Nishimura}, \citenamefont {Komaki},\ and\ \citenamefont
  {Amano}}]{Nishimura1999}%
  \BibitemOpen
  \bibfield  {author} {\bibinfo {author} {\bibfnamefont {C.}~\bibnamefont
  {Nishimura}}, \bibinfo {author} {\bibfnamefont {M.}~\bibnamefont {Komaki}},\
  and\ \bibinfo {author} {\bibfnamefont {M.}~\bibnamefont {Amano}},\ }\bibfield
   {title} {\bibinfo {title} {{Hydrogen permeation through magnesium}},\ }\href
  {https://doi.org/10.1016/S0925-8388(99)00373-4} {\bibfield  {journal}
  {\bibinfo  {journal} {Journal of Alloys and Compounds}\ }\textbf {\bibinfo
  {volume} {293-295}},\ \bibinfo {pages} {329} (\bibinfo {year}
  {1999})}\BibitemShut {NoStop}%
\bibitem [{\citenamefont {Ebisuzaki}\ \emph {et~al.}(1968)\citenamefont
  {Ebisuzaki}, \citenamefont {Kass},\ and\ \citenamefont
  {O'Keeffe}}]{Ebisuzaki1968}%
  \BibitemOpen
  \bibfield  {author} {\bibinfo {author} {\bibfnamefont {Y.}~\bibnamefont
  {Ebisuzaki}}, \bibinfo {author} {\bibfnamefont {W.~J.}\ \bibnamefont
  {Kass}},\ and\ \bibinfo {author} {\bibfnamefont {M.}~\bibnamefont
  {O'Keeffe}},\ }\bibfield  {title} {\bibinfo {title} {{Solubility and
  Diffusion of Hydrogen and Deuterium in Platinum}},\ }\href
  {https://doi.org/10.1063/1.1670604} {\bibfield  {journal} {\bibinfo
  {journal} {The Journal of Chemical Physics}\ }\textbf {\bibinfo {volume}
  {49}},\ \bibinfo {pages} {3329} (\bibinfo {year} {1968})}\BibitemShut
  {NoStop}%
\bibitem [{\citenamefont {Katsuta}\ and\ \citenamefont
  {McLellan}(1979)}]{Katsuta1979}%
  \BibitemOpen
  \bibfield  {author} {\bibinfo {author} {\bibfnamefont {H.}~\bibnamefont
  {Katsuta}}\ and\ \bibinfo {author} {\bibfnamefont {R.}~\bibnamefont
  {McLellan}},\ }\bibfield  {title} {\bibinfo {title} {{Diffusivity
  permeability and solubility of hydrogen in platinum}},\ }\href
  {https://doi.org/10.1016/0038-1098(79)90687-2} {\bibfield  {journal}
  {\bibinfo  {journal} {Solid State Communications}\ }\textbf {\bibinfo
  {volume} {29}},\ \bibinfo {pages} {ii} (\bibinfo {year} {1979})}\BibitemShut
  {NoStop}%
\bibitem [{\citenamefont {Kearns}(1972)}]{Kearns1972}%
  \BibitemOpen
  \bibfield  {author} {\bibinfo {author} {\bibfnamefont {J.}~\bibnamefont
  {Kearns}},\ }\bibfield  {title} {\bibinfo {title} {{Diffusion coefficient of
  hydrogen in alpha zirconium, Zircaloy-2 and Zircaloy-4}},\ }\href
  {https://doi.org/10.1016/0022-3115(72)90065-7} {\bibfield  {journal}
  {\bibinfo  {journal} {Journal of Nuclear Materials}\ }\textbf {\bibinfo
  {volume} {43}},\ \bibinfo {pages} {330} (\bibinfo {year} {1972})}\BibitemShut
  {NoStop}%
\bibitem [{\citenamefont {Gelezunas}\ \emph {et~al.}(1963)\citenamefont
  {Gelezunas}, \citenamefont {Conn},\ and\ \citenamefont
  {Price}}]{Gelezunas1963}%
  \BibitemOpen
  \bibfield  {author} {\bibinfo {author} {\bibfnamefont {V.~L.}\ \bibnamefont
  {Gelezunas}}, \bibinfo {author} {\bibfnamefont {P.~K.}\ \bibnamefont
  {Conn}},\ and\ \bibinfo {author} {\bibfnamefont {R.~H.}\ \bibnamefont
  {Price}},\ }\bibfield  {title} {\bibinfo {title} {{The Diffusion Coefficients
  for Hydrogen in $\beta$-Zirconium}},\ }\href
  {https://doi.org/10.1149/1.2425875} {\bibfield  {journal} {\bibinfo
  {journal} {Journal of The Electrochemical Society}\ }\textbf {\bibinfo
  {volume} {110}},\ \bibinfo {pages} {799} (\bibinfo {year}
  {1963})}\BibitemShut {NoStop}%
\bibitem [{\citenamefont {Gulbransen}\ and\ \citenamefont
  {Andrew}(1954)}]{Gulbransen1954}%
  \BibitemOpen
  \bibfield  {author} {\bibinfo {author} {\bibfnamefont {E.~A.}\ \bibnamefont
  {Gulbransen}}\ and\ \bibinfo {author} {\bibfnamefont {K.~F.}\ \bibnamefont
  {Andrew}},\ }\bibfield  {title} {\bibinfo {title} {{Diffusion of Hydrogen and
  Deuterium in High Purity Zirconium}},\ }\href
  {https://doi.org/10.1149/1.2781154} {\bibfield  {journal} {\bibinfo
  {journal} {Journal of The Electrochemical Society}\ }\textbf {\bibinfo
  {volume} {101}},\ \bibinfo {pages} {560} (\bibinfo {year}
  {1954})}\BibitemShut {NoStop}%
\bibitem [{\citenamefont {Someno}(1960)}]{Someno1960}%
  \BibitemOpen
  \bibfield  {author} {\bibinfo {author} {\bibfnamefont {M.}~\bibnamefont
  {Someno}},\ }\bibfield  {title} {\bibinfo {title} {{Solubility and diffusion
  of hydrogen in zirconium}},\ }\href
  {https://www.osti.gov/biblio/4112367-solubility-diffusion-hydrogen-zirconium}
  {\bibfield  {journal} {\bibinfo  {journal} {Nippon Kinzoku Gakkaishi
  (Japan)}\ }\textbf {\bibinfo {volume} {24}} (\bibinfo {year}
  {1960})}\BibitemShut {NoStop}%
\bibitem [{\citenamefont {Han}\ \emph {et~al.}(1987)\citenamefont {Han},
  \citenamefont {Chang}, \citenamefont {Torgeson}, \citenamefont {Seymour},\
  and\ \citenamefont {Barnes}}]{Han1987}%
  \BibitemOpen
  \bibfield  {author} {\bibinfo {author} {\bibfnamefont {J.-W.}\ \bibnamefont
  {Han}}, \bibinfo {author} {\bibfnamefont {C.-T.}\ \bibnamefont {Chang}},
  \bibinfo {author} {\bibfnamefont {D.~R.}\ \bibnamefont {Torgeson}}, \bibinfo
  {author} {\bibfnamefont {E.~F.~W.}\ \bibnamefont {Seymour}},\ and\ \bibinfo
  {author} {\bibfnamefont {R.~G.}\ \bibnamefont {Barnes}},\ }\bibfield  {title}
  {\bibinfo {title} {{Proton and Sc45 nuclear-magnetic-resonance study of
  hydrogen diffusive hopping in hcp scandium}},\ }\href
  {https://doi.org/10.1103/PhysRevB.36.615} {\bibfield  {journal} {\bibinfo
  {journal} {Physical Review B}\ }\textbf {\bibinfo {volume} {36}},\ \bibinfo
  {pages} {615} (\bibinfo {year} {1987})}\BibitemShut {NoStop}%
\bibitem [{\citenamefont {Raczy{\'{n}}ski}(1978)}]{Raczynski1978}%
  \BibitemOpen
  \bibfield  {author} {\bibinfo {author} {\bibfnamefont {W.}~\bibnamefont
  {Raczy{\'{n}}ski}},\ }\bibfield  {title} {\bibinfo {title} {{Permeability,
  diffusivity, and solubility of hydrogen and deuterium in pure iron at 10 to
  60 °C}},\ }\href {https://doi.org/10.1002/pssa.2210480143} {\bibfield
  {journal} {\bibinfo  {journal} {Physica Status Solidi (a)}\ }\textbf
  {\bibinfo {volume} {48}},\ \bibinfo {pages} {K27} (\bibinfo {year}
  {1978})}\BibitemShut {NoStop}%
\bibitem [{\citenamefont {Quick}\ and\ \citenamefont
  {Johnson}(1978)}]{Quick1978}%
  \BibitemOpen
  \bibfield  {author} {\bibinfo {author} {\bibfnamefont {N.}~\bibnamefont
  {Quick}}\ and\ \bibinfo {author} {\bibfnamefont {H.}~\bibnamefont
  {Johnson}},\ }\bibfield  {title} {\bibinfo {title} {{Hydrogen and deuterium
  in iron, 49–506°C}},\ }\href
  {https://doi.org/10.1016/0001-6160(78)90041-X} {\bibfield  {journal}
  {\bibinfo  {journal} {Acta Metallurgica}\ }\textbf {\bibinfo {volume} {26}},\
  \bibinfo {pages} {903} (\bibinfo {year} {1978})}\BibitemShut {NoStop}%
\bibitem [{\citenamefont {Ishikawa}\ and\ \citenamefont
  {Mclellan}(1985)}]{Ishikawa1985}%
  \BibitemOpen
  \bibfield  {author} {\bibinfo {author} {\bibfnamefont {T.}~\bibnamefont
  {Ishikawa}}\ and\ \bibinfo {author} {\bibfnamefont {R.~B.}\ \bibnamefont
  {Mclellan}},\ }\bibfield  {title} {\bibinfo {title} {{The low-temperature
  diffusion of Hydrogen through annealed, quenched and aged gold}},\ }\href
  {https://doi.org/10.1016/0022-3697(85)90078-2} {\bibfield  {journal}
  {\bibinfo  {journal} {Journal of Physics and Chemistry of Solids}\ }\textbf
  {\bibinfo {volume} {46}},\ \bibinfo {pages} {1393} (\bibinfo {year}
  {1985})}\BibitemShut {NoStop}%
\bibitem [{\citenamefont {Kunz}\ \emph {et~al.}(1983)\citenamefont {Kunz},
  \citenamefont {M{\"{u}}nzel}, \citenamefont {Helfrich},\ and\ \citenamefont
  {Horneff}}]{Kunz1983}%
  \BibitemOpen
  \bibfield  {author} {\bibinfo {author} {\bibfnamefont {W.}~\bibnamefont
  {Kunz}}, \bibinfo {author} {\bibfnamefont {H.}~\bibnamefont {M{\"{u}}nzel}},
  \bibinfo {author} {\bibfnamefont {U.}~\bibnamefont {Helfrich}},\ and\
  \bibinfo {author} {\bibfnamefont {H.}~\bibnamefont {Horneff}},\ }\bibfield
  {title} {\bibinfo {title} {{Bestimmung der Diffusionskoeffizienten f{\"{u}}r
  Tritium in den $\alpha$-Phasen von Titan und Hafnium}},\ }\href
  {https://doi.org/10.1515/ijmr-1983-740505} {\bibfield  {journal} {\bibinfo
  {journal} {International Journal of Materials Research}\ }\textbf {\bibinfo
  {volume} {74}},\ \bibinfo {pages} {289} (\bibinfo {year} {1983})}\BibitemShut
  {NoStop}%
\bibitem [{\citenamefont {Naito}\ \emph {et~al.}(1990)\citenamefont {Naito},
  \citenamefont {Yamamoto},\ and\ \citenamefont {Hashino}}]{Naito1990}%
  \BibitemOpen
  \bibfield  {author} {\bibinfo {author} {\bibfnamefont {S.}~\bibnamefont
  {Naito}}, \bibinfo {author} {\bibfnamefont {M.}~\bibnamefont {Yamamoto}},\
  and\ \bibinfo {author} {\bibfnamefont {T.}~\bibnamefont {Hashino}},\
  }\bibfield  {title} {\bibinfo {title} {{Solubility and diffusivity of
  hydrogen and deuterium in alpha -hafnium at high temperatures}},\ }\href
  {https://doi.org/10.1088/0953-8984/2/8/002} {\bibfield  {journal} {\bibinfo
  {journal} {Journal of Physics: Condensed Matter}\ }\textbf {\bibinfo {volume}
  {2}},\ \bibinfo {pages} {1963} (\bibinfo {year} {1990})}\BibitemShut
  {NoStop}%
\bibitem [{\citenamefont {Bell}\ and\ \citenamefont {Redman}(1983)}]{Bell1983}%
  \BibitemOpen
  \bibfield  {author} {\bibinfo {author} {\bibfnamefont {J.~T.}\ \bibnamefont
  {Bell}}\ and\ \bibinfo {author} {\bibfnamefont {J.~D.}\ \bibnamefont
  {Redman}},\ }\bibfield  {title} {\bibinfo {title} {{Tritium permeability of
  nickel plated stainless steel 21-6-9 and of gold plated aluminum}},\ }\href
  {https://doi.org/10.1007/BF02833442} {\bibfield  {journal} {\bibinfo
  {journal} {Journal of Materials for Energy Systems}\ }\textbf {\bibinfo
  {volume} {4}},\ \bibinfo {pages} {217} (\bibinfo {year} {1983})}\BibitemShut
  {NoStop}%
\bibitem [{\citenamefont {Herro}(1982)}]{Herro1982}%
  \BibitemOpen
  \bibfield  {author} {\bibinfo {author} {\bibfnamefont {H.~M.}\ \bibnamefont
  {Herro}},\ }\emph {\bibinfo {title} {{Hydrogen and deuterium diffusion in
  vanadium alloys}}},\ \href {https://doi.org/10.31274/rtd-180816-5362} {Ph.D.
  thesis},\ \bibinfo  {school} {Iowa State University, Digital Repository},
  \bibinfo {address} {Ames} (\bibinfo {year} {1982})\BibitemShut {NoStop}%
\bibitem [{\citenamefont {Qi}\ \emph {et~al.}(1983)\citenamefont {Qi},
  \citenamefont {V{\"{o}}lkl}, \citenamefont {Lasser},\ and\ \citenamefont
  {Wenzl}}]{Qi1983}%
  \BibitemOpen
  \bibfield  {author} {\bibinfo {author} {\bibfnamefont {Z.}~\bibnamefont
  {Qi}}, \bibinfo {author} {\bibfnamefont {J.}~\bibnamefont {V{\"{o}}lkl}},
  \bibinfo {author} {\bibfnamefont {R.}~\bibnamefont {Lasser}},\ and\ \bibinfo
  {author} {\bibfnamefont {H.}~\bibnamefont {Wenzl}},\ }\bibfield  {title}
  {\bibinfo {title} {{Tritium diffusion in V, Nb and Ta}},\ }\href
  {https://doi.org/10.1088/0305-4608/13/10/015} {\bibfield  {journal} {\bibinfo
   {journal} {Journal of Physics F: Metal Physics}\ }\textbf {\bibinfo {volume}
  {13}},\ \bibinfo {pages} {2053} (\bibinfo {year} {1983})}\BibitemShut
  {NoStop}%
\bibitem [{\citenamefont {Schaumann}\ \emph {et~al.}(1970)\citenamefont
  {Schaumann}, \citenamefont {V{\"{o}}lkl},\ and\ \citenamefont
  {Alefeld}}]{Schaumann1970}%
  \BibitemOpen
  \bibfield  {author} {\bibinfo {author} {\bibfnamefont {G.}~\bibnamefont
  {Schaumann}}, \bibinfo {author} {\bibfnamefont {J.}~\bibnamefont
  {V{\"{o}}lkl}},\ and\ \bibinfo {author} {\bibfnamefont {G.}~\bibnamefont
  {Alefeld}},\ }\bibfield  {title} {\bibinfo {title} {{The diffusion
  coefficients of hydrogen and deuterium in vanadium, niobium, and tantalum by
  gorsky-effect measurements}},\ }\href
  {https://doi.org/10.1002/pssb.19700420141} {\bibfield  {journal} {\bibinfo
  {journal} {physica status solidi (b)}\ }\textbf {\bibinfo {volume} {42}},\
  \bibinfo {pages} {401} (\bibinfo {year} {1970})}\BibitemShut {NoStop}%
\bibitem [{\citenamefont {Bauer}\ \emph {et~al.}(1978)\citenamefont {Bauer},
  \citenamefont {V{\"{o}}lkl}, \citenamefont {Tretkowski},\ and\ \citenamefont
  {Alefeld}}]{Bauer1978}%
  \BibitemOpen
  \bibfield  {author} {\bibinfo {author} {\bibfnamefont {H.~C.}\ \bibnamefont
  {Bauer}}, \bibinfo {author} {\bibfnamefont {J.}~\bibnamefont {V{\"{o}}lkl}},
  \bibinfo {author} {\bibfnamefont {J.}~\bibnamefont {Tretkowski}},\ and\
  \bibinfo {author} {\bibfnamefont {G.}~\bibnamefont {Alefeld}},\ }\bibfield
  {title} {\bibinfo {title} {{Diffusion of hydrogen and deuterium in Nb and Ta
  at high concentrations}},\ }\href {https://doi.org/10.1007/BF01354833}
  {\bibfield  {journal} {\bibinfo  {journal} {Zeitschrift f{\"{u}}r Physik B
  Condensed Matter and Quanta}\ }\textbf {\bibinfo {volume} {29}},\ \bibinfo
  {pages} {17} (\bibinfo {year} {1978})}\BibitemShut {NoStop}%
\bibitem [{\citenamefont {Hampele}\ \emph {et~al.}(1989)\citenamefont
  {Hampele}, \citenamefont {Messer},\ and\ \citenamefont
  {Seeger}}]{Hampele1989}%
  \BibitemOpen
  \bibfield  {author} {\bibinfo {author} {\bibfnamefont {M.}~\bibnamefont
  {Hampele}}, \bibinfo {author} {\bibfnamefont {R.}~\bibnamefont {Messer}},\
  and\ \bibinfo {author} {\bibfnamefont {A.}~\bibnamefont {Seeger}},\
  }\bibfield  {title} {\bibinfo {title} {{Measurements of the Hydrogen
  Diffusivity in Tantalum by the Pulsed-Field Gradient NMR Technique*}},\
  }\href {https://doi.org/10.1524/zpch.1989.164.Part_1.0879} {\bibfield
  {journal} {\bibinfo  {journal} {Zeitschrift f{\"{u}}r Physikalische Chemie}\
  }\textbf {\bibinfo {volume} {164}},\ \bibinfo {pages} {879} (\bibinfo {year}
  {1989})}\BibitemShut {NoStop}%
\bibitem [{\citenamefont {Heidemann}\ \emph {et~al.}(1976)\citenamefont
  {Heidemann}, \citenamefont {Kaindl}, \citenamefont {Salomon}, \citenamefont
  {Wipf},\ and\ \citenamefont {Wortmann}}]{Heidemann1976}%
  \BibitemOpen
  \bibfield  {author} {\bibinfo {author} {\bibfnamefont {A.}~\bibnamefont
  {Heidemann}}, \bibinfo {author} {\bibfnamefont {G.}~\bibnamefont {Kaindl}},
  \bibinfo {author} {\bibfnamefont {D.}~\bibnamefont {Salomon}}, \bibinfo
  {author} {\bibfnamefont {H.}~\bibnamefont {Wipf}},\ and\ \bibinfo {author}
  {\bibfnamefont {G.}~\bibnamefont {Wortmann}},\ }\bibfield  {title} {\bibinfo
  {title} {{Diffusion of Hydrogen in Tantalum Studied by Motional Narrowing of
  M{\"{o}}ssbauer Lines}},\ }\href {https://doi.org/10.1103/PhysRevLett.36.213}
  {\bibfield  {journal} {\bibinfo  {journal} {Physical Review Letters}\
  }\textbf {\bibinfo {volume} {36}},\ \bibinfo {pages} {213} (\bibinfo {year}
  {1976})}\BibitemShut {NoStop}%
\bibitem [{\citenamefont {{Wasilewski, R J;
  Kehl}}(1954)}]{WasilewskiRJ;Kehl1954}%
  \BibitemOpen
  \bibfield  {author} {\bibinfo {author} {\bibfnamefont {G.~L.}\ \bibnamefont
  {{Wasilewski, R J; Kehl}}},\ }\bibfield  {title} {\bibinfo {title}
  {{Diffusion of Hydrogen in Titanium}},\ }\href
  {https://www.osti.gov/biblio/4398780%0A} {\bibfield  {journal} {\bibinfo
  {journal} {Metallurgia (England)}\ }\textbf {\bibinfo {volume} {50}}
  (\bibinfo {year} {1954})}\BibitemShut {NoStop}%
\bibitem [{\citenamefont {Katsuta}\ \emph {et~al.}(1982)\citenamefont
  {Katsuta}, \citenamefont {McLellan},\ and\ \citenamefont
  {Furukawa}}]{Katsuta1982}%
  \BibitemOpen
  \bibfield  {author} {\bibinfo {author} {\bibfnamefont {H.}~\bibnamefont
  {Katsuta}}, \bibinfo {author} {\bibfnamefont {R.~B.}\ \bibnamefont
  {McLellan}},\ and\ \bibinfo {author} {\bibfnamefont {K.}~\bibnamefont
  {Furukawa}},\ }\bibfield  {title} {\bibinfo {title} {{Diffusivity and
  permeability of hydrogen in molybdenum}},\ }\href
  {https://doi.org/10.1016/0022-3697(82)90104-4} {\bibfield  {journal}
  {\bibinfo  {journal} {Journal of Physics and Chemistry of Solids}\ }\textbf
  {\bibinfo {volume} {43}},\ \bibinfo {pages} {533} (\bibinfo {year}
  {1982})}\BibitemShut {NoStop}%
\bibitem [{\citenamefont {Tanabe}\ \emph {et~al.}(1992)\citenamefont {Tanabe},
  \citenamefont {Yamanishi},\ and\ \citenamefont {Imoto}}]{Tanabe1992}%
  \BibitemOpen
  \bibfield  {author} {\bibinfo {author} {\bibfnamefont {T.}~\bibnamefont
  {Tanabe}}, \bibinfo {author} {\bibfnamefont {Y.}~\bibnamefont {Yamanishi}},\
  and\ \bibinfo {author} {\bibfnamefont {S.}~\bibnamefont {Imoto}},\ }\bibfield
   {title} {\bibinfo {title} {{Hydrogen permeation and diffusion in
  molybdenum}},\ }\href {https://doi.org/10.1016/S0022-3115(09)80083-4}
  {\bibfield  {journal} {\bibinfo  {journal} {Journal of Nuclear Materials}\
  }\textbf {\bibinfo {volume} {191-194}},\ \bibinfo {pages} {439} (\bibinfo
  {year} {1992})}\BibitemShut {NoStop}%
\bibitem [{\citenamefont {Katsuta}\ \emph {et~al.}(1964)\citenamefont
  {Katsuta}, \citenamefont {Iwai},\ and\ \citenamefont {Ohno}}]{Katsuta1964}%
  \BibitemOpen
  \bibfield  {author} {\bibinfo {author} {\bibfnamefont {H.}~\bibnamefont
  {Katsuta}}, \bibinfo {author} {\bibfnamefont {T.}~\bibnamefont {Iwai}},\ and\
  \bibinfo {author} {\bibfnamefont {H.}~\bibnamefont {Ohno}},\ }\bibfield
  {title} {\bibinfo {title} {{Mass spectrometer investigations of the degassing
  of Molybdenum, Tungsten and Niobium on heating them in vacuo}},\ }\href
  {https://doi.org/10.1016/0042-207X(64)91918-9} {\bibfield  {journal}
  {\bibinfo  {journal} {Vacuum}\ }\textbf {\bibinfo {volume} {14}},\ \bibinfo
  {pages} {455} (\bibinfo {year} {1964})}\BibitemShut {NoStop}%
\bibitem [{\citenamefont {Katsuta}\ \emph {et~al.}(1983)\citenamefont
  {Katsuta}, \citenamefont {Iwai},\ and\ \citenamefont {Ohno}}]{Katsuta1983}%
  \BibitemOpen
  \bibfield  {author} {\bibinfo {author} {\bibfnamefont {H.}~\bibnamefont
  {Katsuta}}, \bibinfo {author} {\bibfnamefont {T.}~\bibnamefont {Iwai}},\ and\
  \bibinfo {author} {\bibfnamefont {H.}~\bibnamefont {Ohno}},\ }\bibfield
  {title} {\bibinfo {title} {{Diffusivity and permeability of hydrogen in
  neutron irradiated molybdenum and platinum}},\ }\href
  {https://doi.org/10.1016/0022-3115(83)90311-2} {\bibfield  {journal}
  {\bibinfo  {journal} {Journal of Nuclear Materials}\ }\textbf {\bibinfo
  {volume} {115}},\ \bibinfo {pages} {206} (\bibinfo {year}
  {1983})}\BibitemShut {NoStop}%
\bibitem [{\citenamefont {Zvezdin}\ and\ \citenamefont
  {Belyakov}(1968)}]{Zvezdin1968}%
  \BibitemOpen
  \bibfield  {author} {\bibinfo {author} {\bibfnamefont {Y.~I.}\ \bibnamefont
  {Zvezdin}}\ and\ \bibinfo {author} {\bibfnamefont {Y.~I.}\ \bibnamefont
  {Belyakov}},\ }\bibfield  {title} {\bibinfo {title} {{Hydrogen permeability
  of some transition metals and metals of group I of the periodic system}},\
  }\href {https://doi.org/10.1007/BF00714787} {\bibfield  {journal} {\bibinfo
  {journal} {Soviet Materials Science}\ }\textbf {\bibinfo {volume} {3}},\
  \bibinfo {pages} {255} (\bibinfo {year} {1968})}\BibitemShut {NoStop}%
\bibitem [{\citenamefont {Caskey}\ and\ \citenamefont
  {Derrick}(1977)}]{Caskey1977}%
  \BibitemOpen
  \bibfield  {author} {\bibinfo {author} {\bibfnamefont {G.~R.}\ \bibnamefont
  {Caskey}}\ and\ \bibinfo {author} {\bibfnamefont {R.~G.}\ \bibnamefont
  {Derrick}},\ }\bibfield  {title} {\bibinfo {title} {{Hydrogen permeability
  through alpha-brass}},\ }\href {https://doi.org/10.1007/BF02661764}
  {\bibfield  {journal} {\bibinfo  {journal} {Metallurgical Transactions A}\
  }\textbf {\bibinfo {volume} {8}},\ \bibinfo {pages} {511} (\bibinfo {year}
  {1977})}\BibitemShut {NoStop}%
\bibitem [{\citenamefont {Magnusson}\ and\ \citenamefont
  {Frisk}(2017)}]{Magnusson2017}%
  \BibitemOpen
  \bibfield  {author} {\bibinfo {author} {\bibfnamefont {H.}~\bibnamefont
  {Magnusson}}\ and\ \bibinfo {author} {\bibfnamefont {K.}~\bibnamefont
  {Frisk}},\ }\bibfield  {title} {\bibinfo {title} {{Diffusion, Permeation and
  Solubility of Hydrogen in Copper}},\ }\href
  {https://doi.org/10.1007/s11669-017-0518-y} {\bibfield  {journal} {\bibinfo
  {journal} {Journal of Phase Equilibria and Diffusion}\ }\textbf {\bibinfo
  {volume} {38}},\ \bibinfo {pages} {65} (\bibinfo {year} {2017})}\BibitemShut
  {NoStop}%
\bibitem [{\citenamefont {Hagi}(1986)}]{Hagi1986}%
  \BibitemOpen
  \bibfield  {author} {\bibinfo {author} {\bibfnamefont {H.}~\bibnamefont
  {Hagi}},\ }\bibfield  {title} {\bibinfo {title} {{Diffusion Coefficients of
  Hydrogen in Ni-Cu and Ni-Co Alloys}},\ }\href
  {https://doi.org/10.2320/matertrans1960.27.233} {\bibfield  {journal}
  {\bibinfo  {journal} {Transactions of the Japan Institute of Metals}\
  }\textbf {\bibinfo {volume} {27}},\ \bibinfo {pages} {233} (\bibinfo {year}
  {1986})}\BibitemShut {NoStop}%
\bibitem [{\citenamefont {Katz}\ \emph {et~al.}(1971)\citenamefont {Katz},
  \citenamefont {Guinan},\ and\ \citenamefont {Borg}}]{Katz1971}%
  \BibitemOpen
  \bibfield  {author} {\bibinfo {author} {\bibfnamefont {L.}~\bibnamefont
  {Katz}}, \bibinfo {author} {\bibfnamefont {M.}~\bibnamefont {Guinan}},\ and\
  \bibinfo {author} {\bibfnamefont {R.~J.}\ \bibnamefont {Borg}},\ }\bibfield
  {title} {\bibinfo {title} {{Diffusion of H2, D2, and T2 in Single-Crystal Ni
  and Cu}},\ }\href {https://doi.org/10.1103/PhysRevB.4.330} {\bibfield
  {journal} {\bibinfo  {journal} {Physical Review B}\ }\textbf {\bibinfo
  {volume} {4}},\ \bibinfo {pages} {330} (\bibinfo {year} {1971})}\BibitemShut
  {NoStop}%
\bibitem [{\citenamefont {Yamakawa}(1979)}]{Yamakawa1979}%
  \BibitemOpen
  \bibfield  {author} {\bibinfo {author} {\bibfnamefont {K.}~\bibnamefont
  {Yamakawa}},\ }\bibfield  {title} {\bibinfo {title} {{Diffusion of Deuterium
  and Isotope Effect in Nickel}},\ }\href {https://doi.org/10.1143/JPSJ.47.114}
  {\bibfield  {journal} {\bibinfo  {journal} {Journal of the Physical Society
  of Japan}\ }\textbf {\bibinfo {volume} {47}},\ \bibinfo {pages} {114}
  (\bibinfo {year} {1979})}\BibitemShut {NoStop}%
\bibitem [{\citenamefont {Cummings}\ and\ \citenamefont
  {Blackburn}(1987)}]{Cummings1987}%
  \BibitemOpen
  \bibfield  {author} {\bibinfo {author} {\bibfnamefont {D.}~\bibnamefont
  {Cummings}}\ and\ \bibinfo {author} {\bibfnamefont {D.}~\bibnamefont
  {Blackburn}},\ }\bibfield  {title} {\bibinfo {title} {{Pressure modulated
  absorption-desorption: Permeation, diffusion and surface rate coefficients
  for hydrogen in nickel and palladium}},\ }\href
  {https://doi.org/10.1016/0022-3115(87)90283-2} {\bibfield  {journal}
  {\bibinfo  {journal} {Journal of Nuclear Materials}\ }\textbf {\bibinfo
  {volume} {144}},\ \bibinfo {pages} {81} (\bibinfo {year} {1987})}\BibitemShut
  {NoStop}%
\bibitem [{\citenamefont {Atrens}\ \emph {et~al.}(1980)\citenamefont {Atrens},
  \citenamefont {Mezzanotte}, \citenamefont {Fiore},\ and\ \citenamefont
  {Genshaw}}]{Atrens1980}%
  \BibitemOpen
  \bibfield  {author} {\bibinfo {author} {\bibfnamefont {A.}~\bibnamefont
  {Atrens}}, \bibinfo {author} {\bibfnamefont {D.}~\bibnamefont {Mezzanotte}},
  \bibinfo {author} {\bibfnamefont {N.}~\bibnamefont {Fiore}},\ and\ \bibinfo
  {author} {\bibfnamefont {M.}~\bibnamefont {Genshaw}},\ }\bibfield  {title}
  {\bibinfo {title} {{Electrochemical studies of hydrogen diffusion and
  permeability in Ni}},\ }\href {https://doi.org/10.1016/0010-938X(80)90102-X}
  {\bibfield  {journal} {\bibinfo  {journal} {Corrosion Science}\ }\textbf
  {\bibinfo {volume} {20}},\ \bibinfo {pages} {673} (\bibinfo {year}
  {1980})}\BibitemShut {NoStop}%
\bibitem [{\citenamefont {Furuya}\ \emph {et~al.}(1984)\citenamefont {Furuya},
  \citenamefont {Hashimoto},\ and\ \citenamefont {Kino}}]{Furuya1984}%
  \BibitemOpen
  \bibfield  {author} {\bibinfo {author} {\bibfnamefont {Y.}~\bibnamefont
  {Furuya}}, \bibinfo {author} {\bibfnamefont {E.}~\bibnamefont {Hashimoto}},\
  and\ \bibinfo {author} {\bibfnamefont {T.}~\bibnamefont {Kino}},\ }\bibfield
  {title} {\bibinfo {title} {{Hydrogen Permeation through Nickel}},\ }\href
  {https://doi.org/10.1143/JJAP.23.1190} {\bibfield  {journal} {\bibinfo
  {journal} {Japanese Journal of Applied Physics}\ }\textbf {\bibinfo {volume}
  {23}},\ \bibinfo {pages} {1190} (\bibinfo {year} {1984})}\BibitemShut
  {NoStop}%
\bibitem [{\citenamefont {Yamanishi}\ \emph {et~al.}(1983)\citenamefont
  {Yamanishi}, \citenamefont {Tanabe},\ and\ \citenamefont
  {Imoto}}]{Yamanishi1983}%
  \BibitemOpen
  \bibfield  {author} {\bibinfo {author} {\bibfnamefont {Y.}~\bibnamefont
  {Yamanishi}}, \bibinfo {author} {\bibfnamefont {T.}~\bibnamefont {Tanabe}},\
  and\ \bibinfo {author} {\bibfnamefont {S.}~\bibnamefont {Imoto}},\ }\bibfield
   {title} {\bibinfo {title} {{Hydrogen Permeation and Diffusion through Pure
  Fe, Pure Ni and Fe-Ni Alloys}},\ }\href
  {https://doi.org/10.2320/matertrans1960.24.49} {\bibfield  {journal}
  {\bibinfo  {journal} {Transactions of the Japan Institute of Metals}\
  }\textbf {\bibinfo {volume} {24}},\ \bibinfo {pages} {49} (\bibinfo {year}
  {1983})}\BibitemShut {NoStop}%
\bibitem [{\citenamefont {Katsuta}\ and\ \citenamefont {{B.
  McLellan}}(1979)}]{Katsuta1979a}%
  \BibitemOpen
  \bibfield  {author} {\bibinfo {author} {\bibfnamefont {H.}~\bibnamefont
  {Katsuta}}\ and\ \bibinfo {author} {\bibfnamefont {R.}~\bibnamefont {{B.
  McLellan}}},\ }\bibfield  {title} {\bibinfo {title} {{Diffusivity of hydrogen
  in silver}},\ }\href {https://doi.org/10.1016/0036-9748(79)90391-0}
  {\bibfield  {journal} {\bibinfo  {journal} {Scripta Metallurgica}\ }\textbf
  {\bibinfo {volume} {13}},\ \bibinfo {pages} {65} (\bibinfo {year}
  {1979})}\BibitemShut {NoStop}%
\bibitem [{\citenamefont {Anderson}\ \emph {et~al.}(1989)\citenamefont
  {Anderson}, \citenamefont {Ross},\ and\ \citenamefont
  {Bonnet}}]{Anderson1989}%
  \BibitemOpen
  \bibfield  {author} {\bibinfo {author} {\bibfnamefont {I.}~\bibnamefont
  {Anderson}}, \bibinfo {author} {\bibfnamefont {D.}~\bibnamefont {Ross}},\
  and\ \bibinfo {author} {\bibfnamefont {J.}~\bibnamefont {Bonnet}},\
  }\bibfield  {title} {\bibinfo {title} {{Long Range Diffusion of Hydrogen in
  Yttrium*}},\ }\href {https://doi.org/10.1524/zpch.1989.164.Part_1.0923}
  {\bibfield  {journal} {\bibinfo  {journal} {Zeitschrift f{\"{u}}r
  Physikalische Chemie}\ }\textbf {\bibinfo {volume} {164}},\ \bibinfo {pages}
  {923} (\bibinfo {year} {1989})}\BibitemShut {NoStop}%
\bibitem [{\citenamefont {Maeda}\ \emph {et~al.}(1993)\citenamefont {Maeda},
  \citenamefont {Naito}, \citenamefont {Yamamoto}, \citenamefont {Mabuchi},\
  and\ \citenamefont {Hashino}}]{Maeda1993}%
  \BibitemOpen
  \bibfield  {author} {\bibinfo {author} {\bibfnamefont {T.}~\bibnamefont
  {Maeda}}, \bibinfo {author} {\bibfnamefont {S.}~\bibnamefont {Naito}},
  \bibinfo {author} {\bibfnamefont {M.}~\bibnamefont {Yamamoto}}, \bibinfo
  {author} {\bibfnamefont {M.}~\bibnamefont {Mabuchi}},\ and\ \bibinfo {author}
  {\bibfnamefont {T.}~\bibnamefont {Hashino}},\ }\bibfield  {title} {\bibinfo
  {title} {{Diffusivity and solubility of hydrogen and deuterium in yttrium}},\
  }\href {https://doi.org/10.1039/ft9938904375} {\bibfield  {journal} {\bibinfo
   {journal} {Journal of the Chemical Society, Faraday Transactions}\ }\textbf
  {\bibinfo {volume} {89}},\ \bibinfo {pages} {4375} (\bibinfo {year}
  {1993})}\BibitemShut {NoStop}%
\bibitem [{\citenamefont {Buxbaum}\ and\ \citenamefont
  {Johnson}(1985)}]{Buxbaum1985}%
  \BibitemOpen
  \bibfield  {author} {\bibinfo {author} {\bibfnamefont {R.~E.}\ \bibnamefont
  {Buxbaum}}\ and\ \bibinfo {author} {\bibfnamefont {E.~F.}\ \bibnamefont
  {Johnson}},\ }\bibfield  {title} {\bibinfo {title} {{Diffusivity of hydrogen
  isotopes in liquid lithium and in solid yttrium}},\ }\href
  {https://doi.org/10.1021/i100018a007} {\bibfield  {journal} {\bibinfo
  {journal} {Industrial \& Engineering Chemistry Fundamentals}\ }\textbf
  {\bibinfo {volume} {24}},\ \bibinfo {pages} {180} (\bibinfo {year}
  {1985})}\BibitemShut {NoStop}%
\bibitem [{\citenamefont {Naito}\ \emph {et~al.}(1998)\citenamefont {Naito},
  \citenamefont {Yamamoto}, \citenamefont {Doi},\ and\ \citenamefont
  {Kimura}}]{Naito1998}%
  \BibitemOpen
  \bibfield  {author} {\bibinfo {author} {\bibfnamefont {S.}~\bibnamefont
  {Naito}}, \bibinfo {author} {\bibfnamefont {M.}~\bibnamefont {Yamamoto}},
  \bibinfo {author} {\bibfnamefont {M.}~\bibnamefont {Doi}},\ and\ \bibinfo
  {author} {\bibfnamefont {M.}~\bibnamefont {Kimura}},\ }\bibfield  {title}
  {\bibinfo {title} {{High‐Temperature Diffusion of Hydrogen and Deuterium in
  Titanium and Ti3Al}},\ }\href {https://doi.org/10.1149/1.1838662} {\bibfield
  {journal} {\bibinfo  {journal} {Journal of The Electrochemical Society}\
  }\textbf {\bibinfo {volume} {145}},\ \bibinfo {pages} {2471} (\bibinfo {year}
  {1998})}\BibitemShut {NoStop}%
\bibitem [{\citenamefont {Westlake}(1959)}]{Westlake1959}%
  \BibitemOpen
  \bibfield  {author} {\bibinfo {author} {\bibfnamefont {D.~G.}\ \bibnamefont
  {Westlake}},\ }\emph {\bibinfo {title} {{Absorption and diffusion of hydrogen
  in thorium}}},\ \href {https://doi.org/10.31274/rtd-180813-2095} {Ph.D.
  thesis},\ \bibinfo  {school} {Iowa State University, Digital Repository},
  \bibinfo {address} {Ames} (\bibinfo {year} {1959})\BibitemShut {NoStop}%
\bibitem [{\citenamefont {V{\"{o}}lkl}\ \emph {et~al.}(1987)\citenamefont
  {V{\"{o}}lkl}, \citenamefont {Wipf}, \citenamefont {Beaudry},\ and\
  \citenamefont {Gschneidner}}]{Volkl1987}%
  \BibitemOpen
  \bibfield  {author} {\bibinfo {author} {\bibfnamefont {J.}~\bibnamefont
  {V{\"{o}}lkl}}, \bibinfo {author} {\bibfnamefont {H.}~\bibnamefont {Wipf}},
  \bibinfo {author} {\bibfnamefont {B.~J.}\ \bibnamefont {Beaudry}},\ and\
  \bibinfo {author} {\bibfnamefont {K.~A.}\ \bibnamefont {Gschneidner}},\
  }\bibfield  {title} {\bibinfo {title} {{Diffusion of H and D in Lutetium}},\
  }\href {https://doi.org/10.1002/pssb.2221440128} {\bibfield  {journal}
  {\bibinfo  {journal} {physica status solidi (b)}\ }\textbf {\bibinfo {volume}
  {144}},\ \bibinfo {pages} {315} (\bibinfo {year} {1987})}\BibitemShut
  {NoStop}%
\bibitem [{\citenamefont {Outlaw}\ \emph {et~al.}(1982)\citenamefont {Outlaw},
  \citenamefont {Peterson},\ and\ \citenamefont {Schmidt}}]{Outlaw1982}%
  \BibitemOpen
  \bibfield  {author} {\bibinfo {author} {\bibfnamefont {R.}~\bibnamefont
  {Outlaw}}, \bibinfo {author} {\bibfnamefont {D.}~\bibnamefont {Peterson}},\
  and\ \bibinfo {author} {\bibfnamefont {F.}~\bibnamefont {Schmidt}},\
  }\bibfield  {title} {\bibinfo {title} {{Diffusion of hydrogen in pure large
  grain aluminum}},\ }\href {https://doi.org/10.1016/0036-9748(82)90354-4}
  {\bibfield  {journal} {\bibinfo  {journal} {Scripta Metallurgica}\ }\textbf
  {\bibinfo {volume} {16}},\ \bibinfo {pages} {287} (\bibinfo {year}
  {1982})}\BibitemShut {NoStop}%
\bibitem [{\citenamefont {Papp}\ and\ \citenamefont
  {Kov{\'{a}}cs-Cset{\'{e}}nyi}(1981)}]{Papp1981}%
  \BibitemOpen
  \bibfield  {author} {\bibinfo {author} {\bibfnamefont {K.}~\bibnamefont
  {Papp}}\ and\ \bibinfo {author} {\bibfnamefont {E.}~\bibnamefont
  {Kov{\'{a}}cs-Cset{\'{e}}nyi}},\ }\bibfield  {title} {\bibinfo {title}
  {{Diffusion of hydrogen in high purity aluminium}},\ }\href
  {https://doi.org/10.1016/0036-9748(81)90321-5} {\bibfield  {journal}
  {\bibinfo  {journal} {Scripta Metallurgica}\ }\textbf {\bibinfo {volume}
  {15}},\ \bibinfo {pages} {161} (\bibinfo {year} {1981})}\BibitemShut
  {NoStop}%
\bibitem [{\citenamefont {Kizu}\ \emph {et~al.}(1995)\citenamefont {Kizu},
  \citenamefont {Miyazaki},\ and\ \citenamefont {Tanabe}}]{Kizu1995}%
  \BibitemOpen
  \bibfield  {author} {\bibinfo {author} {\bibfnamefont {K.}~\bibnamefont
  {Kizu}}, \bibinfo {author} {\bibfnamefont {K.}~\bibnamefont {Miyazaki}},\
  and\ \bibinfo {author} {\bibfnamefont {T.}~\bibnamefont {Tanabe}},\
  }\bibfield  {title} {\bibinfo {title} {{Hydrogen Permeation and Diffusion in
  Beryllium}},\ }\href {https://doi.org/10.13182/FST95-A30573} {\bibfield
  {journal} {\bibinfo  {journal} {Fusion Technology}\ }\textbf {\bibinfo
  {volume} {28}},\ \bibinfo {pages} {1205} (\bibinfo {year}
  {1995})}\BibitemShut {NoStop}%
\bibitem [{\citenamefont {Frauenfelder}(1969)}]{Frauenfelder1969}%
  \BibitemOpen
  \bibfield  {author} {\bibinfo {author} {\bibfnamefont {R.}~\bibnamefont
  {Frauenfelder}},\ }\bibfield  {title} {\bibinfo {title} {{Solution and
  Diffusion of Hydrogen in Tungsten}},\ }\href
  {https://doi.org/10.1116/1.1492699} {\bibfield  {journal} {\bibinfo
  {journal} {Journal of Vacuum Science and Technology}\ }\textbf {\bibinfo
  {volume} {6}},\ \bibinfo {pages} {388} (\bibinfo {year} {1969})}\BibitemShut
  {NoStop}%
\bibitem [{\citenamefont {Mehrer}(1990)}]{Mehrer1990}%
  \BibitemOpen
  \bibinfo {editor} {\bibfnamefont {H.}~\bibnamefont {Mehrer}},\ ed.,\ \href
  {https://doi.org/10.1007/b37801} {\emph {\bibinfo {title} {{Diffusion in
  Solid Metals and Alloys}}}},\ \bibinfo {series} {Landolt-B{\"{o}}rnstein -
  Group III Condensed Matter}, Vol.~\bibinfo {volume} {26}\ (\bibinfo
  {publisher} {Springer-Verlag},\ \bibinfo {address} {Berlin/Heidelberg},\
  \bibinfo {year} {1990})\ p.\ \bibinfo {pages} {529}\BibitemShut {NoStop}%
\bibitem [{\citenamefont {Caskey}\ \emph {et~al.}(1974)\citenamefont {Caskey},
  \citenamefont {Derrick},\ and\ \citenamefont {Louthan}}]{Caskey1974}%
  \BibitemOpen
  \bibfield  {author} {\bibinfo {author} {\bibfnamefont {G.}~\bibnamefont
  {Caskey}}, \bibinfo {author} {\bibfnamefont {R.}~\bibnamefont {Derrick}},\
  and\ \bibinfo {author} {\bibfnamefont {M.}~\bibnamefont {Louthan}},\
  }\bibfield  {title} {\bibinfo {title} {{Hydrogen diffusion in cobalt}},\
  }\href {https://doi.org/10.1016/0036-9748(74)90055-6} {\bibfield  {journal}
  {\bibinfo  {journal} {Scripta Metallurgica}\ }\textbf {\bibinfo {volume}
  {8}},\ \bibinfo {pages} {481} (\bibinfo {year} {1974})}\BibitemShut {NoStop}%
\bibitem [{\citenamefont {Kittel}\ \emph {et~al.}(1996)\citenamefont {Kittel},
  \citenamefont {McEuen},\ and\ \citenamefont {McEuen}}]{kittel1996}%
  \BibitemOpen
  \bibfield  {author} {\bibinfo {author} {\bibfnamefont {C.}~\bibnamefont
  {Kittel}}, \bibinfo {author} {\bibfnamefont {P.}~\bibnamefont {McEuen}},\
  and\ \bibinfo {author} {\bibfnamefont {P.}~\bibnamefont {McEuen}},\
  }\href@noop {} {\emph {\bibinfo {title} {Introduction to solid state
  physics}}},\ Vol.~\bibinfo {volume} {8}\ (\bibinfo  {publisher} {Wiley New
  York},\ \bibinfo {year} {1996})\BibitemShut {NoStop}%
\bibitem [{\citenamefont {Li}\ and\ \citenamefont
  {Wu}(2001)}]{li2001correlation}%
  \BibitemOpen
  \bibfield  {author} {\bibinfo {author} {\bibfnamefont {C.}~\bibnamefont
  {Li}}\ and\ \bibinfo {author} {\bibfnamefont {P.}~\bibnamefont {Wu}},\
  }\bibfield  {title} {\bibinfo {title} {Correlation of bulk modulus and the
  constituent element properties of binary intermetallic compounds},\
  }\href@noop {} {\bibfield  {journal} {\bibinfo  {journal} {Chemistry of
  materials}\ }\textbf {\bibinfo {volume} {13}},\ \bibinfo {pages} {4642}
  (\bibinfo {year} {2001})}\BibitemShut {NoStop}%
\bibitem [{\citenamefont
  {Wikipedia}(2017)}]{wiki:List_of_elements_by_atomic_properties}%
  \BibitemOpen
  \bibfield  {author} {\bibinfo {author} {\bibnamefont {Wikipedia}},\
  }\href@noop {} {\bibinfo {title} {{List of elements by atomic properties} ---
  {W}ikipedia{,} the free encyclopedia}},\ \bibinfo {howpublished}
  {\url{http://en.wikipedia.org/w/index.php?title=List\%20of\%20elements\%20by\%20atomic\%20properties&oldid=1107080949}}
  (\bibinfo {year} {2017})\BibitemShut {NoStop}%
\bibitem [{\citenamefont {Bakker}(1998)}]{bakker1998enthalpies}%
  \BibitemOpen
  \bibfield  {author} {\bibinfo {author} {\bibfnamefont {H.}~\bibnamefont
  {Bakker}},\ }\bibfield  {title} {\bibinfo {title} {Enthalpies in alloys:
  Miedema’s semi-empirical model, vol. 1},\ }\href@noop {} {\bibfield
  {journal} {\bibinfo  {journal} {Materials Science Foundation}\ } (\bibinfo
  {year} {1998})}\BibitemShut {NoStop}%
\bibitem [{kno(2017{\natexlab{a}})}]{knowledgedoor}%
  \BibitemOpen
  \href
  {https://www.knowledgedoor.com/2/elements_handbook/linear_thermal_expansion_coefficient.html}
  {\bibinfo {title} {Linear thermal expansion coefficient: The elements
  handbook at knowledgedoor}} (\bibinfo {year}
  {2017}{\natexlab{a}})\BibitemShut {NoStop}%
\bibitem [{kno(2017{\natexlab{b}})}]{knowledgedoor1}%
  \BibitemOpen
  \href
  {http://www.knowledgedoor.com/2/elements_handbook/debye_temperature.html}
  {\bibinfo {title} {Debye temperature: The elements handbook at
  knowledgedoor}} (\bibinfo {year} {2017}{\natexlab{b}})\BibitemShut {NoStop}%
\bibitem [{\citenamefont {Shang}\ \emph {et~al.}(2016)\citenamefont {Shang},
  \citenamefont {Zhou}, \citenamefont {Wang}, \citenamefont {Ross},
  \citenamefont {Liu}, \citenamefont {Hu}, \citenamefont {Fang}, \citenamefont
  {Wang},\ and\ \citenamefont {Liu}}]{shang2016}%
  \BibitemOpen
  \bibfield  {author} {\bibinfo {author} {\bibfnamefont {S.-L.}\ \bibnamefont
  {Shang}}, \bibinfo {author} {\bibfnamefont {B.-C.}\ \bibnamefont {Zhou}},
  \bibinfo {author} {\bibfnamefont {W.~Y.}\ \bibnamefont {Wang}}, \bibinfo
  {author} {\bibfnamefont {A.~J.}\ \bibnamefont {Ross}}, \bibinfo {author}
  {\bibfnamefont {X.~L.}\ \bibnamefont {Liu}}, \bibinfo {author} {\bibfnamefont
  {Y.-J.}\ \bibnamefont {Hu}}, \bibinfo {author} {\bibfnamefont {H.-Z.}\
  \bibnamefont {Fang}}, \bibinfo {author} {\bibfnamefont {Y.}~\bibnamefont
  {Wang}},\ and\ \bibinfo {author} {\bibfnamefont {Z.-K.}\ \bibnamefont
  {Liu}},\ }\bibfield  {title} {\bibinfo {title} {A comprehensive
  first-principles study of pure elements: vacancy formation and migration
  energies and self-diffusion coefficients},\ }\href@noop {} {\bibfield
  {journal} {\bibinfo  {journal} {Acta Materialia}\ }\textbf {\bibinfo {volume}
  {109}},\ \bibinfo {pages} {128} (\bibinfo {year} {2016})}\BibitemShut
  {NoStop}%
\bibitem [{\citenamefont {Wang}\ \emph {et~al.}(2004)\citenamefont {Wang},
  \citenamefont {Curtarolo}, \citenamefont {Jiang}, \citenamefont {Arroyave},
  \citenamefont {Wang}, \citenamefont {Ceder}, \citenamefont {Chen},\ and\
  \citenamefont {Liu}}]{Wang2004}%
  \BibitemOpen
  \bibfield  {author} {\bibinfo {author} {\bibfnamefont {Y.}~\bibnamefont
  {Wang}}, \bibinfo {author} {\bibfnamefont {S.}~\bibnamefont {Curtarolo}},
  \bibinfo {author} {\bibfnamefont {C.}~\bibnamefont {Jiang}}, \bibinfo
  {author} {\bibfnamefont {R.}~\bibnamefont {Arroyave}}, \bibinfo {author}
  {\bibfnamefont {T.}~\bibnamefont {Wang}}, \bibinfo {author} {\bibfnamefont
  {G.}~\bibnamefont {Ceder}}, \bibinfo {author} {\bibfnamefont {L.-Q.}\
  \bibnamefont {Chen}},\ and\ \bibinfo {author} {\bibfnamefont {Z.-K.}\
  \bibnamefont {Liu}},\ }\bibfield  {title} {\bibinfo {title} {{Ab initio
  lattice stability in comparison with CALPHAD lattice stability}},\ }\href
  {https://doi.org/10.1016/j.calphad.2004.05.002} {\bibfield  {journal}
  {\bibinfo  {journal} {Calphad}\ }\textbf {\bibinfo {volume} {28}},\ \bibinfo
  {pages} {79} (\bibinfo {year} {2004})}\BibitemShut {NoStop}%
\bibitem [{\citenamefont {Villars}\ and\ \citenamefont
  {Hulliger}(1987)}]{Villars1987}%
  \BibitemOpen
  \bibfield  {author} {\bibinfo {author} {\bibfnamefont {P.}~\bibnamefont
  {Villars}}\ and\ \bibinfo {author} {\bibfnamefont {F.}~\bibnamefont
  {Hulliger}},\ }\bibfield  {title} {\bibinfo {title} {{Structural-stability
  domains for single-coordination intermetallic phases}},\ }\href
  {https://doi.org/10.1016/0022-5088(87)90584-4} {\bibfield  {journal}
  {\bibinfo  {journal} {Journal of the Less Common Metals}\ }\textbf {\bibinfo
  {volume} {132}},\ \bibinfo {pages} {289} (\bibinfo {year}
  {1987})}\BibitemShut {NoStop}%
\bibitem [{\citenamefont {Ranganathan}\ and\ \citenamefont
  {Inoue}(2006)}]{Ranganathan2006}%
  \BibitemOpen
  \bibfield  {author} {\bibinfo {author} {\bibfnamefont {S.}~\bibnamefont
  {Ranganathan}}\ and\ \bibinfo {author} {\bibfnamefont {A.}~\bibnamefont
  {Inoue}},\ }\bibfield  {title} {\bibinfo {title} {{An application of Pettifor
  structure maps for the identification of pseudo-binary quasicrystalline
  intermetallics}},\ }\href {https://doi.org/10.1016/j.actamat.2006.01.041}
  {\bibfield  {journal} {\bibinfo  {journal} {Acta Materialia}\ }\textbf
  {\bibinfo {volume} {54}},\ \bibinfo {pages} {3647} (\bibinfo {year}
  {2006})}\BibitemShut {NoStop}%
\bibitem [{\citenamefont {Arnoult}\ and\ \citenamefont
  {McLellan}(1973)}]{Arnoult1973}%
  \BibitemOpen
  \bibfield  {author} {\bibinfo {author} {\bibfnamefont {W.~J.}\ \bibnamefont
  {Arnoult}}\ and\ \bibinfo {author} {\bibfnamefont {R.~B.}\ \bibnamefont
  {McLellan}},\ }\bibfield  {title} {\bibinfo {title} {{Thermodynamics of
  transition metal-hydrogen solid solutions}},\ }\href
  {https://doi.org/10.1016/0001-6160(73)90089-8} {\bibfield  {journal}
  {\bibinfo  {journal} {Acta Metallurgica}\ }\textbf {\bibinfo {volume} {21}},\
  \bibinfo {pages} {1397} (\bibinfo {year} {1973})}\BibitemShut {NoStop}%
\bibitem [{\citenamefont {Vykhodets}\ \emph {et~al.}(2005)\citenamefont
  {Vykhodets}, \citenamefont {Kurennykh}, \citenamefont {Lakhtin},
  \citenamefont {Pastukhov},\ and\ \citenamefont {Fishman}}]{vykhodets2005}%
  \BibitemOpen
  \bibfield  {author} {\bibinfo {author} {\bibfnamefont {V.}~\bibnamefont
  {Vykhodets}}, \bibinfo {author} {\bibfnamefont {T.}~\bibnamefont
  {Kurennykh}}, \bibinfo {author} {\bibfnamefont {A.}~\bibnamefont {Lakhtin}},
  \bibinfo {author} {\bibfnamefont {E.~A.}\ \bibnamefont {Pastukhov}},\ and\
  \bibinfo {author} {\bibfnamefont {A.~Y.}\ \bibnamefont {Fishman}},\
  }\bibfield  {title} {\bibinfo {title} {Activation energy of hydrogen, oxygen
  and nitrogen diffusion in metals},\ }\href@noop {} {\bibfield  {journal}
  {\bibinfo  {journal} {Doklady Akademii Nauk-Rossijskaya Akademiya Nauk}\
  }\textbf {\bibinfo {volume} {401}},\ \bibinfo {pages} {772} (\bibinfo {year}
  {2005})}\BibitemShut {NoStop}%
\bibitem [{\citenamefont {Ward}(1963)}]{ward1963}%
  \BibitemOpen
  \bibfield  {author} {\bibinfo {author} {\bibfnamefont {J.~H.}\ \bibnamefont
  {Ward}},\ }\bibfield  {title} {\bibinfo {title} {{Hierarchical Grouping to
  Optimize an Objective Function}},\ }\href
  {https://doi.org/10.1080/01621459.1963.10500845} {\bibfield  {journal}
  {\bibinfo  {journal} {Journal of the American Statistical Association}\
  }\textbf {\bibinfo {volume} {58}},\ \bibinfo {pages} {236} (\bibinfo {year}
  {1963})}\BibitemShut {NoStop}%
\bibitem [{\citenamefont {Pedregosa}\ \emph {et~al.}(2011)\citenamefont
  {Pedregosa}, \citenamefont {Varoquaux}, \citenamefont {Gramfort},
  \citenamefont {Michel}, \citenamefont {Thirion}, \citenamefont {Grisel},
  \citenamefont {Blondel}, \citenamefont {Prettenhofer}, \citenamefont {Weiss},
  \citenamefont {Dubourg}, \citenamefont {Vanderplas}, \citenamefont {Passos},
  \citenamefont {Cournapeau}, \citenamefont {Brucher}, \citenamefont {Perrot},\
  and\ \citenamefont {Duchesnay}}]{scikit-learn}%
  \BibitemOpen
  \bibfield  {author} {\bibinfo {author} {\bibfnamefont {F.}~\bibnamefont
  {Pedregosa}}, \bibinfo {author} {\bibfnamefont {G.}~\bibnamefont
  {Varoquaux}}, \bibinfo {author} {\bibfnamefont {A.}~\bibnamefont {Gramfort}},
  \bibinfo {author} {\bibfnamefont {V.}~\bibnamefont {Michel}}, \bibinfo
  {author} {\bibfnamefont {B.}~\bibnamefont {Thirion}}, \bibinfo {author}
  {\bibfnamefont {O.}~\bibnamefont {Grisel}}, \bibinfo {author} {\bibfnamefont
  {M.}~\bibnamefont {Blondel}}, \bibinfo {author} {\bibfnamefont
  {P.}~\bibnamefont {Prettenhofer}}, \bibinfo {author} {\bibfnamefont
  {R.}~\bibnamefont {Weiss}}, \bibinfo {author} {\bibfnamefont
  {V.}~\bibnamefont {Dubourg}}, \bibinfo {author} {\bibfnamefont
  {J.}~\bibnamefont {Vanderplas}}, \bibinfo {author} {\bibfnamefont
  {A.}~\bibnamefont {Passos}}, \bibinfo {author} {\bibfnamefont
  {D.}~\bibnamefont {Cournapeau}}, \bibinfo {author} {\bibfnamefont
  {M.}~\bibnamefont {Brucher}}, \bibinfo {author} {\bibfnamefont
  {M.}~\bibnamefont {Perrot}},\ and\ \bibinfo {author} {\bibfnamefont
  {E.}~\bibnamefont {Duchesnay}},\ }\bibfield  {title} {\bibinfo {title}
  {Scikit-learn: Machine learning in {P}ython},\ }\href@noop {} {\bibfield
  {journal} {\bibinfo  {journal} {Journal of Machine Learning Research}\
  }\textbf {\bibinfo {volume} {12}},\ \bibinfo {pages} {2825} (\bibinfo {year}
  {2011})}\BibitemShut {NoStop}%
\bibitem [{\citenamefont {Cover}\ and\ \citenamefont {Hart}(1967)}]{Cover1967}%
  \BibitemOpen
  \bibfield  {author} {\bibinfo {author} {\bibfnamefont {T.}~\bibnamefont
  {Cover}}\ and\ \bibinfo {author} {\bibfnamefont {P.}~\bibnamefont {Hart}},\
  }\bibfield  {title} {\bibinfo {title} {{Nearest neighbor pattern
  classification}},\ }\href {https://doi.org/10.1109/TIT.1967.1053964}
  {\bibfield  {journal} {\bibinfo  {journal} {IEEE Transactions on Information
  Theory}\ }\textbf {\bibinfo {volume} {13}},\ \bibinfo {pages} {21} (\bibinfo
  {year} {1967})}\BibitemShut {NoStop}%
\bibitem [{\citenamefont {Altman}(1992)}]{Altman1992}%
  \BibitemOpen
  \bibfield  {author} {\bibinfo {author} {\bibfnamefont {N.~S.}\ \bibnamefont
  {Altman}},\ }\bibfield  {title} {\bibinfo {title} {{An Introduction to Kernel
  and Nearest-Neighbor Nonparametric Regression}},\ }\href
  {https://doi.org/10.1080/00031305.1992.10475879} {\bibfield  {journal}
  {\bibinfo  {journal} {The American Statistician}\ }\textbf {\bibinfo {volume}
  {46}},\ \bibinfo {pages} {175} (\bibinfo {year} {1992})}\BibitemShut
  {NoStop}%
\bibitem [{\citenamefont {Murphy}(2012)}]{murphy2012}%
  \BibitemOpen
  \bibfield  {author} {\bibinfo {author} {\bibfnamefont {K.~P.}\ \bibnamefont
  {Murphy}},\ }\href@noop {} {\emph {\bibinfo {title} {{Machine learning : a
  probabilistic perspective}}}}\ (\bibinfo  {publisher} {MIT Press},\ \bibinfo
  {address} {Cambridge, MA},\ \bibinfo {year} {2012})\BibitemShut {NoStop}%
\bibitem [{\citenamefont {Rasmussen}\ and\ \citenamefont
  {Williams}(2006)}]{Rasmussen2006}%
  \BibitemOpen
  \bibfield  {author} {\bibinfo {author} {\bibfnamefont {C.~E.}\ \bibnamefont
  {Rasmussen}}\ and\ \bibinfo {author} {\bibfnamefont {C.~K.~I.}\ \bibnamefont
  {Williams}},\ }\href@noop {} {\emph {\bibinfo {title} {Lecture Notes in
  Computer Science (including subseries Lecture Notes in Artificial
  Intelligence and Lecture Notes in Bioinformatics)}}},\ Vol.\ \bibinfo
  {volume} {3176}\ (\bibinfo  {publisher} {the MIT Press},\ \bibinfo {year}
  {2006})\BibitemShut {NoStop}%
\bibitem [{\citenamefont {Deringer}\ \emph {et~al.}(2021)\citenamefont
  {Deringer}, \citenamefont {Bart{\'{o}}k}, \citenamefont {Bernstein},
  \citenamefont {Wilkins}, \citenamefont {Ceriotti},\ and\ \citenamefont
  {Cs{\'{a}}nyi}}]{Deringer2021}%
  \BibitemOpen
  \bibfield  {author} {\bibinfo {author} {\bibfnamefont {V.~L.}\ \bibnamefont
  {Deringer}}, \bibinfo {author} {\bibfnamefont {A.~P.}\ \bibnamefont
  {Bart{\'{o}}k}}, \bibinfo {author} {\bibfnamefont {N.}~\bibnamefont
  {Bernstein}}, \bibinfo {author} {\bibfnamefont {D.~M.}\ \bibnamefont
  {Wilkins}}, \bibinfo {author} {\bibfnamefont {M.}~\bibnamefont {Ceriotti}},\
  and\ \bibinfo {author} {\bibfnamefont {G.}~\bibnamefont {Cs{\'{a}}nyi}},\
  }\bibfield  {title} {\bibinfo {title} {{Gaussian Process Regression for
  Materials and Molecules}},\ }\href
  {https://doi.org/10.1021/acs.chemrev.1c00022} {\bibfield  {journal} {\bibinfo
   {journal} {Chemical Reviews}\ }\textbf {\bibinfo {volume} {121}},\ \bibinfo
  {pages} {10073} (\bibinfo {year} {2021})}\BibitemShut {NoStop}%
\bibitem [{\citenamefont {Breiman}(2001)}]{Breiman2001}%
  \BibitemOpen
  \bibfield  {author} {\bibinfo {author} {\bibfnamefont {L.}~\bibnamefont
  {Breiman}},\ }\bibfield  {title} {\bibinfo {title} {{Random Forests}},\
  }\href {https://doi.org/https://doi.org/10.1023/A:1010933404324} {\bibfield
  {journal} {\bibinfo  {journal} {Machine Learning}\ }\textbf {\bibinfo
  {volume} {45}},\ \bibinfo {pages} {5} (\bibinfo {year} {2001})}\BibitemShut
  {NoStop}%
\bibitem [{\citenamefont {Friedman}(2001)}]{Friedman2001}%
  \BibitemOpen
  \bibfield  {author} {\bibinfo {author} {\bibfnamefont {J.~H.}\ \bibnamefont
  {Friedman}},\ }\bibfield  {title} {\bibinfo {title} {{Greedy function
  approximation: A gradient boosting machine.}},\ }\bibfield  {journal}
  {\bibinfo  {journal} {The Annals of Statistics}\ }\textbf {\bibinfo {volume}
  {29}},\ \href {https://doi.org/10.1214/aos/1013203451}
  {10.1214/aos/1013203451} (\bibinfo {year} {2001})\BibitemShut {NoStop}%
\bibitem [{\citenamefont {Friedman}(2002)}]{Friedman2002}%
  \BibitemOpen
  \bibfield  {author} {\bibinfo {author} {\bibfnamefont {J.~H.}\ \bibnamefont
  {Friedman}},\ }\bibfield  {title} {\bibinfo {title} {{Stochastic gradient
  boosting}},\ }\href {https://doi.org/10.1016/S0167-9473(01)00065-2}
  {\bibfield  {journal} {\bibinfo  {journal} {Computational Statistics \& Data
  Analysis}\ }\textbf {\bibinfo {volume} {38}},\ \bibinfo {pages} {367}
  (\bibinfo {year} {2002})}\BibitemShut {NoStop}%
\bibitem [{\citenamefont {Zou}\ and\ \citenamefont {Hastie}(2005)}]{Zou2005}%
  \BibitemOpen
  \bibfield  {author} {\bibinfo {author} {\bibfnamefont {H.}~\bibnamefont
  {Zou}}\ and\ \bibinfo {author} {\bibfnamefont {T.}~\bibnamefont {Hastie}},\
  }\bibfield  {title} {\bibinfo {title} {{Regularization and variable selection
  via the elastic net}},\ }\href
  {https://doi.org/10.1111/j.1467-9868.2005.00503.x} {\bibfield  {journal}
  {\bibinfo  {journal} {Journal of the Royal Statistical Society: Series B
  (Statistical Methodology)}\ }\textbf {\bibinfo {volume} {67}},\ \bibinfo
  {pages} {301} (\bibinfo {year} {2005})}\BibitemShut {NoStop}%
\bibitem [{\citenamefont {Tipping}(2001)}]{Tipping2001}%
  \BibitemOpen
  \bibfield  {author} {\bibinfo {author} {\bibfnamefont {M.}~\bibnamefont
  {Tipping}},\ }\bibfield  {title} {\bibinfo {title} {{Sparse Bayesian learning
  and the relevance vector machine}},\ }\href@noop {} {\bibfield  {journal}
  {\bibinfo  {journal} {Journal of Machine Learning Research}\ }\textbf
  {\bibinfo {volume} {1}},\ \bibinfo {pages} {211} (\bibinfo {year}
  {2001})}\BibitemShut {NoStop}%
\bibitem [{\citenamefont {Lu}\ \emph {et~al.}(2019)\citenamefont {Lu},
  \citenamefont {Zou}, \citenamefont {Jacobs}, \citenamefont {Afflerbach},
  \citenamefont {Lu},\ and\ \citenamefont {Morgan}}]{Lu2019}%
  \BibitemOpen
  \bibfield  {author} {\bibinfo {author} {\bibfnamefont {H.-J.}\ \bibnamefont
  {Lu}}, \bibinfo {author} {\bibfnamefont {N.}~\bibnamefont {Zou}}, \bibinfo
  {author} {\bibfnamefont {R.}~\bibnamefont {Jacobs}}, \bibinfo {author}
  {\bibfnamefont {B.}~\bibnamefont {Afflerbach}}, \bibinfo {author}
  {\bibfnamefont {X.-G.}\ \bibnamefont {Lu}},\ and\ \bibinfo {author}
  {\bibfnamefont {D.}~\bibnamefont {Morgan}},\ }\bibfield  {title} {\bibinfo
  {title} {{Error assessment and optimal cross-validation approaches in machine
  learning applied to impurity diffusion}},\ }\href
  {https://doi.org/10.1016/j.commatsci.2019.06.010} {\bibfield  {journal}
  {\bibinfo  {journal} {Computational Materials Science}\ }\textbf {\bibinfo
  {volume} {169}},\ \bibinfo {pages} {109075} (\bibinfo {year}
  {2019})}\BibitemShut {NoStop}%
\bibitem [{Note1()}]{Note1}%
  \BibitemOpen
  \bibinfo {note} {See Supplemental Material at [URL will be inserted by
  publisher] for hyperparameter sensitivity analysis, linear models for
  deuterium and tritium activation energies, and the grouping analysis applied
  to the SHAP values}\BibitemShut {NoStop}%
\bibitem [{\citenamefont {Janotti}\ \emph {et~al.}(2004)\citenamefont
  {Janotti}, \citenamefont {Kr{\v{c}}mar}, \citenamefont {Fu},\ and\
  \citenamefont {Reed}}]{Janotti2004}%
  \BibitemOpen
  \bibfield  {author} {\bibinfo {author} {\bibfnamefont {A.}~\bibnamefont
  {Janotti}}, \bibinfo {author} {\bibfnamefont {M.}~\bibnamefont
  {Kr{\v{c}}mar}}, \bibinfo {author} {\bibfnamefont {C.~L.}\ \bibnamefont
  {Fu}},\ and\ \bibinfo {author} {\bibfnamefont {R.~C.}\ \bibnamefont {Reed}},\
  }\bibfield  {title} {\bibinfo {title} {{Solute Diffusion in Metals: Larger
  Atoms Can Move Faster}},\ }\href
  {https://doi.org/10.1103/PhysRevLett.92.085901} {\bibfield  {journal}
  {\bibinfo  {journal} {Physical Review Letters}\ }\textbf {\bibinfo {volume}
  {92}},\ \bibinfo {pages} {085901} (\bibinfo {year} {2004})}\BibitemShut
  {NoStop}%
\bibitem [{\citenamefont {Bishop}(2006)}]{bishop2006}%
  \BibitemOpen
  \bibfield  {author} {\bibinfo {author} {\bibfnamefont {C.~M.}\ \bibnamefont
  {Bishop}},\ }\bibfield  {title} {\bibinfo {title} {{Bayesian Linear
  Regression}},\ }in\ \href@noop {} {\emph {\bibinfo {booktitle} {Pattern
  Recognition and Machine Learning}}}\ (\bibinfo  {publisher} {Springer New
  York, NY},\ \bibinfo {address} {New York},\ \bibinfo {year} {2006})\ \bibinfo
  {edition} {1st}\ ed.,\ Chap.~\bibinfo {chapter} {3}, pp.\ \bibinfo {pages}
  {152--159}\BibitemShut {NoStop}%
\bibitem [{\citenamefont {Lundberg}\ and\ \citenamefont
  {Lee}(2017)}]{NIPS2017_7062}%
  \BibitemOpen
  \bibfield  {author} {\bibinfo {author} {\bibfnamefont {S.~M.}\ \bibnamefont
  {Lundberg}}\ and\ \bibinfo {author} {\bibfnamefont {S.-I.}\ \bibnamefont
  {Lee}},\ }\bibfield  {title} {\bibinfo {title} {A unified approach to
  interpreting model predictions},\ }in\ \href
  {http://papers.nips.cc/paper/7062-a-unified-approach-to-interpreting-model-predictions.pdf}
  {\emph {\bibinfo {booktitle} {Advances in Neural Information Processing
  Systems 30}}},\ \bibinfo {editor} {edited by\ \bibinfo {editor}
  {\bibfnamefont {I.}~\bibnamefont {Guyon}}, \bibinfo {editor} {\bibfnamefont
  {U.~V.}\ \bibnamefont {Luxburg}}, \bibinfo {editor} {\bibfnamefont
  {S.}~\bibnamefont {Bengio}}, \bibinfo {editor} {\bibfnamefont
  {H.}~\bibnamefont {Wallach}}, \bibinfo {editor} {\bibfnamefont
  {R.}~\bibnamefont {Fergus}}, \bibinfo {editor} {\bibfnamefont
  {S.}~\bibnamefont {Vishwanathan}},\ and\ \bibinfo {editor} {\bibfnamefont
  {R.}~\bibnamefont {Garnett}}}\ (\bibinfo  {publisher} {Curran Associates,
  Inc.},\ \bibinfo {year} {2017})\ pp.\ \bibinfo {pages}
  {4765--4774}\BibitemShut {NoStop}%
\bibitem [{\citenamefont {de~Boer}\ \emph {et~al.}(1988)\citenamefont
  {de~Boer}, \citenamefont {Boom}, \citenamefont {Mattens}, \citenamefont
  {Miedema},\ and\ \citenamefont {Niessen}}]{DeBoer1988}%
  \BibitemOpen
  \bibfield  {author} {\bibinfo {author} {\bibfnamefont {F.~R.}\ \bibnamefont
  {de~Boer}}, \bibinfo {author} {\bibfnamefont {R.}~\bibnamefont {Boom}},
  \bibinfo {author} {\bibfnamefont {W.}~\bibnamefont {Mattens}}, \bibinfo
  {author} {\bibfnamefont {A.~R.}\ \bibnamefont {Miedema}},\ and\ \bibinfo
  {author} {\bibfnamefont {A.~K.}\ \bibnamefont {Niessen}},\ }\href@noop {}
  {\emph {\bibinfo {title} {{Cohesion in Metals: Transition Metal Alloys
  (Cohesion and Structure)}}}},\ edited by\ \bibinfo {editor} {\bibfnamefont
  {F.}~\bibnamefont {de~Boer}}\ and\ \bibinfo {editor} {\bibfnamefont
  {D.}~\bibnamefont {Pettifor}}\ (\bibinfo  {publisher} {North-Holland},\
  \bibinfo {address} {New York},\ \bibinfo {year} {1988})\BibitemShut {NoStop}%
\bibitem [{\citenamefont {Bouten}\ and\ \citenamefont
  {Miedema}(1980)}]{Bouten1980}%
  \BibitemOpen
  \bibfield  {author} {\bibinfo {author} {\bibfnamefont {P.}~\bibnamefont
  {Bouten}}\ and\ \bibinfo {author} {\bibfnamefont {A.}~\bibnamefont
  {Miedema}},\ }\bibfield  {title} {\bibinfo {title} {{On the heats of
  formation of the binary hydrides of transition metals}},\ }\href
  {https://doi.org/10.1016/0022-5088(80)90110-1} {\bibfield  {journal}
  {\bibinfo  {journal} {Journal of the Less Common Metals}\ }\textbf {\bibinfo
  {volume} {71}},\ \bibinfo {pages} {147} (\bibinfo {year} {1980})}\BibitemShut
  {NoStop}%
\bibitem [{\citenamefont {{Van Mal}}\ \emph {et~al.}(1974)\citenamefont {{Van
  Mal}}, \citenamefont {Buschow},\ and\ \citenamefont {Miedema}}]{VanMal1974}%
  \BibitemOpen
  \bibfield  {author} {\bibinfo {author} {\bibfnamefont {H.}~\bibnamefont {{Van
  Mal}}}, \bibinfo {author} {\bibfnamefont {K.}~\bibnamefont {Buschow}},\ and\
  \bibinfo {author} {\bibfnamefont {A.}~\bibnamefont {Miedema}},\ }\bibfield
  {title} {\bibinfo {title} {{Hydrogen absorption in LaNi5 and related
  compounds: Experimental observations and their explanation}},\ }\href
  {https://doi.org/10.1016/0022-5088(74)90146-5} {\bibfield  {journal}
  {\bibinfo  {journal} {Journal of the Less Common Metals}\ }\textbf {\bibinfo
  {volume} {35}},\ \bibinfo {pages} {65} (\bibinfo {year} {1974})}\BibitemShut
  {NoStop}%
\bibitem [{\citenamefont {Miedema}\ \emph {et~al.}(1976)\citenamefont
  {Miedema}, \citenamefont {Buschow},\ and\ \citenamefont {{Van
  Mal}}}]{Miedema1976}%
  \BibitemOpen
  \bibfield  {author} {\bibinfo {author} {\bibfnamefont {A.}~\bibnamefont
  {Miedema}}, \bibinfo {author} {\bibfnamefont {K.}~\bibnamefont {Buschow}},\
  and\ \bibinfo {author} {\bibfnamefont {H.}~\bibnamefont {{Van Mal}}},\
  }\bibfield  {title} {\bibinfo {title} {{Which intermetallic compounds of
  transition metals form stable hydrides?}},\ }\href
  {https://doi.org/10.1016/0022-5088(76)90057-6} {\bibfield  {journal}
  {\bibinfo  {journal} {Journal of the Less Common Metals}\ }\textbf {\bibinfo
  {volume} {49}},\ \bibinfo {pages} {463} (\bibinfo {year} {1976})}\BibitemShut
  {NoStop}%
\bibitem [{\citenamefont {Herbst}(2002)}]{Herbst2002}%
  \BibitemOpen
  \bibfield  {author} {\bibinfo {author} {\bibfnamefont {J.}~\bibnamefont
  {Herbst}},\ }\bibfield  {title} {\bibinfo {title} {{On extending Miedema's
  model to predict hydrogen content in binary and ternary hydrides}},\ }\href
  {https://doi.org/10.1016/S0925-8388(01)01939-9} {\bibfield  {journal}
  {\bibinfo  {journal} {Journal of Alloys and Compounds}\ }\textbf {\bibinfo
  {volume} {337}},\ \bibinfo {pages} {99} (\bibinfo {year} {2002})}\BibitemShut
  {NoStop}%
\end{thebibliography}

\newpage


\section*{Supplementary material}

# Explainable Machine Learning for Hydrogen Diffusion in Metals and Random Binary Alloys: Supplemental Information

Grace M. Lu,[1] Matthew Witman,[2] Sapan Agarwal,[2] Vitalie Stavila,[2] and Dallas R. Trinkle[1, *]

[1]*Department of Materials Science and Engineering,*
*University of Illinois at Urbana-Champaign, Urbana, Illinois 61801, USA*

[2]*Sandia National Laboratories, Livermore, California 94551, USA*

(Dated: September 25, 2023)

# Abstract

Section S1 contains a sensitivity analysis of the hyperparameters used in this paper. Section S1 contains the results of a hyperparameter sensitivity test, and Section S2 contains the hyperparameter bias analysis. S3 contains linear fits with one feature for the difference in activation energies for deuterium and tritium diffusion compared to hydrogen. Section S4 contains the SHAP values, another metric of feature analysis, and shows the grouping analysis via feature correlations applied to the SHAP values.

Keywords: hydrogen diffusion; machine-learning; metals; binary alloys

## S1. HYPERPARAMETER SENSITIVITY ANALYSIS

To determine the sensitivity of our results to the chosen hyperparameters, we decrease or increase each hyperparameter by a single step (the step sizes are reported in Table 3 in the main text) and refit our models. The RMSEs reported here are the on the test data for the alloy split. All hyperparameters, except the learning rate for the gradient boosting tree, had a difference of the RMSE of less than 10 meV. By reducing the learning rate by 0.1, the RMSE also decreased

TABLE SI. Sensitivity analysis for the hyperparameters. For each RMSE reported, the hyperparameter was decreased or increased by a single step, and the models were refit. The RMSE reported is for test data for the alloy split.

| Model | Hyperparameter | Step Down (meV) | Optimal (meV) | Step Up (meV) |
|---|---|---|---|---|
| K-Nearest Neighbors | Neighbor of Neighbors | – | 117 | 110 |
|  | Weighting Scheme | – | 117 | 117 |
| Random Forest | Number of Estimators | 119 | 119 | 119 |
|  | Maximum Depth | 120 | 119 | 119 |
| Gradient Boosting Tree | Learning Rate | 100 | 111 | 119 |
|  | Number of Estimators | 104 | 111 | 111 |
|  | Maximum Depth | 104 | 111 | 107 |
| Positive ElasticNet | Ridge Penalty $\alpha$ | 97 | 98 | 101 |
|  | $l_1$ ratio | 98 | 98 | 101 |

---

* dtrinkle@illinois.edu

by 11 meV. This is likely because the learning rate is the hyperparameter that controls the level of overfitting.

## S2. HYPERPARAMETER BIAS ANALYSIS

We also find that using the same data for selecting hyperparameters and testing the performance of our models had a negligible effect on the RMSEs as shown in the table below. This implies that there is little bias, and our models are not overfitting the hyperparameters to the individual test sets used. We calculated these values by using a nested cross-validation scheme where we removed a single alloy group from the dataset, then using 5-fold cross-validation (80-20% random split) to choose the hyperparameters. Note that the Bayesian Ridge model is not included because it has no hyperparameters. The final RMSEs reported below are the RMSEs calculated solely from data not used to choose the hyperparameters. We found that the unbiased RMSEs were, at most, 7 meV larger than the ones reported in the paper, implying that little to no positive bias or overfitting of the hyperparameters is present for our models. This is likely because the number of hyperparameters is low, and our results match with previous work done by Wainer and Cawlyer [1] that has found that nested cross-validation (with no repeated data used to determine the hyperparameter and the performance on the test sets) is overzealous compared to the much cheaper, flat cross-validation (where the same data is used to determine the best hyperparameters and the performance on the test set) for classification problems.

TABLE SII. Original RMSEs for the alloy split compared to unbiased nested cross-validation RMSEs. The nested RMSEs were calculated solely on data not used to fit the hyperparameters for the models and differed by at most 7 meV from the RMSEs obtained from the alloy split implying that little overfitting of the hyperparameters is present.

| Model | Original RMSE | Nested RMSE |
|---|---|---|
| K-Nearest Neighbors | 117 | 116 |
| Gaussian Process | 106 | 105 |
| Random Forest | 119 | 112 |
| Gradient Boosting Tree | 111 | 115 |
| Positive ElasticNet | 98 | 100 |

## S3. DEUTERIUM AND TRITIUM DIFFUSION

The isotope effect on the activation energies is significantly smaller than the RMSE on the machine learning models, so we fit linear models in Fig. S1 with only the feature with the strongest correlation to the difference between the activation energies. The 23 entries of deuterium come from 13 elements (Cr [2], Cu [3], Fe [4, 5], Hf [6], Lu [7], Nb [8–11], Ni [12, 13], Pd [14, 15], Ta [8, 9, 11], Ti [16], V [8, 10, 11], Y [17], and Zr [18]), and Nb-V [10] and Ti-V binary alloys [10]. The tritium data data was much more limited with only 8 elemental entries (Cu [12], Hf [19], Nb [11], Ni [12], Ta [11], Ti [19], V [11], and Y [20]) and no alloy data. While the linear models for both models are poor and have $R^2$ values smaller than 0.75 due to the lack of data, the most correlated features to the isotope differences are the ground state volume $\sigma$ and the electronegativity difference respectively.

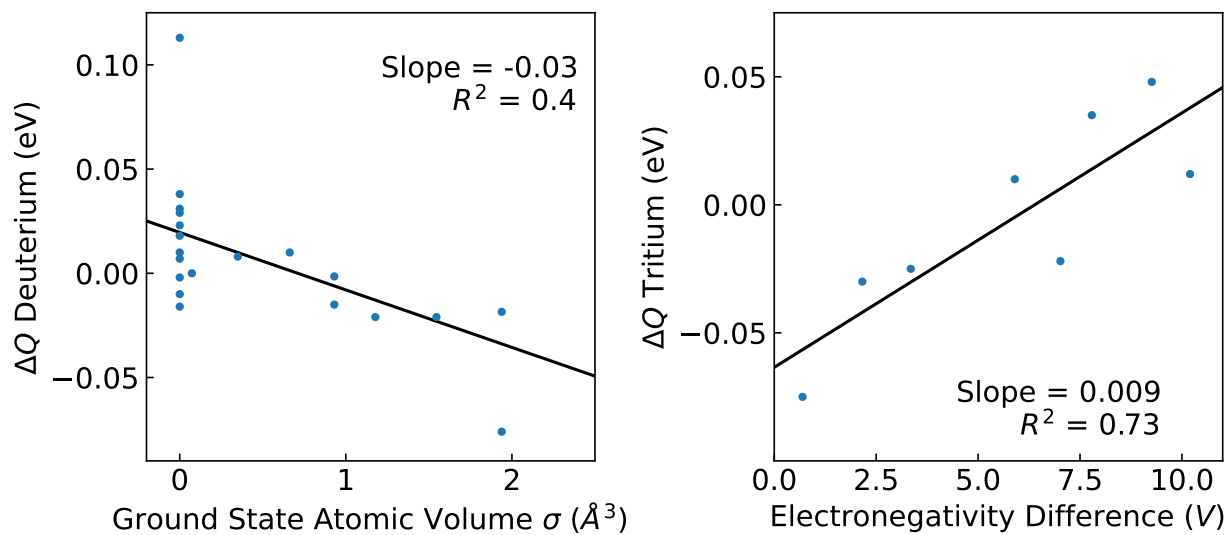

FIG. S1. Linear models for the difference in activation energy between deuterium (left) or tritium (right) with hydrogen. Only the single feature with the strongest correlation to the isotope differences is used to fit these models.

## S4. SHAP IMPORTANCE

In Fig. S2, the Shapley additive explanations (SHAP) values serve as a different metric of the feature importances, which measures how much each feature directly impacts the prediction. The SHAP values [21] are calculated by fitting a simpler model that is an interpretable, additive

4

approximation of the original model. These values measure how much each prediction changes after training with a feature and take into account multicollinearity by taking into account all possible subsets of features. The SHAP values for the packing factor, for all except the *k*-nearest neighbors model, show that the packing factor by itself entails a 0.1 eV difference between the BCC and close-packed metals, and that it is strongly important for all database entries.

In Fig. S3, the SHAP values can undergo the same grouping analysis as in Fig. 5 in the main text, as the permutation importances and lead to similar relative rankings but a lower level of sparsity. In Fig. S2, the packing factor is the most important feature for all models except for the *k*-nearest neighbors where it is second. The top three features of the models contain a total of 9 different features, the same as the permutation importances from the main text -even though the SHAP values are more robust with regards to correlations between features because it uses combinations of features. As a result, the SHAP values, which required fitting multiple simpler models, took 1000 times longer to calculate than the permutation importances. Like the permutation importances, the individual average SHAP values are a poor measure of feature importance and are hard to interpret. The grouped SHAP values and the permutation importances agree that the packing factor and electronic specific heat groups are the two most important feature groups, but the SHAP values are spread out over more features. This may be due to the high amount of correlation between the features in our dataset. Because the permutation importance does not take correlations into account, the reported permutation importance will be smaller than the actual feature importance when similar information is being encoded by other features. This lack of sparsity led to increased importances mainly for the alloy group, which is the largest in terms of feature count.

---

[1] J. Wainer and G. Cawley, Nested cross-validation when selecting classifiers is overzealous for most practical applications., Expert Systems with Applications **182**, 10.1016/j.eswa.2021.115222 (2021).

[2] D. Stöver, H. Buchkremer, and R. Hecker, Hydrogen and deuterium permeation through metallic and surface-oxidized chromium, Surface and Coatings Technology **28**, 281 (1986).

[3] G. R. Caskey and R. G. Derrick, Hydrogen permeability through alpha-brass, Metallurgical Transactions A **8**, 511 (1977).

[4] W. Raczyński, Permeability, diffusivity, and solubility of hydrogen and deuterium in pure iron at 10 to

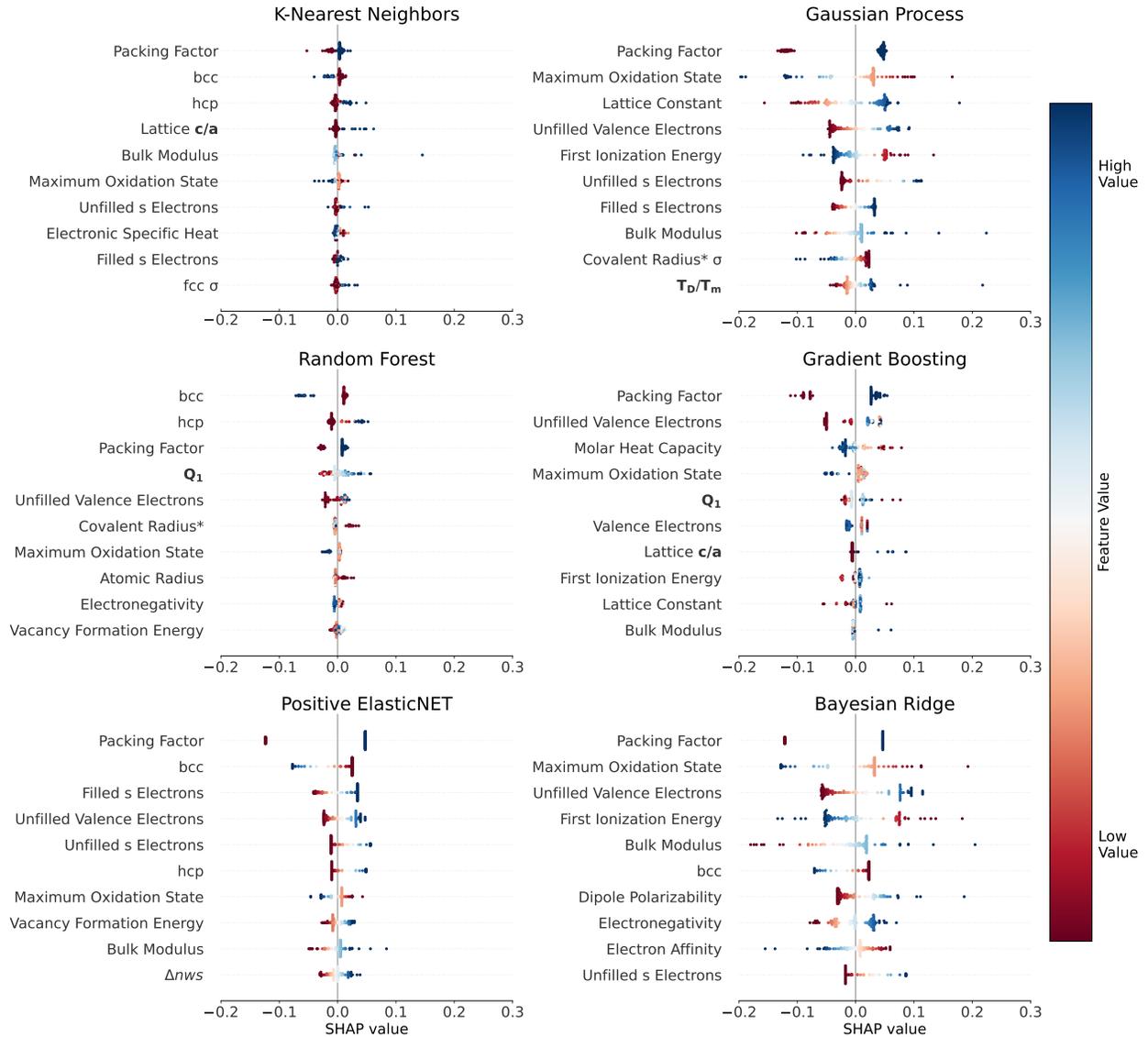

FIG. S2. 10 most important features based on the SHAP values ordered by average importance. For each row, we plot the SHAP value, or how much the overall predicted $Q$ changes, as a result of the feature value, shown via the colors. The packing factor is the most important feature for all six models, and for all except the $k$-nearest neighbor model, entails a 0.1 eV change in the activation energy by itself between close-packed and bcc crystal structures.

60 °C, Physica Status Solidi (a) **48**, K27 (1978).

[5] N. Quick and H. Johnson, Hydrogen and deuterium in iron, 49–506°C, Acta Metallurgica **26**, 903 (1978).

[6] S. Naito, M. Yamamoto, and T. Hashino, Solubility and diffusivity of hydrogen and deuterium in alpha

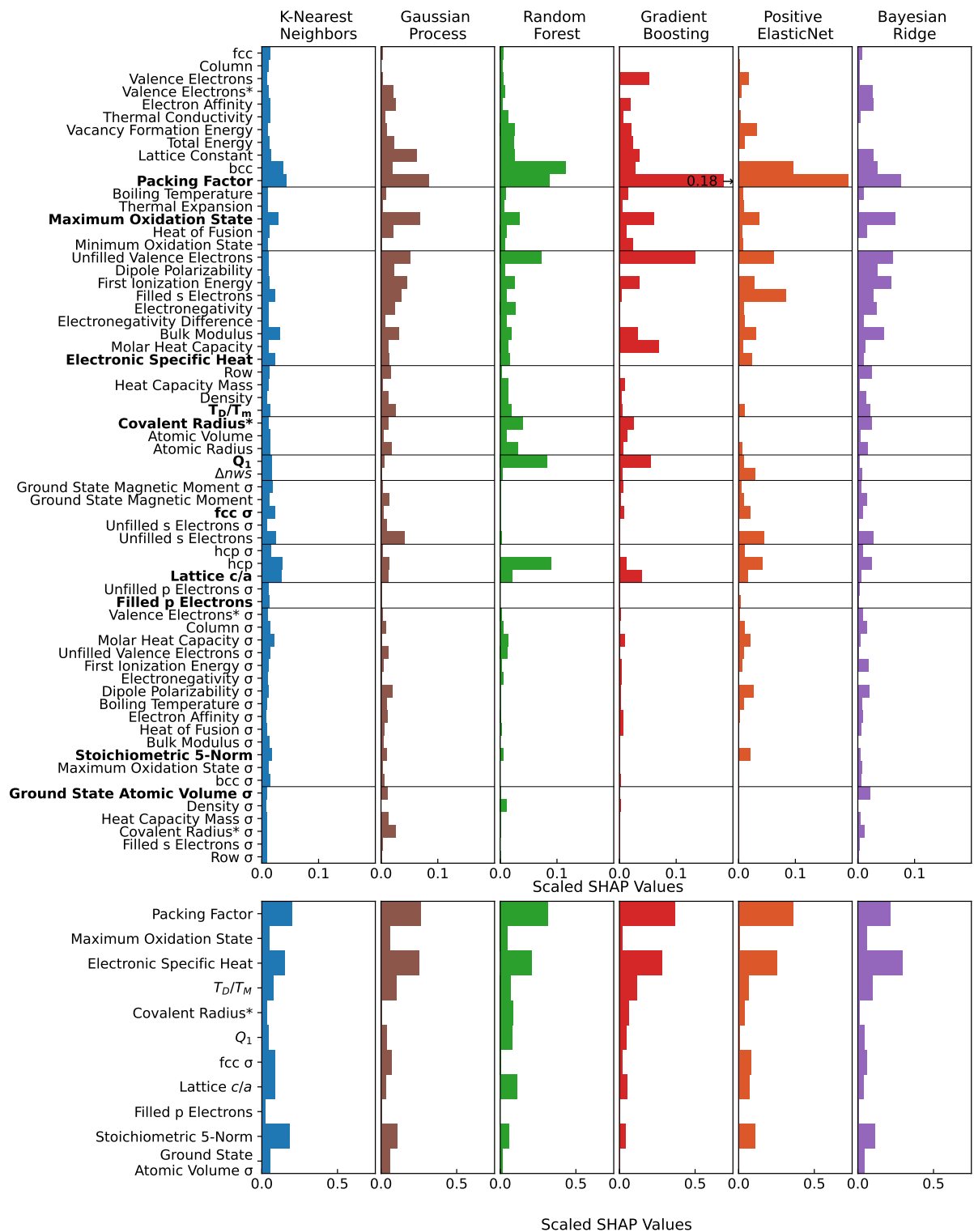

FIG. S3. Absolute values of the SHAP values for all features averaged over the database entries. While more features have a non-negligible importance compared to the permutation importance, the relative rankings of the feature groups remains similar. The alloy group has a larger importance here due to its large size coupled with the lower level of sparsity. The permutation importances are more sparse because the permutation importances for correlated features may be smaller than their true importances, as less information is lost when the values are shuffled. The SHAP values are more robust when working with correlated features

-hafnium at high temperatures, Journal of Physics: Condensed Matter **2**, 1963 (1990).

[7] J. Völkl, H. Wipf, B. J. Beaudry, and K. A. Gschneidner, Diffusion of H and D in Lutetium, physica status solidi (b) **144**, 315 (1987).

[8] G. Schaumann, J. Völkl, and G. Alefeld, The diffusion coefficients of hydrogen and deuterium in vanadium, niobium, and tantalum by gorsky-effect measurements, physica status solidi (b) **42**, 401 (1970).

[9] H. C. Bauer, J. Völkl, J. Tretkowski, and G. Alefeld, Diffusion of hydrogen and deuterium in Nb and Ta at high concentrations, Zeitschrift für Physik B Condensed Matter and Quanta **29**, 17 (1978).

[10] H. M. Herro, *Hydrogen and deuterium diffusion in vanadium alloys*, Ph.D. thesis, Iowa State University, Digital Repository, Ames (1982).

[11] Z. Qi, J. Völkl, R. Lasser, and H. Wenzl, Tritium diffusion in V, Nb and Ta, Journal of Physics F: Metal Physics **13**, 2053 (1983).

[12] L. Katz, M. Guinan, and R. J. Borg, Diffusion of H2, D2, and T2 in Single-Crystal Ni and Cu, Physical Review B **4**, 330 (1971).

[13] K. Yamakawa, Diffusion of Deuterium and Isotope Effect in Nickel, Journal of the Physical Society of Japan **47**, 114 (1979).

[14] W. Jost and A. Widmann, Über die Diffusion von Wasserstoff und von Deuterium in Palladium. II, Zeitschrift für Physikalische Chemie **45B**, 285 (1940).

[15] J. Völkl, G. Wollenweber, K.-H. Klatt, and G. Alefeld, Notizen: Reversed Isotope Dependence for Hydrogen Diffusion in Palladium, Zeitschrift für Naturforschung A **26**, 922 (1971).

[16] S. Naito, M. Yamamoto, M. Doi, and M. Kimura, High-Temperature Diffusion of Hydrogen and Deuterium in Titanium and Ti3Al, Journal of The Electrochemical Society **145**, 2471 (1998).

[17] T. Maeda, S. Naito, M. Yamamoto, M. Mabuchi, and T. Hashino, Diffusivity and solubility of hydrogen and deuterium in yttrium, Journal of the Chemical Society, Faraday Transactions **89**, 4375 (1993).

[18] E. A. Gulbransen and K. F. Andrew, Diffusion of Hydrogen and Deuterium in High Purity Zirconium, Journal of The Electrochemical Society **101**, 560 (1954).

[19] W. Kunz, H. Münzel, U. Helfrich, and H. Horneff, Bestimmung der Diffusionskoeffizienten für Tritium in den $\alpha$-Phasen von Titan und Hafnium, International Journal of Materials Research **74**, 289 (1983).

[20] R. E. Buxbaum and E. F. Johnson, Diffusivity of hydrogen isotopes in liquid lithium and in solid yttrium, Industrial & Engineering Chemistry Fundamentals **24**, 180 (1985).

[21] S. M. Lundberg and S.-I. Lee, A unified approach to interpreting model predictions, in *Advances in*

*Neural Information Processing Systems 30*, edited by I. Guyon, U. V. Luxburg, S. Bengio, H. Wallach, R. Fergus, S. Vishwanathan, and R. Garnett (Curran Associates, Inc., 2017) pp. 4765–4774.


\end{document}